\documentclass[PhD,english]{dfaunictthesis}
\usepackage{lipsum}
\usepackage[Lenny]{fncychap}
\usepackage[utf8]{inputenc}
\usepackage{amsmath}
\usepackage{physics}
\usepackage{amsthm}
\usepackage{amssymb}
\usepackage{amsfonts}
\usepackage{latexsym}
\usepackage{graphicx}
\usepackage{float}
\usepackage{cancel}
\usepackage{graphicx}
\usepackage{mathtools}
\usepackage{fancyhdr}
\usepackage{siunitx}
\usepackage{comment}
\usepackage{systeme}
\usepackage{bm}
\usepackage{tikz}
\usepackage{hyperref}
\usepackage{booktabs}
\usetikzlibrary{arrows,matrix}
\usepackage[scale=2]{ccicons}
\usepackage{lipsum}
\usepackage{braket}
\usepackage{bbold}
\usepackage[mathscr]{eucal}
\begin{document}
\author{Andrea Spina}
\title{Perturbations and phenomenology of compact objects in quantum gravity theories}
\aayear{2025-2026}

\begin{supervisors}
   \supervisor{Chiar.mo}{Prof.}{A. Bonanno}
   \supervisor{Chiar.mo}{Prof.}{M.L. Pumo}
\end{supervisors}


\phdname{physics} 

\maketitlepage

\tableofcontents

	\chapter*{List of publications}

    This Thesis is based on the following papers:
\begin{description}

\item[-] \cite{Bonanno:2025dry} A. Bonanno, R. A. Konoplya, G. Oglialoro, A. Spina, \textit{"Regular Black Holes from
Proper-Time Flow in Quantum Gravity and their Quasinormal Modes, Shadow and
Hawking Radiation"}, JCAP 12 (2025) 042,\\ arXiv:2509.12469 [gr-qc].

\item[-] \cite{Konoplya:2025afm} R. A. Konoplya, A. Spina, A. Zhidenko, \textit{"Time Evolution of Black Hole Perturbations
in Quadratic Gravity"}, Phys. Rev. D 112, 024060 (2025), arXiv:2505.01128 [gr-qc].

\item[-] \cite{Bonanno:2025ksz} A. Bonanno, S. Silveravalle, A. Spina, \textit{"Veiled Singularities in Einstein-Weyl Gravity:
Stability and Physical Interpretation of Horizonless Solutions"} , arXiv: 2511.10151
[gr-qc], Under review of Phys. Rev. D.

\item[-] \cite{Spina:2025wxb} A. Spina, \textit{"Black Holes in Asymptotic Safety: A Review of Solutions and Phenomenology"}, Theor. Phys. IJGTP 1, 8 (2025),\\ doi:10.53941/ijgtp.2025.100008,
arXiv:2510.14552 [gr-qc].

\end{description}
\newpage
Other works:
\begin{description}

\item[-] \cite{Spina:2024npx} A. Spina, S. Silveravalle, A. Bonanno, \textit{"Scalar Perturbations of Regular Black Holes
derived from a Non-Singular Collapse Model in Asymptotic Safety"}, in Proceedings
of the 17th Marcel Grossmann Meeting, doi:10.1142/14814, July 2026,arXiv:2410.05936 [gr-qc].

\end{description}

\chapter*{Abstract}

Understanding the behavior of gravity in the strong-curvature regime is one of the central challenges in the search for a consistent quantum theory of gravity. While General Relativity has achieved extraordinary observational success, its prediction of spacetime singularities and its perturbative non-renormalizability indicate that it cannot represent the ultimate description of gravitational interactions.

This thesis investigates the dynamics of perturbations and phenome\-nology of compact objects in two complementary approaches to quantum gravity: quadratic gravity and asymptotic safety. The central theme of this work is the use of linear perturbations as a probe of both the physical viability of quantum-corrected spacetimes and their potential observational signatures. Through the analysis of quasinormal modes, late-time dynamics, black hole shadows, and Hawking radiation, we explore how modifications of the underlying gravitational theory manifest themselves in astrophysically relevant observables.

The first part of the thesis is devoted to quadratic gravity, focusing in particular on the Einstein–Weyl sector. The theory contains an additional massive spin-2 degree of freedom, whose presence profoundly modifies the perturbative structure of compact objects. We derive and analyze the tensor perturbation equations for static and spherically symmetric solutions, including both black holes and horizonless compact configurations. In the monopole sector, the dynamics is governed by the massive mode and reveals the existence of Gregory–Laflamme-type instabilities for specific regions of the parameter space, providing important constraints on the physical viability of black holes. For higher multipoles, we study the coupled dynamics of the massless and massive gravitational sectors, obtaining the corresponding time-domain evolution.
Our results show that these geometries possess distinctive dynamical features, including the persistence of universal massive-field tails and a breakdown of the standard eikonal correspondence between quasinormal modes and null geodesics. Particular attention is also devoted to veiled singularities, a class of horizonless solutions satisfying a form of weak cosmic censorship. We investigate their physical interpretation and show that they can mimic several observational properties of black holes, including the existence of a shadow, while their perturbative dynamics raises novel challenges due to the singular structure of the effective potential.

The second part of the thesis focuses on black holes arising in the asymptotic safety program. Within this framework, quantum effects can regularize the spacetime geometry at high curvatures while preserving agreement with General Relativity at large distances. We construct and study a novel dynamical regular black hole solution generated by gravitational collapse within the proper-time flow approach. The resulting spacetime provides a self-consistent non-singular description of collapse and is shown to depend only weakly on gauge choices and parametrization ambiguities, demonstrating a remarkable degree of robustness. We derive the corresponding perturbation equations and perform a systematic analysis of their dynamical response. The resulting quasinormal mode spectrum exhibits characteristic deviations from the Schwarzschild case while remaining stable under tensor perturbations. We further investigate the observational consequences of these quantum-corrected geometries through the study of shadows, geodesic properties, and Hawking radiation. While shadow observables remain largely compatible with current experimental constraints, quantum corrections significantly modify the evaporation process, leading to lower temperatures and suppressed emission rates. These effects have direct implications for primordial black hole phenomenology and for their possible contribution to the dark matter content of the Universe.

Taken together, the results presented in this thesis show that perturbative dynamics constitutes a powerful bridge between fundamental theories of quantum gravity and observational astrophysics. 

\chapter{{Introduction}}

General Relativity (GR) is one of the most successful physical theories ever developed. It provides extremely accurate predictions for the motion of planets and stars and has successfully described phenomena that cannot be accounted for within Newtonian gravity, such as the bending of light, gravitational redshift, and the precession of the perihelion of Mercury. Moreover, Einstein’s theory predicts two fundamentally new phenomena: the existence of black holes \cite{Kerr:1963ud}, i.e. compact objects whose escape velocity exceeds the speed of light, and gravitational waves, whose direct detection has opened a new observational window on the Universe \cite{Misner:1973prb}. 

Despite its remarkable success, GR is a classical theory and exhibits fundamental limitations. In particular, it predicts the formation of spacetime singularities, both in the interior of black holes and at the beginning of the Universe \cite{Penrose:1964wq,Hawking:1970zqf,Hawking:1973uf}. In these regions, curvature invariants diverge and the predictive power of the theory breaks down, signaling the incompleteness of Einstein’s equations at very high energies or in regimes of extreme curvature. Furthermore, when treated as a quantum field theory, attempts to quantize General Relativity following the same perturbative procedures employed for the other fundamental interactions encounter severe difficulties. They lead to ultraviolet divergences that cannot be absorbed into a finite number of counterterms, preventing the construction of a predictive perturbative quantum theory, so that GR is perturbatively non-renormalizable \cite{Goroff:1985sz}. These issues suggest that General Relativity should be regarded as an effective low-energy description of gravity and consequently, strongly motivate the search for a consistent theory of quantum gravity, which is expected to provide a well-defined mechanism to resolve classical singularities and extend the validity of gravitational physics to the ultraviolet regime. A large variety of approaches have been proposed in order to extend our understanding of gravitational interactions beyond the classical regime. Examples include string theory \cite{Polchinski:1998rq}, loop quantum gravity \cite{Rovelli:1997yv}, asymptotic safety \cite{Weinberg:1979}, non-local gravity \cite{Modesto:2017sdr}, and higher-derivative extensions of Einstein gravity \cite{Stelle:1976gc}. Although these frameworks differ substantially in their fundamental assumptions, they share the common goal of providing a consistent ultraviolet completion of gravity and resolving the singularity problem.\\

Among the various approaches to quantum gravity listed above, an important direction is provided by extensions of GR including higher-order curvature corrections. In particular, quadratic gravity, obtained by adding curvature-squared terms to the Einstein-Hilbert action, represents the simplest theoretically consistent extension of GR and naturally arises as the leading correction in the effective action of several quantum gravity frameworks \cite{tHooft:1974toh,Benedetti:2013jk,Zwiebach:1985uq}. Remarkably, when interpreted as a fundamental quantum field theory, quadratic gravity is perturbatively renormalizable \cite{Stelle:1976gc}. However, this property comes at the price of a loss of unitarity, due to the presence of a massive spin-2 ghost associated with a wrong-sign kinetic term in the propagator and several hypotheses have been formulated to overcome this problem \cite{Platania:2020knd,Anselmi:2018tmf}. Despite this issue, quadratic gravity provides a rich phenomenology, admitting a variety of compact object solutions, including black holes and more exotic configurations \cite{Lu:2015psa}.\\

Another promising approach is offered by the asymptotic safety scenario \cite{Reuter:2019byg,Reuter:2001ag}, originally proposed by Weinberg. In this framework, gravity is described as a quantum field theory that remains well-defined at arbitrarily high energies due to the presence of a non-trivial ultraviolet fixed point \cite{Weinberg:1979}.Evidence for such a fixed point has been obtained within the functional renormalization group approach, particularly in four dimensions \cite{Reuter:1996cp}. Although a complete formulation is still under development, this non-perturbative scenario provides a consistent setting to explore quantum-corrected spacetimes \cite{Koch:2014cqa}. In particular, renormalization group improvements \cite{Bonanno:2000ep} of classical solutions often lead to geometries in which curvature singularities are smoothed out, giving rise to regular black hole models.\\

Given these theoretical developments, it becomes essential to identify observational signatures capable of distinguishing between different gravitational scenarios. The strong-field regime, realized in the vicinity of compact objects, with phenomena connected such as gravitational waves, shadows, accretion and emissions, represents a natural testing ground for this purpose. For many decades the study of strong gravitational fields remained largely theoretical, owing to the scarcity of observational data capable of probing the immediate vicinity of compact objects. This situation has dramatically changed in recent years.

The advent of gravitational-wave astronomy and horizon-scale imaging has inaugurated a new era of precision tests of gravity in the strong-field regime. Observations that were once considered inaccessible now provide direct information about the geometry of spacetime near compact objects and allow increasingly stringent tests of General Relativity and its alternatives.

Specifically, the detection of gravitational waves by the LIGO and Virgo collaborations \cite{LIGOScientific:2016aoc,LIGOScientific:2017vwq}, together with future experiments such as LISA \cite{LISA:2017pwj} and the Einstein Telescope, offers direct access to the dynamics of compact object mergers. At the same time, the imaging of black hole shadows by the Event Horizon Telescope \cite{EventHorizonTelescope:2019dse} provides complementary information on the geometry of spacetime near the horizon. 

In this context, the analysis of the ringdown phase—the stage following the merger of compact objects—is of particular interest. During this phase, the emitted gravitational signal is dominated by a superposition of damped oscillations known as quasinormal modes (QNM) \cite{Nollert:1999ji,Konoplya:2011qq,Kokkotas:1999bd,Berti_2009}. These modes depend exclusively on the parameters of the background spacetime and therefore has motivated the development of black-hole spectroscopy, whose objective is to infer the properties of compact objects directly from the frequencies and damping times of their oscillation modes, providing a direct probe of the underlying theory of gravity. 

More generally, linear perturbation theory offers a powerful framework to investigate both the stability and the dynamical properties of compact objects. The late-time evolution of perturbations is largely insensitive to the details of the initial excitation and is instead dictated by the geometry and the parameters of the background spacetime. After an initial transient, the system undergoes a damping oscillation dominated by QNM, which the real part of the complex frequency describes the actual oscillation frequency, while the imaginary part is the decay rate of the particular oscillation, followed at very late times by asymptotic tails.

Because of their universality, quasinormal modes can be regarded as footprint of a given solution, allowing for meaningful comparisons between theoretical predictions and observational data.
Additionally, studying QNM in alternative gravity theories or in the presence of quantum corrections can reveal potential deviations from classical black hole predictions. For these reasons, the study of QNMs in modified gravity and quantum-corrected scenarios has attracted considerable attention in recent years \cite{Bolokhov:2023ozp,Dubinsky:2024fvi,Malik:2024nhy,Konoplya:2024lch,Taylor:2024duw}.

In addition to their observational relevance, perturbations also provide essential information about the stability of compact objects, which is a necessary requirement for their physical viability \cite{Kodama:2003jz,Ishibashi:2003ap}. \\

 On the one hand, perturbation theory is employed to assess the stability of compact objects arising in quadratic gravity and to explore previously unexplored configurations such as veiled singularities. On the other hand, quantum-corrected solutions obtained within the asymptotic safety framework are constructed and analyzed through a variety of observables, including quasinormal modes, black-hole shadows, Hawking radiation and geodesic properties.

The overarching goal is to establish whether quantum-gravity effects can leave potentially observable imprints on astrophysical phenomena and to identify robust signatures capable of discriminating between different theoretical scenarios.

The primary objective of this thesis is to investigate how quantum-gravity inspired corrections affect the dynamics, stability and observational properties of compact objects. The work combines theoretical developments, numerical methods and phenomenological analyses and particular emphasis is placed on the role of linear perturbations as a diagnostic tool capable of simultaneously probing the mathematical consistency of a solution and its phenomenological implications.

We analyze stability through tensor perturbations in quadratic gravity, considering both black hole solutions \cite{Konoplya:2025afm} and more unconventional configurations such as naked (veiled) singularities \cite{Bonanno:2025ksz}. This problem has remained largely unexplored until recently due to the technical challenges associated with deriving suitable equations to study solutions and perturbation.

In parallel, we explore the asymptotic safety framework by constructing and studying a novel dynamical black hole solution obtained from gravitational collapse within the proper-time approach \cite{Bonanno:2025dry}. We then examine its phenomenological implications, including its quasinormal mode spectrum, Hawking radiation and his potential connections with primordial black holes and dark matter,  shadows, and ISCO (Innermost Stable Circular Orbit), with the aim of identifying observable deviations from classical predictions due to quantum corrections.\\

This thesis is organized as follows. In Chapter 2, we presents the general formalism of linear perturbations of compact objects, together with the numerical techniques employed in this work, such as WKB approximation, continued fraction method, time domain integration and Prony's method. Chapter 3 introduce quadratic gravity, Einstein Weyl restriction and review its main theoretical properties and known solutions. In Chapter 4, we perform a detailed perturbative analysis of solutions in quadratic gravity, with particular emphasis on the properties and stability of veiled singularities. Chapter 5 we introduce the asymptotic safety scenario, his main regular black holes solution and including the construction and analysis of a new dynamical solution arising from collapsing matter and his gravitational perturbations. Chapter 6 discusses the phenomenological implications of the models considered, focusing on black hole shadows, and his observational constrains, Hawking radiation, and constraints related to primordial black holes. Finally, Chapter 7 summarizes the main results of this work and outlines possible directions for future research. 

\section{Conventions, notation and units}

Throughout this thesis we adopt the metric signature $(-,+,+,+)$. Greek indices $\mu,\nu,\rho,\sigma,\ldots$ run over spacetime coordinates, while Latin indices $i,j,k,\ldots$ denote spatial coordinates.

Unless otherwise stated, we adopt natural units
\begin{equation}
c=\hbar=1.
\end{equation}
For numerical calculations and perturbative analyses, Newton's constant $G_0$ is also absorbed into the choice of units, so that masses, lengths and times are expressed in terms of a single fundamental scale. Consequently, frequencies have dimensions of inverse length (or inverse mass), and quantities are often presented through dimensionless combinations such as
\begin{equation}
\omega M,
\qquad
\omega r_h,
\end{equation}
where $M$ denotes the mass of the compact object and $r_h$ the event-horizon radius.
However, in several theoretical developments, particularly within the asymptotic-safety framework, Newton's constant $G_0$ and factors $8\pi G_0$ or $16\pi G_0$ are retained whenever convenient in the presentation of the field equations and action principles, while they may be absorbed into the choice of units in later stages of the analysis.

In the Einstein--Weyl sector of quadratic gravity, the massive spin--2 degree of freedom introduces the characteristic mass scale $m_2$ which is frequently used to define dimensionless quantities. In particular, the parameter
\begin{equation}
p=r_h m_2
\end{equation}
plays a central role in the analysis of black hole solutions and their perturbations. Depending on the context, radial coordinates, frequencies, coupling and masses may therefore be expressed in units of $m_2$.

For numerical computations, it is often convenient to normalize the characteristic scale of the background geometry by fixing either the black-hole mass parameter or the horizon radius. Consequently, quantities are frequently reported in dimensionless form.

Moreover, in the discussion of Hawking radiation and primordial black holes, theoretical calculations are performed in natural units and subsequently converted into physical units whenever phenomenological predictions are presented. In those cases, masses are expressed in grams, energies in GeV, and emission rates in the corresponding physical units.

The curvature tensors are defined as follows; The Riemann tensor is
\begin{equation}
R^{\rho}_{\;\sigma\mu\nu}
=
\partial_\mu \Gamma^\rho_{\nu\sigma}
-
\partial_\nu \Gamma^\rho_{\mu\sigma}
+
\Gamma^\rho_{\mu\lambda}\Gamma^\lambda_{\nu\sigma}
-
\Gamma^\rho_{\nu\lambda}\Gamma^\lambda_{\mu\sigma},
\end{equation}
and for the Ricci tensor and scalar
\begin{equation}
R_{\mu\nu}=R^{\rho}_{\;\mu\rho\nu},
\qquad
R=g^{\mu\nu}R_{\mu\nu}.
\end{equation}

\chapter{Linear perturbations of compact objects}
Black holes always have complex distributions of matter around them, such as accretion disks, strong magnetic fields, neutron stars, other stars, planets, etc., and thus actively interact with the surrounding matter. \\ Therefore, a real black hole can never be totally described by its few parameters and is always in its perturbed state. The concept of perturbations, or the linearization of the dynamical field equations, is straightforward: one considers the background value of the field under consideration, which acquires a small perturbation. Higher-order terms beyond the first of these perturbations are neglected \cite{Chandrasekhar:1985kt}. The metric of a perturbed black hole can be decomposed as
\begin{equation}
    g_{\mu\nu}=g^0_{\mu\nu}+\delta g_{\mu\nu},
\end{equation}
where $g^0_{\mu\nu}$ denotes the background metric (for example Schwarzschild or Kerr metric) and $\delta g_{\mu\nu}$ its linear perturbation. Throughout this chapter, we assume that the perturbation is sufficiently small so that terms of order $O(\delta g^2)$ and higher can be neglected.

The linearization procedure leads to a set of differential equations governing the evolution of the perturbations on the given background spacetime. In general, perturbations can be classified according to their tensorial nature, namely scalar, vector, and tensor perturbations, each satisfying different dynamical equations. In the case of gravitational perturbations, the perturbation $\delta g_{\mu\nu}$ can be decomposed into modes with definite parity under spatial inversion, commonly referred to as axial (odd-parity) and polar (even-parity) perturbations \cite{Regge:1957td,Kodama:2003jz}.

For static and spherically symmetric background spacetimes, the line element can be written in the general form
\begin{equation}
    ds^2=-h(r)dt^2+\frac{dr^2}{f(r)}+r^2d\Omega^2,
\end{equation} 
where $t$ and $r$ denote the temporal and radial coordinates, respectively, while $h(r)$ and $f(r)$ are arbitrary metric functions characterizing the background geometry and $ d\Omega^2=d\theta^2+\sin^2\theta\, d\phi^2 $ is the metric of a unit two-sphere.

The perturbation equations can be reduced to a set of decoupled master equations for suitably defined gauge-invariant variables. More precisely, after decomposing the metric perturbations into general (scalar, vector, tensor) spherical harmonics, the dependence on the angular coordinates can be separated, while the remaining radial and temporal dependence is encoded in a master variable. We denote this variable by $\Psi(t,r)$; its precise definition depends on the parity of the perturbation, and the underlying gravitational theory. The introduction of such a master variable allows the original system of coupled linearized equations to be recast, when possible, into a single second-order wave equation.

For static and spherically symmetric backgrounds, it is convenient to introduce the tortoise coordinate $r_*$ according to
\begin{equation}
    dr_*=\frac{dr}{\sqrt{h(r)f(r)}}.
\end{equation}
This could allow the radial part of the perturbation equations to be reduced in a Schr\"odinger-like form. After separation of the angular variables and the introduction of the tortoise coordinate, the master equation typically takes the form
\begin{equation} \label{td_equation}
    -\frac{d^2}{dt^2}\Psi+\frac{d^2}{dr_*^2}{\Psi}-V_{\text{eff}}(r)\Psi=0.
\end{equation}
Here, $\Psi=\Psi(t,r)$ is the master variable describing the dynamical perturbation, while $V_{\text{eff}}(r)$ is the effective potential governing its radial propagation.

The effective potential is obtained by substituting the perturbation ansatz into the linearized field equations and choosing a gauge eliminating the redundant perturbation functions. It depends on the background metric functions $h(r)$ and $f(r)$, their derivatives, the multipole number $l$, and, in modified theories of gravity, on the additional parameters and propagating degrees of freedom of the theory \cite{Konoplya:2011qq}. Thus, $V_{\rm eff}$ contains the information about how the background geometry and the gravitational dynamics affect the propagation of the perturbation. Its explicit form is therefore theory and perturbation dependent. In General Relativity, for example, gravitational perturbations of the Schwarzschild black hole lead to the Regge--Wheeler \cite{Regge:1957td} and Zerilli \cite{Zerilli:1970se} potentials for the axial and polar sectors, respectively. In modified theories of gravity, the corresponding potential can acquire additional terms and, in some cases, the perturbation equations can involve additional dynamical degrees of freedom, it could be seen in details in \ref{Pert} the derivation of the effective potential in quadratic gravity and in \ref{sec:perturbations} the procedure we used for gravitational perturbations in asymptotic safety.

\section{Time evolution of perturbations}
\begin{figure}[h]
\centering
\includegraphics[width=0.8\columnwidth]{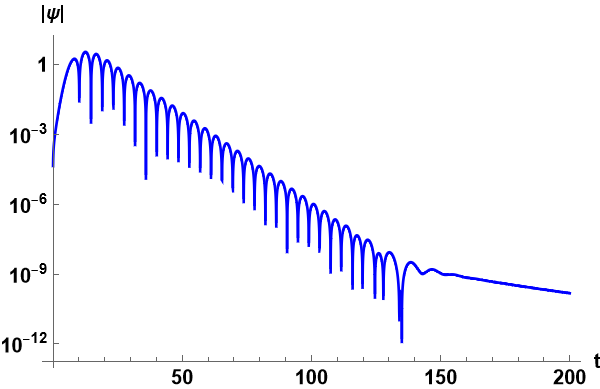}
\caption{Semi-logarithmic plot of the time-domain profile of the axial gravitational perturbation for the Schwarzschild black hole (for $l=2$). We can identify after the initial outburst, the ringdown phase and a power law tail at late time} 
\label{td Schw}
\end{figure}

Integrating equation \eqref{td_equation} we obtain the full evolution of the perturbation in the time domain, giving information about the stability and the response over time to a disturbance. Let us consider, as a qualitative example, the time-domain profile of the axial gravitational perturbation for the Schwarzschild black hole (for $l=2$) in fig. \ref{td Schw}.

We can divide the evolution of the perturbations into three stages. The first stage depends on the initial conditions and on the point where we look at the signal and corresponds to an initial outburst. At later time we see exponential damping of the amplitude of the perturbations, which is followed by the so-called tails at asymptotically late time.

One can observe, that, after the initial outburst at the first stage, the behavior of the amplitude of the perturbations does not depend on the initial conditions. Being independent on the source of the perturbations, it depends only on the parameters of the field and the black hole. Therefore, the damping stage appears to be an important characteristic of the black hole \cite{Konoplya:2011qq,Kokkotas:1999bd,Berti_2009}. The exponential damping of the perturbations is called quasi-normal ringing. It can be split to the superposition of exponentially damping oscillations, that can be represented as a set of complex frequencies, which are called quasi-normal modes. The real part of a quasi-normal frequency describes the actual frequency of the oscillation, while the imaginary part is its damping rate.

The last part of the signal is characterized by a power-law decay which is a manifestation of the noncompleteness of the set of quasinormal modes, i.e. at sufficiently late time the quasinormal modes are suppressed by the exponential or power-law tails. Physically, these tails are due to backscattering off the background curvature, and therefore they depend on the asymptotics of the spacetime \cite{Price:1971fb,Ching:1994bd}.

A unified and mathematical rigorous framework to describe all these contributions is provided by the retarded Green's function associated with the perturbation equation \cite{Ching:1995tj,Leaver:1986gd}.

Starting from eq.\eqref{td_equation}, the solution for $t>0$ can be expressed in terms of the retarded Green's function $G(r_*,r_*';t)$ \cite{Morse1955MethodsOT} as
\begin{equation}
\Psi(r_*,t) =
\int dr_*'
\left[
G(r_*,r_*';t)\,\partial_t\Psi(r_*',0)
+
\partial_t G(r_*,r_*';t)\,\Psi(r_*',0)
\right],
\label{eq:green_td}
\end{equation}
where $G$ satisfies
\begin{equation}
\left[
\frac{d^2}{dr_*^2}
-
\frac{d^2}{dt^2}
-
V(r_*)
\right] G(r_*,r_*';t)
=
\delta(t)\delta(r_*-r_*').
\end{equation}

It is convenient to introduce the one-sided Fourier transform of the Green's function,
\begin{equation}
\tilde G(r_*,r_*';\omega)
=
\int_0^{\infty} dt\, G(r_*,r_*';t)\, e^{i\omega t},
\label{eq:fourier_transform}
\end{equation}
where $\omega$ is allowed to be complex. The transformed function satisfies the inhomogeneous ordinary differential equation
\begin{equation}
\left[
\frac{d^2}{dr_*^2}
+
\omega^2
-
V(r_*)
\right]\tilde G(r_*,r_*';\omega)
=
\delta(r_*-r_*').
\label{eq:green_fd}
\end{equation}

The solution of Eq.~\eqref{eq:green_fd} can be constructed from two linearly independent solutions of the corresponding homogeneous equation. We denote by $f(\omega,r_*)$ the solution satisfying the appropriate boundary condition at the left boundary (event horizon or the origin), and by $g(\omega,r_*)$ the solution satisfying the outgoing-wave condition at spatial infinity. The frequency-domain Green's function can then be written as
\begin{equation}
\tilde G(r_*,r_*';\omega) =
\begin{cases}
\dfrac{f(\omega,r_*)\, g(\omega,r_*')}{W(\omega)}, & r_* < r_*', \\
\dfrac{f(\omega,r_*')\, g(\omega,r_*)}{W(\omega)}, & r_* > r_*',
\end{cases}
\label{eq:green_wronskian}
\end{equation}
where
\begin{equation}
W(\omega) = g\,\partial_{r_*} f - f\,\partial_{r_*} g
\end{equation}
is the Wronskian, which is independent of $r_*$.

The time-domain Green's function is recovered by inverting the Fourier transform,
\begin{equation}
G(r_*,r_*';t)
=
\frac{1}{2\pi}
\int_{-\infty+i c}^{\infty+i c}
d\omega\,
\tilde G(r_*,r_*';\omega)\,
e^{-i\omega t}.
\label{eq:inverse_transform}
\end{equation}

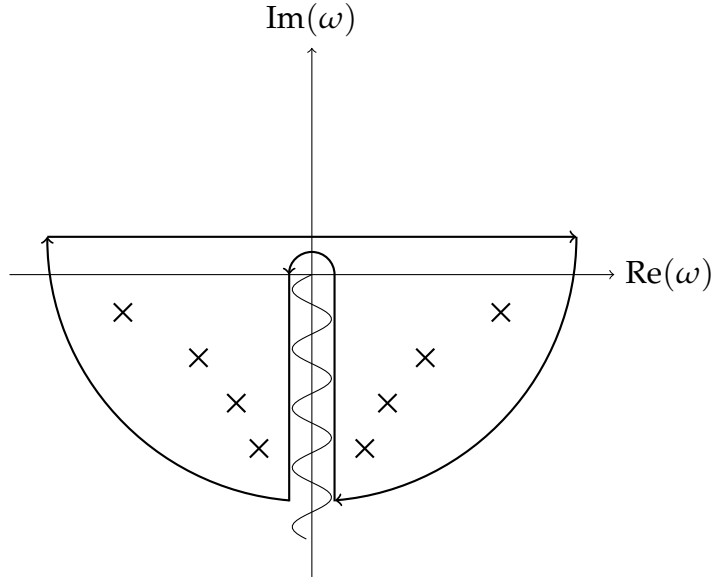
\begin{figure}[h] 
\centering
\begin{tikzpicture}[scale=1]

\draw[->] (-4,0) -- (4,0) node[right] {$\Re(\omega)$};
\draw[->] (0,-4) -- (0,3) node[above] {$\Im(\omega)$};

\draw[thick] (0.3,-3) -- (0.3,0);
\draw[thick] (-0.3,-3) -- (-0.3,0);
\draw[thick,->] (-3.5,0.5) -- (3.5,0.5);

\draw[thick,->] (3.5,0.5) arc (0:-85:3.5);
\draw[thick,<-] (-3.5,0.5) arc (-180:-95:3.5);
\draw[thick,->] (0.3,0) arc (0:180:0.3);
\foreach \x/\y in {-2.5/-0.5, -1.5/-1.1, -1/-1.7, -0.7/-2.3, 2.5/-0.5, 1.5/-1.1, 1/-1.7, 0.7/-2.3} {
  \draw[thick] (\x-0.12,\y-0.12) -- (\x+0.12,\y+0.12);
  \draw[thick] (\x-0.12,\y+0.12) -- (\x+0.12,\y-0.12);
}
\draw[thin, domain=0:-3.5, samples=100]
plot ({0.26*sin(8*\x r)}, {\x});
\end{tikzpicture}
\caption{Integration contour in the complex $\omega$-plane. 
Quasinormal modes correspond to poles of the Green's function in the lower half-plane, 
while the branch cut along the negative imaginary axis gives rise to late-time tails.}
\label{Green}
\end{figure}

The integration contour is taken, as shown in figure \ref{Green}, in the complex $\omega$-plane. For $t>0$, the contour can be closed in the lower half-plane, and the integral can be evaluated by standard complex analysis techniques.

The resulting signal naturally decomposes into three distinct contributions \cite{Konoplya:2011qq,Berti_2009}: (i) a prompt response associated with the large-frequency behavior of the integrand, (ii) a sum over residues at the poles of the Green's function, which are the zeros of the Wronskian, these poles correspond to perturbations satisfying both left and right boundary conditions, so  they represent the QNM contribution to the response \cite{Andersson:1995zk}. This contribution decays exponentially as $t$ increases, so that for asymptotically late-time the contribution is negligible for most cases, and (iii) depending on the asymptotic structure of the potential, are present singularities on the imaginary axis, thereby, arise a branch-cut contribution related to these non-analyticities of the Green's function, which gives the late-time tails contribution \cite{Ching:1994bd}. The importance of these contributions depends on the asymptotic structure of the effective potential and on the boundary conditions imposed by the background spacetime.

\subsection{Late-time tails}
Before starting with the more notorious QNM, we want to focus on the last part of the signal, which shows a peculiar and very interesting characteristic, as we will show.

The first systematic study of late-time tails was carried out by Price \cite{Price:1971fb}, who analyzed the asymptotic behavior of the linearized wave equation in the Schwarzschild background. He showed
that perturbations of massless scalar and gravitational fields decay at late times according to a power law of the form \begin{equation}
|\Psi| \sim t^{-(2\ell+3)},
\label{eq:PriceLaw}
\end{equation}
where $\ell$ denotes the multipole number. It was shown that the asymptotic tails in the Kerr
spacetime coincide with those of the Schwarzschild black hole, a
behavior shared by several other asymptotically flat solutions \cite{Ching:1995tj,Berti_2009,Ching:1994bd,Bonanno:2025dry,Rosato:2025rtr}.
This confirms that, for massless fields, the late-time decay is
primarily determined by the asymptotic flatness of the spacetime
rather than by its detailed structure in the strong-field region.

In contrast, shortly thereafter Price's paper, Bi\v{c}\'{a}k demonstrated that, in the Reissner--Nordstr\"{o}m spacetime, the late-time tails of scalar-field perturbations differ between the non-extremal and extremal cases \cite{Bičák1972}.

Moreover, a qualitatively different late-time behavior arises for AdS black holes, which tails are absent \cite{Horowitz:1999jd}, or for black holes with de Sitter asymptotics. In this case, the decay of perturbations is no longer governed by a power law, but instead exhibits an exponential falloff, reflecting the presence of a cosmological horizon and the corresponding modification of the asymptotic boundary conditions \cite{Brady:1996za,Brady:1999wd}.

The late-time dynamics of massive fields is drastically different
from that of massless ones. In this case, the asymptotic signal is
characterized by long-lived oscillations rather than a purely
monotonic decay. At asymptotically late times, which can be defined
as $t/M \gtrsim (\mu M)^{-3}$, where $\mu$ is the inverse Compton wavelength of the field, the field behaves as \cite{Koyama:2001ee,Konoplya:2006gq}
\begin{equation}
|\Psi| \sim t^{-5/6}\sin(F(\mu) t), \qquad t \to \infty,
\label{eq:MassiveTail}
\end{equation}
Remarkably, this asymptotic decay is independent of the multipole
number $l$ and this behavior is quite universal not only because of independence on the multipole number but also because emerges for a wide class of black holes and compact objects \cite{Konoplya:2025afm,Bonanno:2025ksz,Koyama:2000hj,Konoplya:2023fmh}.

\section{Quasinormal modes}

It is possible to determine the time evolution of any initial perturbation simply by integrating \eqref{td_equation} numerically. However, we often want to study the quasi-normal spectrum of a black hole. Following the standard procedure for such a normal-mode analysis, we will assume a solution of the perturbation equation as a Fourier transform for the function
\begin{equation}
    \Psi(r,t)=\sum_{n=0}^\infty e^{-i\omega_n t}Q_n(r)
\end{equation}
Let us note that generally the functions $e^{-i\omega_n}$, where $\omega_n$ is the quasi-normal spectrum, do not form basis in the vector space of the solutions of the equation \eqref{td_equation}. Therefore, the signal cannot be expanded in terms of these functions for all time. Indeed, as we discussed previously, the exponential decay is replaced by slower power-law decay at asymptotically late time. Substituting this ansatz in the equation we obtain the following expression  
\begin{equation} \label{master_eq}
    \frac{d^2 Q}{d r_*^2} + \left[\omega^2 - V_{\text{eff}}(r)\right]Q = 0 ,
\end{equation}
with $\omega$ is the complex frequency of the quasi-normal spectrum. Thus, in a spherically symmetric background, the study of BH perturbations due to linearized fields can be reduced to the study of the differential equation \eqref{master_eq}, and the searching of the quasi-normal modes is reduced to the problem of finding eigenfrequencies of the equation \eqref{master_eq}.

\subsection{Boundary conditions}
To determine the free oscillation modes of a black hole, which correspond to ‘natural’ solutions of this unforced equation, we must impose physically appropriate boundary conditions. These conditions are fixed by the causal structure of the spacetime and by the requirement that the perturbation describes a self-sustained response of the system, without external radiation injected from the boundaries \cite{Berti_2009,Konoplya:2011qq}. For asymptotically flat black holes, the event horizon and spatial infinity correspond to the asymptotic regions $r_*\to -\infty$ and $r_*\to +\infty$, respectively. In most cases of interest, the effective potential vanishes in both limits $V_{eff}\to 0$, and the solutions reduce asymptotically to free waves, \begin{equation}
Q\sim e^{\pm i\omega r_*} . 
\end{equation} 
At the event horizon, physical regularity requires that no radiation emer\-ges from inside the black hole. One therefore imposes a purely ingoing condition at the horizon, 
\begin{equation}
Q(r)\sim e^{-i\omega r_*}, \qquad r_*\to -\infty . \label{eq:qnm_bc_horizon} \end{equation} 
Indeed, the full perturbation behaves as \begin{equation} \Psi\sim e^{-i\omega(t+r_*)} , \end{equation} which corresponds to a wave propagating towards decreasing $r_*$, namely into the black hole. At spatial infinity, instead, quasinormal modes are defined by requiring that no radiation enters the spacetime from infinity. In asymptotically flat spacetimes, where the metric at spatial infinity tends to the Minkowski one, this selects the purely outgoing solution, \begin{equation} Q(r)\sim e^{+i\omega r_*}, \qquad r_*\to +\infty . \label{eq:qnm_bc_infinity} \end{equation} The full perturbation then behaves as \begin{equation} \Psi\sim e^{-i\omega(t-r_*)} , \end{equation} which represents an outgoing wave propagating towards spatial infinity.

Quasinormal modes frequencies (QNM) are defined as the solutions of the equation \eqref{master_eq} with the above boundary conditions, and represents an infinite discrete set of complex frequencies, 
\begin{equation}
\omega=\omega_R+i\omega_I . 
\end{equation} 
Stable modes satisfy $\omega_I<0$. The real part $\omega_R$ determines the oscillation frequency, while the imaginary part governs the damping of the signal.

It should be noted that for other types of compact objects the boundary conditions must be reconsidered on the merits of the individual case. In fact, the boundary conditions must be modified when the asymptotic structure or the nature of the central object changes. For asymptotically de Sitter black holes, the outer boundary is the cosmological horizon, where one again imposes purely outgoing waves. For asymptotically anti-de Sitter spacetimes, the potential typically diverges at infinity and one usually imposes normalizability, Dirichlet, Neumann, or more general boundary conditions depending on the field and on the physical setup \cite{Horowitz:1999jd,Kovtun:2005ev}. For massive fields, one may also consider quasi-bound states, for which the field is required to decay exponentially at spatial infinity rather than being purely outgoing. Finally, for horizonless compact objects such as stars, wormholes, naked singularities, gravastars or exotic compact objects, the inner boundary condition is no longer an ingoing condition at an event horizon. It must instead be replaced by regularity at the center, matching across a stellar surface, or a model-dependent reflective condition.

\subsection{Stability}

Gravitational stability of any static or stationary solution describing a compact object is a fundamental requirement for its physical viability. If a background is unstable under small perturbations, they will grow with time and inevitably destroyed the system and, thereby, cannot continue to exist. A compact object can be regarded as physically viable only if the corresponding metric is proven to be stable.

One can study stability against linear perturbations in the following way. For a well-posed initial value problem, we have to analyze the complete spectrum of quasinormal modes, if all the modes are decaying in time, i.e. $\omega_I<0$, the solution is evidently stable. If there is at least one growing mode, the space-time is unstable with a characteristic growth timescale set by $1/\omega_I$ of the QNM.

In some cases, stability can be established without computing the complete quasinormal spectrum. In particular, for a Schr\"odinger-like equation with a real effective potential and, a sufficient condition for stability is that the potential be positive throughout the physical domain \cite{Konoplya:2011qq,Kodama:2003jz,Ishibashi:2003ap},
\begin{equation}
    V_{\mathrm{eff}}(r)\geq 0.
\end{equation}
Indeed, a non negative definite potential excludes bound states with negative $\omega^2$, which would correspond to purely imaginary frequencies with $\omega_I>0$ and hence to exponentially growing perturbations. This criterion is sufficient rather than necessary: if the potential becomes negative in some region, the background is not necessarily unstable, and the quasinormal spectrum or other stability techniques must be analyzed explicitly. This is particularly relevant for effective potentials in modified theories of gravity, where negative regions can arise without necessarily implying the existence of an unstable mode

\section{Numerical methods} \label{num_met}

In this section we present the numerical methods used for the calculation of the perturbation. Specifically, we will introduce the WKB approximation with Padè approximant, Leaver method and Prony's method for the evaluation of the QNM frequencies, while we present two schematization for the time-domain integration for the time evolution analysis. It is worth noting that exist several different methods for the calculation of the quasinormal frequencies \cite{Konoplya:2011qq}, such as spectral methods \cite{Jansen:2017oag,Fortuna:2020obg}, or direct integration \cite{Chandrasekhar:1975zza,Cho:2011sf}.

\subsection{WKB approximation}
\begin{figure}[h]
\centering
\includegraphics[width=0.8\columnwidth]{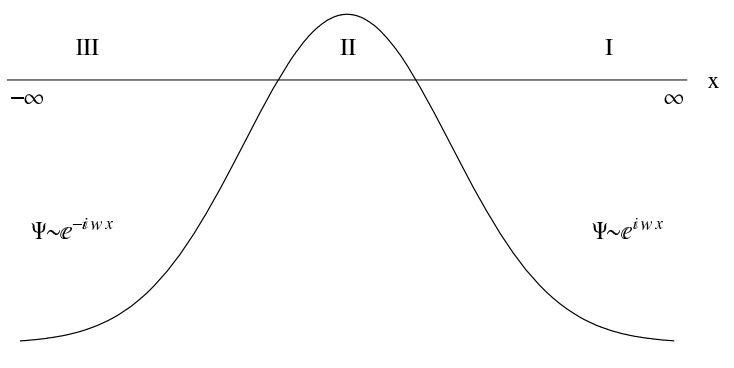}
\caption{The three regions separated by the two turning points of the potential considerated \cite{Konoplya:2011qq}} 
\label{fig:WKB}
\end{figure}

The WKB (Wentzel, Kramers, Brillouin) technique was applied to finding of the quasi-normal modes of black holes for the first time by Schutz and Will \cite{Schutz:1985km}. The method is used for effective potentials that have the form of a barrier potential so that have constant value at the boundaries, and a peak in a specific zone \ref{fig:WKB}.
The method is based on a matching procedure between the asymptotic solutions, expressed as a superposition of plane waves, and the solution in the region close to the maximum of the potential barrier. In particular, assuming that the effective potential has a single maximum, it can be expanded in a Taylor series around it, such as
\begin{equation} \label{Taylor}
    \Psi=e^{\sum\frac{S_n(x)\epsilon^n}{\epsilon}},
\end{equation}
with $x=r_*$ and $\epsilon$ is a formal bookkeeping parameter introduced to organize the expansion orders, which is set to $\epsilon=1$ at the end of the calculation. Defining the two turning points by $ V_{\rm eff}(r_*)=\omega^2,$
and matching the solutions across these turning points provides the results through the WKB approximation.

\noindent The expansion at successive orders were developed through the years (up to third order in \cite{Iyer:1986np}, up to six order in \cite{Konoplya:2003ii}, and through thirteen in \cite{Matyjasek:2017psv}) and the general expression for the quasinormal frequencies is formulated as an expansion around the eikonal regime with this form \cite{Konoplya:2019hlu}:
\begin{align} \label{Wkb formula}
    &
    \omega^2 = V_0 + A_2(K^2) + A_4(K^2) + A_6(K^2) + \ldots \nonumber 
    \\
&- iK\sqrt{-2V_2} \left[1 + A_3(K^2) + A_5(K^2) + A_7(K^2) + \ldots \right],
\end{align}
where, due to the boundary condition, $ K = n + \tfrac{1}{2} $, with $ n = 0, 1, 2, \ldots $ denoting the overtone number. In this expression, $ V_0 $ is the peak value of the effective potential, and $ V_2 $ is its second derivative at that point. The coefficients $ A_i $ (for $ i = 2, 3, 4, \ldots$) are higher-order correction terms, which are polynomials of the derivatives of the potential up to order $ 2i $ and of $K^2$\cite{Konoplya:2019hlu}.
\subsection{Padè approximant}

In order to increase accuracy of the higher-order WKB formula, it has been proposed to use Padé approximants \cite{Matyjasek:2017psv,Konoplya:2019hlu} for the usual WKB formula. Within this approach one has to define a polynomial $P_k(\epsilon)$ by introducing powers of the order parameter $\epsilon$ in the righthand side of the WKB formula as follow:
\begin{align}
    P_k(\epsilon) = & V_0 + A_2(K^2)\epsilon^2 + A_4(K^2)\epsilon^4 + A_6(K^2)\epsilon^6 + \ldots \nonumber 
    \\
&- iK\sqrt{-2V_2} \left[\epsilon + A_3(K^2)\epsilon^3 + A_5(K^2)\epsilon^5 + \ldots \right],
\end{align}
where $k$ is the polynomial order which coincides with the order of the WKB formula.
We consider a family of Pad\'e approximants $P_{\tilde{n}/\tilde{m}}(\epsilon)$ associated with the polynomial $P_k(\epsilon)$ in the vicinity of $\epsilon = 0$, with $\tilde{n} + \tilde{m}= k$. The Pad\'e approximants are defined as rational functions of the form
\begin{equation}
P_{\tilde{n}/\tilde{m}}(\epsilon) =
\frac{Q_0 + Q_1 \epsilon + \cdots + Q_{\tilde{n}} \epsilon^{\tilde{n}}}
{R_0 + R_1 \epsilon + \cdots + R_{\tilde{m}} \epsilon^{\tilde{m}}},
\label{eq:pade}
\end{equation}
constructed such that the difference between the Pad\'e approximant and the original polynomial satisfies
\begin{equation}
P_{\tilde{n}/\tilde{m}}(\epsilon) - P_k(\epsilon) = \mathcal{O}\left(\epsilon^{k+1}\right).
\end{equation}

This condition ensures that the rational representation in Eq.~\eqref{eq:pade} reproduces the polynomial expansion up to order $k$. Consequently, the coefficients $Q_0, Q_1, \ldots, Q_{\tilde{n}}$ and $R_0, R_1, \ldots, R_{\tilde{m}}$ can, in principle, be determined analogously to the procedure used to compute the coefficients $A_2, A_3, \ldots, A_k$. In practice, however, it is more convenient to determine the coefficients $Q_i$ and $R_j$ numerically. 

The resulting Pad\'e approximant is then employed to approximate the squared quasinormal mode frequency according to
\begin{equation}
\omega^2 = P_{\tilde{n}/\tilde{m}}(1).
\end{equation}
In practice, it turns out that when $\tilde{n} = \tilde{m}$, the approximation is
much better compared to the initial WKB formula \cite{Konoplya:2019hlu}.

Recently the WKB method with Padè approximant has been optimized for speed \cite{Matyjasek:2026yiu} and generalized for very higher orders with the use of recursive terms, and has been used for calculation with previous limitation \cite{Konoplya:2026rjh} as higher overtones with $n>l$ and for massive perturbations. 

\subsection{Leaver method}

An accurate and widely used approach for computing quasinormal modes (QNMs) of black holes is the method of continued fractions developed by Leaver~\cite{Leaver:1985ax}. This technique is particularly powerful for spherically and axially symmetric spacetimes where the perturbation equations reduce to second-order ordinary differential equations with appropriate boundary conditions at the event horizon and at spatial infinity.

The solution $\Psi(r)$ is then factorized to extract the known asymptotic behavior at the boundaries:
\begin{equation}
\Psi(r) = e^{i \omega r_{*}} \left( \frac{r - r_h}{r} \right)^{\alpha} \sum_{n=0}^{\infty} a_n \left( \frac{r - r_h}{r} \right)^n,
\end{equation}
where the exponent  $\alpha$ is chosen to satisfy the ingoing condition at the event horizon and the outgoing condition at infinity. Substituting this ansatz into the wave equation yields a three-term or higher terms (reducible via Gaussian elimination) recurrence relation for the coefficients $a_n$, depending on the values of the resultant frequencies and parameters of the wave equation, where (for the three-term reccurence relation) are:
\begin{equation}
\alpha_0 a_1 + \beta_0 a_0 = 0,
\end{equation}
\begin{equation}
\alpha_n a_{n+1} + \beta_n a_n + \gamma_n a_{n-1} = 0, \quad n \geq 1,
\end{equation}
where $\alpha_n$, $\beta_n$, and $\gamma_n$ are functions of the parameters of the background metric and the frequency $\omega$. The condition for convergence of the infinite series corresponds to the vanishing of a continued fraction:
\begin{equation}
\beta_0 - \frac{\alpha_0 \gamma_1}{\beta_1 -} \, \frac{\alpha_1 \gamma_2}{\beta_2 -} \, \frac{\alpha_2 \gamma_3}{\beta_3 -} \cdots = 0.
\label{eq:cfcondition}
\end{equation}
This equation is solved numerically to determine the quasinormal frequencies $\omega$. The method is known for its excellent convergence properties and high accuracy, even for higher overtones.

In cases where the wave equation possesses multiple regular singular points along the complex $r$-plane contour connecting the event horizon and infinity, the standard implementation of the continued fraction method may fail due to the breakdown of convergence of the Frobenius series. To overcome this limitation, is is possible to use the integration-through-midpoints approach proposed in~\cite{Rostworowski:2006bp}. The core idea is to start with a Frobenius expansion around a convenient regular singular point, evaluate this solution at an intermediate (midpoint) value of the radial coordinate lying within the convergence radius of the initial series, and then re-expand the solution into a new Frobenius series centered at this midpoint. This process is repeated iteratively as needed until the domain reaches a point where the standard Leaver continued fraction condition can be imposed. The method is particularly effective for spacetimes with complex potential structures. It can be useful also to further employ the Nollert improvement~\cite{Nollert:1993zz} in its generalized form that accommodates an $N$-term recurrence relation \cite{Zhidenko:2006rs} to optimize the computation.

\subsection{Time-domain integration}\label{TDI}
Starting from the general second-order equation governing the time evolution of the perturbations \eqref{td_equation}, our goal is to compute the time evolution of $\Psi$. To do so, we discretize  the equation on a grid.  The discretization can be operated in different ways, here we present two different methods used in thesis. First, the wave function and the potential matrices as follows:
$$
\begin{array}{rclcl}
\Psi(r_*, t) &=& \Psi(j \Delta r_*, i \Delta t) &=& \Psi_{j,i}, \\
    V(r(r_*)) &=& V(j \Delta r_*) &=& V_j.
\end{array}
$$
We then apply our discretization method to approximate the derivatives using the following scheme \cite{Zhu_2014}:
\begin{align}\notag
   &
    - \frac{\Psi_{j,i+1} - 2\Psi_{j,i} + \Psi_{j,i-1}}{\Delta t^2} + \frac{\Psi_{j+1,i} - 2\Psi_{j,i} + \Psi_{j-1,i}}{\Delta r_*^2}
    \\
   & - V_j \Psi_{j,i} +O({\Delta t^2,\Delta r_*^2}) = 0.
\end{align}
To solve this equation, we first need to express the potentials as functions of the tortoise coordinate $r_*$. This requires determining the dependence of the radial coordinate $r$ on $r_*$, which we obtain by solving the following differential equation numerically:
\begin{equation}
    \frac{dr}{dr_*}=\sqrt{h(r) f(r)}.
\end{equation}
Once this transformation is obtained and substituted into the potential, we can proceed solving the equation. Imposing the initial condition as followed
\begin{equation} \label{gaussTD}
    \Psi(r_*,0)=e^{-\frac{(r_*-r_0)^2}{2\sigma}},
\end{equation}
we choose a Gaussian distribution for all components, centering it at $r_*=r_0$ to ensure it is located near the peak of the potential. This is also the point where we monitor the time evolution of the perturbation. This scheme could be useful also for more complex equation, for example when also first derivative terms are present.

Instead, for the second method we followed the technique explained in~\cite{Gundlach:1993tp,Konoplya:2011qq}. Introducing the light-cone coordinates $v=t+r_*$ and $u=t-r_*$, the equation~\eqref{td_equation} takes the form
\begin{equation}
    4\frac{\mathrm{d}^2}{\mathrm{d}v \mathrm{d}u}\Psi(u,v)+V(u,v)\Psi(u,v)=0.
\end{equation}
To integrate the wave equation numerically, we discretize the \((u,v)\) plane on a uniform grid with step size \(h\). The field \(\Psi(u,v)\) is then evolved using the standard scheme
\begin{equation}\label{pertgrid}
    \Psi(N)=\Psi(W)+\Psi(E)-\Psi(S)+\frac{h^2}{8}V(S)(\Psi(W)+\Psi(E)),
\end{equation}
where the grid points are defined as
$$
S = (u, v), \qquad 
W = (u + h, v), \qquad 
E = (u, v + h), \qquad 
N = (u + h, v + h).
$$
With appropriate initial data and with procedure similar to the previous one for the potential, the integration can be performed iteratively, row by row, to obtain the time-domain evolution of the perturbation $\Psi$.

\subsection{Prony's method} \label{Prony}

To extract the frequencies values from time evolution we used Prony's method \cite{Konoplya:2011qq}, assuming to fit the data by a combination of complex dam\-ped exponential
\begin{equation} \label{prony}
    \Psi(t)\simeq \sum_{i=1}^p C_i e^{-i\omega_i t}
\end{equation}
and at the end, you can obtain the frequencies as \cite{Konoplya:2011qq,Berti:2007dg},
\begin{equation}
    \omega_j=\frac{i}{h}\log z_j
\end{equation}
where $h$ is the time sampling interval, and $z_j$ are the roots of polynomials obtained from the exponential in eq.~\eqref{prony}.

Because the quasi-normal stage is not a precisely defined time interval, in practice, it is difficult to determine when the quasi-normal ringing begins. In fact, when we observe explicitly damped oscillations, we usually see only the fundamental mode, and first few overtones, while higher overtones, which damp quickly, are already exponentially suppressed.

\chapter{Quadratic gravity}

Generally, the first-order corrections to General Relativity have to be in the form of the addition of terms quadratic in the curvature tensors to the Einstein-Hilbert action \cite{tHooft:1974toh,Benedetti:2013jk,Zwiebach:1985uq}.In fact, in perturbative quantum gravity, the metric is expanded around flat spacetime as
\begin{equation}
g_{\mu\nu} = \eta_{\mu\nu} + h_{\mu\nu},
\end{equation}
where $h_{\mu\nu}$ represents a perturbation of the graviton field. In Einstein gravity, the Einstein--Hilbert action contains two derivatives of the metric, and therefore the graviton propagator behaves at large momenta as
\begin{equation}
D(k)\sim \frac{1}{k^2}.
\end{equation}
At the same time, interaction vertices contain two powers of momenta. As a consequence, the ultraviolet divergences worsen at higher loop orders. Indeed, for a generic Feynman diagram with $L$ loops, $I$ internal propagators and $V$ vertices, the superficial degree of divergence is
\begin{equation}
\omega = 4L -2I +2V .
\end{equation}
Using the topological relation
\begin{equation}
L = I - V +1 ,
\end{equation}
one finds
\begin{equation}
\omega = 2 + 2L .
\end{equation}
Therefore, the degree of divergence increases with the number of loops. This implies that quantum corrections generate divergences with increasingly large powers of momenta, requiring counterterms with more and more derivatives of the metric.
The forms of the one-loop counterterms were found rigorously by ’t Hooft and Veltman in \cite{tHooft:1974toh} to be proportional to
\begin{equation}
\sqrt{-g}R,
\qquad
\sqrt{-g}R^2,
\qquad
\sqrt{-g}R_{\mu\nu}R^{\mu\nu}.
\end{equation}
The appearance of these counterterms motivates the study of the most general four-derivative gravitational action. Specifically, if we restrict ourselves to a theory where only curvature tensors and not their derivatives are present and if only the metric tensor appears as the dynamical field, we could expand in power series the action, in principle, with all terms which are invariants under diffeomorphism. Stopping at first order correction, the most general action containing quadratic terms of the curvature is:
\begin{equation}\label{action}
\mathcal{S} = \int \mathrm{d}^4x \, \sqrt{-g} 
\left[
  \gamma R 
  + a R^2 
  + b R^{\mu\nu} R_{\mu\nu}
  + c R^{\mu\nu\rho\sigma} R_{\mu\nu\rho\sigma}
\right],
\end{equation}
where $\gamma = 1/(16\pi G_0)$. We can recast this action in a different way, to show some peculiar aspects, as shown by Stelle. 
\section{Stelle's action}
Using the four-dimensional Gauss-Bonnet identity 
\begin{equation}\label{GB}
I_{GB} = \int \mathrm{d}^4x \, \sqrt{-g} 
\left[R^2 
  -4 R^{\mu\nu} R_{\mu\nu}
  + R^{\mu\nu\rho\sigma} R_{\mu\nu\rho\sigma}
\right],
\end{equation}
whose integrand is a total derivative in four dimensions, one finds that the quantity $R^{\mu\nu\rho\sigma} R_{\mu\nu\rho\sigma}$ is not independent of $R^{\mu\nu} R_{\mu\nu}$ and $R^2$. Since the Gauss--Bonnet term is purely topological in four dimensions, it does not contribute to the equations of motion. Therefore, the action \eqref{action} is equivalent, up to a total derivative, to
\begin{equation}
\mathcal{S}_{\text{Stelle}} =
\int d^4x \, \sqrt{-g}
\left[
\gamma R
+
\left(
\beta + \frac{2}{3}\alpha
\right) R^2
- 2\alpha R^{\mu\nu} R_{\mu\nu} 
\right].
\end{equation}
  where we have introduced the redefined couplings
  \begin{equation}
  \alpha = \frac{b+4c}{2},
  \qquad
  \beta = a-\frac{b+7c}{3}.
  \end{equation}
These relations follow directly from the four-dimensional Gauss--Bonnet identity, after eliminating the Riemann-squared term in favour of the Ricci-squared and Ricci-scalar terms. 

Starting from this action, Stelle demonstrated in \cite{Stelle:1976gc} that the propagator of the theory takes the form
\begin{equation}
D_{\mu\nu\rho\sigma}(k)=
\frac{1}{(2\pi)^4 i}
\left[
\frac{2 P^{(2)}_{\mu\nu\rho\sigma}(k)}
{k^2 \left( 2\alpha k^2 + \gamma \right)}
-\frac{4 P^{(0\text{-}s)}_{\mu\nu\rho\sigma}(k)}
{k^2 \left( 6\beta k^2 + \gamma \right)}
+\text{gauge fixing terms}
  \right]
  \label{eq:propagator_quadratic}
  \end{equation}
where $P^{(2)}_{\mu\nu\rho\sigma}(k)$, $P^{(0\text{-}s)}_{\mu\nu\rho\sigma}(k)$ are projectors for symmetric rank-two tensors. At large momenta, the propagator behaves as
\begin{equation}
D_{\mu\nu\rho\sigma}(k)\sim \frac{1}{k^4},
\end{equation}
which represents an improvement with respect to the Einstein-Hilbert case, where the propagator scales as $k^{-2}$. This ultraviolet behaviour is the key ingredient behind the power-counting renormalizability of quadratic gravity. Since the interaction vertices also contain at most four powers of momenta, the superficial degree of divergence now becomes
\begin{equation}
\omega = 4L -4I +4V .
\end{equation}
Using again the relation
\begin{equation}
L = I -V +1 ,
\end{equation}
one obtains
\begin{equation}
\omega = 4 .
\end{equation}
Therefore, the degree of divergence no longer increases with the loop order. Quantum corrections can only generate operators containing up to four derivatives of the metric, namely terms already present in the action itself. The divergences can thus be absorbed into a finite redefinition of the couplings, and the theory becomes power-counting renormalizable. This is precisely the motivation for considering an action containing all operators up to four derivatives of the metric. As a consequence, all divergences can be absorbed into a redefinition of the couplings, and the theory is power-counting renormalizable.

The propagator in Eq.~\eqref{eq:propagator_quadratic} can also be rewritten in a form where the pole structure becomes explicit:
\begin{equation}
\begin{split}
D_{\mu\nu\rho\sigma}(k)
=
&\frac{1}{\gamma (2\pi)^4 i}
\Bigg[
\frac{2 P^{(2)}_{\mu\nu\rho\sigma}(k)
- 4 P^{(0\text{-}s)}_{\mu\nu\rho\sigma}(k)}
{k^2}
-
\frac{2 P^{(2)}_{\mu\nu\rho\sigma}(k)}
{k^2 + m_2^2}
+\\
&\frac{4 P^{(0\text{-}s)}_{\mu\nu\rho\sigma}(k)}
{k^2 + m_0^2}
+ \text{g.f.t.}
\Bigg],
\label{eq:propagator_poles}
\end{split}
\end{equation}
  where
  \begin{equation}
  m_2^2=\frac{\gamma}{2\alpha},
  \qquad
  m_0^2=\frac{\gamma}{6\beta}.
  \end{equation}

The first term in Eq.~\eqref{eq:propagator_poles} corresponds to the usual massless graviton propagator. The propagator contains two additional massive poles, corresponding respectively to a massive spin-$2$ mode and a massive scalar mode.
The nature of these excitations can be understood by studying the residues at the poles. For the massless pole at $k^2=0$, one finds
\begin{equation}
\underset{k^2=0}{\mathrm{Res}}
, D_{\mu\nu\rho\sigma}(k)=
\frac{1}{\gamma (2\pi)^4 i}
\left[
2 P^{(2)}_{\mu\nu\rho\sigma}-
4 P^{(0\text{-}s)}_{\mu\nu\rho\sigma}
\right].
\end{equation}
The massive spin-$2$ pole is located at
\begin{equation}
k^2=-m_2^2,
\end{equation}
with residue
\begin{equation}
\underset{k^2=-m_2^2}{\mathrm{Res}}
D_{\mu\nu\rho\sigma}(k)=-
\frac{
2 P^{(2)}_{\mu\nu\rho\sigma}
}{
\gamma (2\pi)^4 i
}.
\end{equation}
The minus sign relative to the massless graviton pole signals that this degree of freedom is a ghost. Such a state can be interpreted either as carrying negative kinetic energy, leading to a Hamiltonian unbounded from below and severe instabilities, or equivalently as a negative-norm state, implying a violation of perturbative unitarity \cite{Hawking:2001yt}. Finally, the scalar pole is located at
\begin{equation}
k^2=-m_0^2,
\end{equation}
with residue
\begin{equation}
\underset{k^2=-m_0^2}{\mathrm{Res}}
D_{\mu\nu\rho\sigma}(k)=
\frac{
4 P^{(0\text{-}s)}_{\mu\nu\rho\sigma}
}{
\gamma (2\pi)^4 i
}.
\end{equation}
For positive $m_0^2$, this pole corresponds to a physical massive scalar excitation.

Therefore, the spectrum of quadratic gravity around flat spacetime contains three propagating degrees of freedom: the usual massless spin-$2$ graviton, a massive scalar field, and a massive spin-$2$ ghost.

We prefer to recast the action as
\begin{equation}
\mathcal{S}=
\int d^4x \, \sqrt{-g}
\left[
\gamma R
+ \beta R^2
- \alpha C_{\mu\nu\rho\sigma} C^{\mu\nu\rho\sigma}
+ \chi \, \mathcal{G}
\right],
\end{equation}
where $C_{\mu\nu\rho\sigma}$ is the Weyl tensor, defined as
\begin{equation}
\begin{split}
C_{\mu\nu\rho\sigma}
=&
R_{\mu\nu\rho\sigma}
- \frac{1}{2}
\left(
g_{\mu\rho} R_{\nu\sigma}
- g_{\mu\sigma} R_{\nu\rho}
- g_{\nu\rho} R_{\mu\sigma}
+ g_{\nu\sigma} R_{\mu\rho}
\right)\\&
+ \frac{1}{6}
R \left(
g_{\mu\rho} g_{\nu\sigma}
- g_{\mu\sigma} g_{\nu\rho}
\right),
\label{eq:weyl_tensor}
\end{split}
\end{equation}
which is traceless under any contraction of its indices. From now on we do not take into account tha Gauss-Bonnet term $\mathcal{G}$ as it does not contribute to the equation of motion, as said before, and we can write the Stelle action of Quadratic gravity we will work with:  
\begin{equation}
    \mathcal{S}_{\text{Stelle}} =\int d^4x \, \sqrt{-g}
\left[
\gamma R
+ \beta R^2
- \alpha C_{\mu\nu\rho\sigma} C^{\mu\nu\rho\sigma} \right].
\label{Stelle_action}
\end{equation}
We can see again from this action, that at quantum level three fundamental particles are present:
\begin{itemize}
    \item a massless spin-2 tensor mode, which can be identified with the
    standard graviton;

    \item a massive scalar mode with spin $S = 0$ and mass
    $m_0^2 = \gamma/(6\beta)$, originating from the $R^2$ term;
    
    \item a massive spin-2 tensor mode with mass
    $m_2^2 = \gamma/(2\alpha)$, associated with the
    $C_{\mu\nu\rho\sigma} C^{\mu\nu\rho\sigma}$ term, which represents
    the ghost degree of freedom introduced in the theory.
\end{itemize}
Minimizing eq.\eqref{Stelle_action} we obtain the equation of motion
\begin{equation}
\begin{split}
G_{\mu\nu} =&
 \gamma \left(
R_{\mu\nu}
- \frac{1}{2} R \, g_{\mu\nu}
\right)
+ 2\beta \left(
R_{\mu\nu}
- \frac{1}{4} R \, g_{\mu\nu}
- \nabla_\mu \nabla_\nu
+ g_{\mu\nu} \Box
\right) R\\&
- 4\alpha \left(
\nabla^\rho \nabla^\sigma
+ \frac{1}{2} R^{\rho\sigma}
\right)
C_{\mu\rho\nu\sigma}
= 0.
\end{split}
\end{equation}
It is trivial to derive that all the vacuum solutions of General Relativity $R_{\mu\nu}=0$ are also solution of quadratic gravity. Indeed, using second Bianchi identity the equations take the form
\begin{equation}
   \begin{split}
G_{\mu\nu} =&
 \gamma \left(
R_{\mu\nu}
- \frac{1}{2} R \, g_{\mu\nu}
\right)
+ 2\beta \left(
R_{\mu\nu}
- \frac{1}{4} R \, g_{\mu\nu}
- \nabla_\mu \nabla_\nu
+ g_{\mu\nu} \Box
\right) R\\&
- \alpha \bigg(4 R^{\rho\sigma} R_{\mu\rho\nu\sigma}
- R^{\rho\sigma} R_{\rho\sigma} g_{\mu\nu}
- \frac{4}{3} R R_{\mu\nu}
+ \frac{1}{3} R^2 G_{\mu\nu}
+ 2 \Box R_{\mu\nu}+\\&
- \frac{1}{3} \Box R \, g_{\mu\nu}
- \frac{2}{3} \nabla_\mu \nabla_\nu R\bigg)
= 0.
\end{split} 
\end{equation}
It was also demonstrated \cite{Lu:2015cqa} that all the static and asymptotically flat black hole solutions of quadratic gravity have $R=0$. To arrive at this affirmation, first we need to demonstrate the following theorem:

Let $(\mathcal{M},g_{\mu\nu})$ be a static spacetime solving the vacuum field equations of quadratic gravity, and let $\Sigma$ be a spacelike hypersurface. Suppose that at the boundary $\partial\Sigma$ at least one of the following conditions holds:
\begin{enumerate}
    \item $g_{\mu\nu}\kappa^\mu\kappa^\nu \rightarrow 0$, where $\kappa^\mu$ is the timelike Killing vector;
    \item $R \rightarrow 0$;
    \item $D_a R \rightarrow 0$,
\end{enumerate}
where $D_a$ denotes the covariant derivative compatible with the induced metric on $\Sigma$. Then the Ricci scalar must vanish identically throughout $\Sigma$,
\begin{equation}
    R=0.
\end{equation}

\begin{proof}
Consider a generic static line element of the form
\begin{equation}
ds^2=-\tau^2(x)\,dt^2+\xi_{ij}(x)\,dx^i dx^j .
\end{equation}
In vacuum, the trace equation of quadratic gravity reduces to
\begin{equation}
\Box R-m_0^2 R=0 ,
\end{equation}
which, using the metric above, can be written as
\begin{equation}
D^iD_iR+\frac{1}{\tau}D^i\tau\,D_iR-m_0^2R=0,
\end{equation}
where $D_i$ is the covariant derivative associated with the spatial metric $\xi_{ij}$.

Multiplying the equation by $\tau R$ and integrating over the hypersurface $\Sigma$, one obtains
\begin{equation}
\int_\Sigma d^3x\,\sqrt{\xi}\,
\left(
\tau R D^iD_iR
+
R D^i\tau\,D_iR
-
m_0^2\tau R^2
\right)=0.
\end{equation}
Integrating the first term by parts yields
\begin{equation}
\left[\tau R D_iR\right]_{\partial\Sigma}
-
\int_\Sigma d^3x\,\sqrt{\xi}\,\tau
\left(
D_iR D^iR
+
m_0^2R^2
\right)
=0.
\end{equation}

Under any of the assumptions stated above, the boundary contribution vanishes. Therefore,
\begin{equation}
\int_\Sigma d^3x\,\sqrt{\xi}\,\tau
\left(
D_iR D^iR
+
m_0^2R^2
\right)
=0.
\end{equation}
Since the spatial metric $\xi_{ij}$ is positive definite for a static spacetime and $\tau>0$, both terms in the integrand are non-negative. The integral can therefore vanish only if each term vanishes identically:
\begin{equation}
D_iR=0,
\qquad
m_0^2R^2=0.
\end{equation}
Assuming $m_0^2\neq0$, it follows immediately that
\begin{equation}
R=0
\end{equation}
everywhere on $\Sigma$.
\end{proof}

As consequence of the theorem, asymptotically flat black holes, which have as boundary $\partial\Sigma$ the spatial infinity where $R\to 0,$, will have $R=0$, as said at the beginning. Furthermore, remembering that the only term in the equations of motions proportional to $\beta$ is 
$$\left(
R_{\mu\nu}
- \frac{1}{4} R \, g_{\mu\nu}
- \nabla_\mu \nabla_\nu
+ g_{\mu\nu} \Box
\right) R,$$
and when $R=0$, it goes to zero, whereby these are also solutions of a theory with $\beta = 0$.

\section{Einstein-Weyl gravity}

Einstein-Weyl gravity studies the sector of quadratic gravity with $\beta=0$ and defined by the action 
\begin{equation}\label{E-W Action}
S = \int \mathrm{d}^4 x \sqrt{-g} \left[\gamma R -\alpha C_{\mu \nu \rho \sigma}C^{\mu \nu \rho \sigma}\right].
\end{equation}
This theory includes all solutions of the full quadratic gravity theory in the sector with vanishing Ricci scalar. In particular, every static and asymptotically flat black hole solution in quadratic gravity also satisfies the equations of the Einstein-Weyl theory~\cite{Bonanno:2019rsq,Lu:2015cqa,Lu:2015psa}. 
The equations of motion are obtained by minimizing the action~\eqref{E-W Action}, yielding
\begin{equation}\label{eomham}
\mathcal{H}_{\mu\nu}=\gamma\left(R_{\mu\nu}-\frac{1}{2}R\,g_{\mu\nu}\right)-4\alpha\left(\nabla^\rho\nabla^\sigma+\frac{1}{2}R^{\rho\sigma}\right)C_{\mu\rho\nu\sigma}=0.
\end{equation}
Due to the tracelessness of the Weyl tensor, taking the trace of~\eqref{eomham} gives
\begin{equation}\label{trace}
    \mathcal{H}{^\mu_\mu}=\gamma R=0.
\end{equation}    
In general, the fourth-order nature of the field equations can be handled in different ways. First method, which we used in \cite{Bonanno:2025ksz} and first introduced in \cite{perkinsthesis}, uses the constraint in ~\eqref{trace} allowing one to eliminate higher-derivative contributions, effectively reducing the system to a set of second-order equations for the metric functions. A second commonly adopted approach\cite{Held:2022abx,Antoniou:2024jku}, that we used in \cite{Konoplya:2025afm} and we review with more details in \ref{Pert}, consists in introducing suitable auxiliary fields, which allow one to rewrite the theory in an equivalent second-order form at the level of the action and varying it with respect to them. Specifically, inserting the auxiliary field \( f_{\mu\nu} \) into the action, we obtain:

\begin{equation}
    S = \int \mathrm{d}^4 x \sqrt{-g} \left[R +2f_{\mu\nu}G_{\mu\nu}+m_2^2(f_{\mu\nu}f^{\mu\nu}-f^2) \right].
\end{equation}
Varying the modified action with respect to \( g_{\mu\nu} \) and \( f_{\mu\nu} \), we obtain the following equations of motion. For \( g_{\mu\nu} \), we have:
\begin{equation}
     \mathcal{E}^{(g)}_{\mu\nu}\equiv G_{\mu\nu}+ m_2^2 \left(f_{\mu\nu}-f g_{\mu\nu}\right)=0\, ,
\label{eq:field_g}
\end{equation}
and consequently, for $f_{\mu\nu}$ \cite{Antoniou:2024jku} 
\begin{equation}
\begin{split}
    &
    \mathcal{E}^{(f)}_{\mu\nu}\equiv
    G_{\mu\nu}+G_{(\mu}{}^\rho f_{\nu)\rho}-g_{\mu\nu}G^{\rho\sigma}f_{\rho\sigma}-fR_{\mu\nu}
    \\
    &
    +R f_{\mu\nu}+\Box f_{\mu\nu}+(\nabla_\mu\nabla_\nu-g_{\mu\nu})f
    -2\nabla_\rho\nabla_{(\mu}f_{\nu)}{}^\rho\\
    &
    +g_{\mu\nu}\nabla_\rho\nabla_\sigma f^{\rho\sigma}+m_2^2\big[\big(f^2-f_{\rho\sigma}f^{\rho\sigma}\big)/2\\
    &
    +2\big(f_\mu{}^\rho f_{\nu\rho}-f\,f_{\mu\nu}\big)\big]=0.
 \end{split}
\label{eq:field_f}
\end{equation}
We note that these equations resemble those of classical General Relativity, with this formulation, the additional degrees of freedom associated with higher-derivative terms are made explicit and the additional terms that vanish in the limit when $\alpha$ goes to zero, recovering Einstein’s equations.  Both approaches are physically equivalent, as they describe the same propagating degrees of freedom, but differ in their practical implementation. Practically, taking a static and spherically symmetric metric
\begin{equation}\label{metricQG}
\mathrm{d}s^2=-h(r)\mathrm{d}t^2+\frac{\mathrm{d}r^2}{f(r)}+r^2\mathrm{d}\Omega^2.
\end{equation}
the equations of motion reduce to a system of two second order ordinary differential equations~\cite{Lu:2015psa}:
\begin{equation}\label{eom}
\begin{split}
&4 h(r)^2 \big(r f'(r)+f(r)-1\big)+r h(r) \big(r f'(r) h'(r)
+2 f(r) \big(r h''(r)+2 h'(r)\big)\big)\\
&-r^2 f(r) h'(r)^2=0,\\[0.3cm]
&\alpha  r^2 f(r) h(r) \big(r f'(r)+3 f(r)\big) h'(r)^2+2 r^2 f(r) h(r)^2 h'(r) \big(\alpha  r f''(r)+\\&+\alpha  f'(r)-\gamma  r\big)+h(r)^3 \big(r \big(3 \alpha  r f'(r)^2-4 \alpha  f'(r)+2 \gamma  r\big)-2 f(r) \big(4 \alpha+\\& +2 \alpha  r^2 f''(r)-2 \alpha  r f'(r)+\gamma  r^2\big)+8 \alpha  f(r)^2\big)-\alpha  r^3 f(r)^2 h'(r)^3=0.
\end{split}
\end{equation}
Since these equations cannot be solved analytically, we must resort to numerical methods and approximations to extract meaningful results.
\section{Solutions in quadratic gravity}
In this section we will exploit the analytical and numerical approximation to obtain the solutions of equations \eqref{eom}. We can divide our analysis in different parts, a linear regime where an analytic approximation was found, and a non linear regime, near the origin or the horizons where we need some type of expansion and pure numerical calculations.
\subsection{Linear regime}
The weak field limit is fundamental for the physics of the theory, because it contains all the information about the classical limit, from GR to Newtonian gravity. It is important also for geometrical reason, in fact gives the asymptotic properties of the solutions. Furthermore, since its solutions are analytical it is used as boundary condition for numerical and semi-analytical calculations of the other regimes.

To derive the linearized equations of motion we consider a perturbation around a flat
metric as
\begin{equation}
h(r)=1+\epsilon V(r), \quad \quad f(r)=1+\epsilon W(r),    
\end{equation}
and expand the equations of motion at linear order of $\epsilon$, getting
\begin{equation}\begin{split}
    G_\mu^\mu=&\gamma (\nabla^2 V(r)+2Y(r))=0,\\
    G_i^i-G_t^t=&\frac{4}{3}\alpha \nabla^2 Y(r)-\frac{4}{3}\alpha \nabla^2 \nabla^2 V(r)+\gamma\nabla^2 V(r)=0,
\end{split} \end{equation}
where $Y(r)=r^{-2}(r W(r))'$ and having imposed $\epsilon=1$. Using the Fourier transform it is possible to solve the system analytically, obtaining \cite{Stelle:1977ry,Lu:2015cqa,Silveravalle:2023lnl}:
\begin{equation}\begin{split}
   h(r)=\, &1+C_t-\frac{2\,M}{r}+2S^-_2 \frac{e^{-m_2\, r}}{r}+2S^+_2 \frac{e^{m_2\, r}}{r}, \\
f(r)=\, &1-\frac{2\,M}{r}+S^-_2 \frac{e^{-m_2\, r}}{r}(1+m_2\, r)+S^+_2 \frac{e^{m_2\, r}}{r}(1-m_2\, r), 
\end{split}
\end{equation}
with $m_2^2=\gamma/ 2\alpha$ the mass of the tensor mode. Remarkably, we obtain the
exact Schwarzschild solution plus corrections. Imposing asymptotic flatness, i. e. Requiring that $h(r),f(r)\to 1$ as $r\to\infty$, set $C_t , S_2^+=0$, leading to the weak field solutions:
\begin{equation}\begin{split} \label{eq|weakfield}
h(r)=\, &1-\frac{2\,M}{r}+2S^-_2 \frac{e^{-m_2\, r}}{r}, \\
f(r)=\, &1-\frac{2\,M}{r}+S^-_2 \frac{e^{-m_2\, r}}{r}(1+m_2\, r),
\end{split}\end{equation}
where $M$ is $G$ times the ADM mass. These solutions are exponentially suppressed corrections to the Schwarzschild metric; in particular, the correction to the gravitational potential in the Newtonian limit $\phi(r)=\frac{1}{2}(h(r)-1)$ takes precisely the form of a Yukawa potential, as expected from the presence of a massive mediator. For this reason, we define the parameter $S_2^-$ as the Yukawa charge of the solution. Once the parameters $M$ and $S_2^-$ are specified, the gravitational properties of the solutions at large distances are completely determined.

\subsection{Expansion at fixed radius} \label{st_classification}
To classify different families of solutions, it is crucial to analyze the behavior of the metric near the origin or a metric singularity which characterizes their specific physical nature. It is possible to use a variation of the Frobenius method to find the possible behaviors and to determine different families of solutions. Taking an expansion of the metric functions in the form
\begin{equation} \label{eq|frobenius}
\begin{split}
h(r)&=\left(r-r_0\right)^t\!\left[\displaystyle\sum_{n=0}^N h_{t+\frac{n}{\Delta}}\left(r-r_0\right)^{\frac{n}{\Delta}}\!+O\!\left(\left(r-r_0\right)^{\frac{N\!+\!1}{\Delta}}\right)\!\right],\\
f(r)&=\left(r-r_0\right)^s\!\left[\displaystyle\sum_{n=0}^N f_{s+\frac{n}{\Delta}}\left(r-r_0\right)^{\frac{n}{\Delta}}\!+\!O\left(\left(r-r_0\right)^{\frac{N\!+\!1}{\Delta}}\right)\!\right],
\end{split}
\end{equation}
with $h_0, f_0 \ne 0$, and the parameter $\Delta$ which determines whether the power series contains integer or fractional powers, it is possible to solve the equations of motion order by order in $(r-r_0)$. The equation with lowest order determine the admissible values for $s$ and $t$, while the higher orders establish the given free parameters. The families of solutions can then be classified by the label $(s,t)_{r_0}^\Delta$~\cite{Lu:2015psa,Podolsky:2019gro,Silveravalle:2023lnl,Silveravalle:2022wij}. The relevant families of solutions are \ref{tab:solution_families}:
\begin{table}[h]
\centering
\begin{tabular}{c c }
\hline
\textbf{Family} & \textbf{Physical interpretation} \\
\hline

$(0,0)_{0}^{1}$ 
& Vacuum \\

$(1,1)_{r_H}^{1}$ 
& Black holes \\

$(1,0)_{r_T}^{2}$ 
& Non-symmetric wormholes \\

$(\text{-}1,\text{-}1)_{0}^{1}$ 
& Repulsive naked singularities \\

$(\text{-}2,2)_{0}^{1}$ 
& Attractive naked singularities  \\

\hline
\end{tabular}
\caption{Main families of static and spherically symmetric solutions in Einstein--Weyl gravity classified according to the near-origin (or near-throat/horizon) behavior of the metric functions.}
\label{tab:solution_families}
\end{table}

 \paragraph{$\mathbf{(0,0)_{0}^1}$ - family.} This family contains all solutions whose curvature invariants remain finite at the origin. Since physically acceptable compact stars are expected to be regular everywhere, their interior geometry belongs to this class. In the presence of matter, solving the Einstein--Weyl field equations coupled to a perfect-fluid stress-energy tensor \cite{Silveravalle:2023lnl} yields the following near-origin expansions for the metric functions and matter variables:
 \begin{equation}
     \begin{split}     
h(r) &= h_0 \Big[
1 
+ h_2 r^2
+ h_4(h_2, f_2, \rho_0, p_0)\, r^4
+ \mathcal{O}(r^6)
\Big], \\
f(r) &= 1
+ f_2 r^2
+ f_4(h_2, f_2, \rho_0, p_0)\, r^4
+ \mathcal{O}(r^6), \\
\rho(r) &= \rho_0
+ \rho_2(h_2, f_2, \rho_0, p_0)\, r^2
+ \mathcal{O}(r^4), \\
p(r) &= p_0
+ p_2(h_2, f_2, \rho_0, p_0)\, r^2
+ \mathcal{O}(r^4)\,.
\end{split}
 \end{equation}
where the parameters $\rho_0$ and $p_0$ are related by the equation of state of the fluid. For further details, see \cite{Bonanno:2021zoy,Silveravalle:2023lnl,perkinsthesis}.

 \paragraph{$\mathbf{(1,1)_{r_H}^1}$ - black holes.} A simultaneous root of the functions $h(r)$ and $f(r)$ defines an event horizon. In Einstein-Weyl gravity, black hole solutions have standard horizons, with the metric functions expandable in integer powers starting at the linear term. Besides the Schwarzschild solution, there are also black holes with non-Schwarzschild metrics. Both types of solutions depend on a single free parameter and, in the $(M,S_2^-)$ plane, they form two distinct lines: one for Schwarzschild black holes and one for the non-Schwarzschild ones, which intersect at a specific mass. We will resume the analysis of black holes with the further description of some characteristics and their perturbations in \ref{BH}. Moreover, for further details, see~\cite{Bonanno:2019rsq,Lu:2015cqa,Goldstein:2017rxn}.

\paragraph{$\mathbf{(1,0)_{r_T}^2}$ - non-symmetric wormholes.}  Curvature invariants are regular at $r = r_0,$ so we can extend spacetime after this point, and a root of the function $f(r)$ that is not a root of $h(r)$ defines a wormhole throat. In Einstein–Weyl gravity, the metric near the throat must be expanded in half-integer powers.
\begin{equation}
\begin{split}
h(r) =& h_0 \Bigg[
1 
+ h_{1/2}(r - r_0)^{1/2}
+ h_1 (r - r_0)+\\&
+ h_{3/2}(h_{1/2}, f_1, h_1, f_{3/2}, r_0)\,(r - r_0)^{3/2}+
\\&+ \mathcal{O}\big((r - r_0)^2\big)
\Bigg], \\
f(r) =& f_1 (r - r_0)
+ f_{3/2} (r - r_0)^{3/2}
+ f_2(h_{1/2}, f_1, h_1, f_{3/2}, r_0)\,(r - r_0)^2+
\\&+ \mathcal{O}\big((r - r_0)^{5/2}\big)\,,
\end{split}
\end{equation}
with $h_0$, $h_{1/2}$, $f_1$, $h_1$, $f_{3/2}$ and $r_0$ as free parameters to be fixed from constrains.
This indicates that the throat is non-symmetric, and it connects an asymptotically flat spacetime to one where the metric vanishes asymptotically and a singularity lies at infinity. These solutions at the end depend on two free parameters and, in the $(M,S_2^-)$ plane, they fill most of the region with positive mass and an attractive Yukawa charge ($S_2^-<0$). For further details, see~\cite{Lu:2015psa,Bonanno:2022ibv,perkinsthesis}.

\paragraph{$\mathbf{(\text{-}1,\text{-}1)_{0}^1}$ - repulsive (standard) naked singularities.} One of the allowed behaviors near the origin for the metric of vacuum solutions is that both $h(r)$ and $f(r)$ diverge as $r^{-1}$ 
\begin{align}
h(r) &= h_{-1} \left(
\frac{1}{r}
+ \frac{1}{f_{-1}}
+ h_2 r^2
+ h_3(f_{-1},h_2,f_2)\, r^3
+ \mathcal{O}(r^4)
\right), \\
f(r) &= \frac{f_{-1}}{r}
+ 1
+ f_2 r^2
+ f_4(f_{-1},h_2,f_2)\, r^3
+ \mathcal{O}(r^4)\,.
\end{align}
This is the same behavior as in the Schwarzschild metric, but in Einstein–Weyl gravity it can also occur for naked singularities with a repulsive gravitational potential at the origin. In fact, this family is singular at the origin, with the curvature invariants behaving as
\begin{align}
R^2 &\sim C(f_{-1},h_2,f_2) + \mathcal{O}(r), \\
R_{\mu\nu}R^{\mu\nu} &\sim C(f_{-1},h_2,f_2) + \mathcal{O}(r), \\
R_{\mu\nu\rho\sigma}R^{\mu\nu\rho\sigma} &\sim r^{-6} + \mathcal{O}(r^{-5})\,.
\end{align}
It appears that for such naked singularities, the power series expansion must be refined with logarithmic corrections, which, however, do not alter the qualitative features of the solutions~\cite{Silveravalle:2023lnl}. These solutions depend on two free parameters and, in the $(M,S_2^-)$ plane, they fill most of the region with negative mass, as well as the region with positive mass and repulsive Yukawa charge ($S_2^->0$). For further details, see~\cite{Lu:2015psa,Silveravalle:2022wij,perkinsthesis}.

\paragraph{$\mathbf{(\text{-}2,2)_{0}^1}$ - attractive (veiled) naked singularities.} The other allowed behavior near the origin for the metric of vacuum solutions is that $h(r)$ vanishes as $r^2$ and $f(r)$ diverge as $r^{-2}$, making all terms in the metric vanish as $r^2$. This behavior is specific to quadratic gravity and corresponds to naked singularities with an attractive gravitational potential. These solutions depend on two free parameters and, in the $(M,S_2^-)$ plane, they occupy small regions near the black hole lines. This family of solutions has been extensively studied by Holdom as black hole mimickers, and we make a general in-depth analysis of the solution in \ref{VS} \cite{Bonanno:2025ksz}. For further details, see~\cite{Lu:2015psa,perkinsthesis,Holdom:2002xy,Holdom:2016nek,Holdom:2019bdv,Holdom:2020onl}. 

\subsubsection{Phase diagram}
\begin{figure}[h]
\centering
\includegraphics[width=0.9\columnwidth]{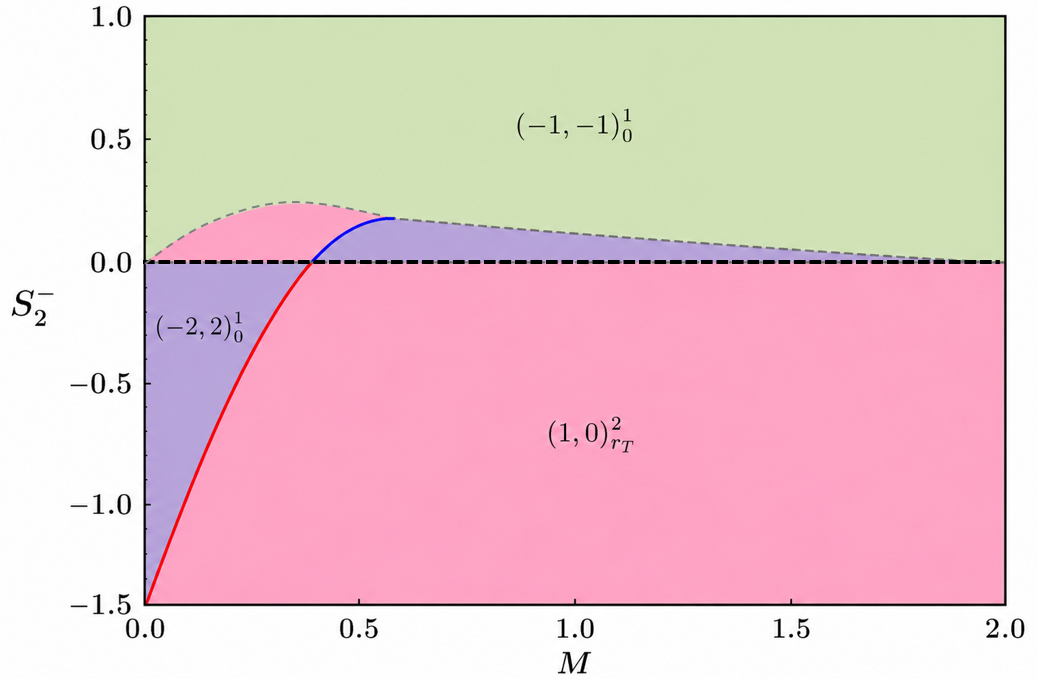}
\caption{Phase diagram of vacuum solutions in Einstein-Weyl theory. The dashed black line indicates Schwarzschild black holes, while the solid red and blue lines indicate non-Schwarzschild black holes with respectively attractive and repulsive Yukawa charge.} 
\label{Phase diagram}
\end{figure}

After classifying the solution respect their behavior near the origin, singularities or fixed radius, we can plot a phase diagram, as done in \cite{Silveravalle:2022wij}, of these solutions spanned by the parameters $M$ and $S_2^-$, in units of $m_2$, to observe the regions which associate the weak fields parameters to their characteristic near the origin. In our work we restricted on the attractive mass ($M>0$) region, since they are the most phenomenologically and astrophysical intuitive and probable solutions. The diagram is shown in \ref{Phase diagram};

We can note that the diagram is divided in five zones, one characterized by the $\mathbf{(\text{-}1,\text{-}1)_{0}^1}$ solutions, which populate a big area with $S_2^->0$, two regions described by $\mathbf{(1,0)_{r_T}^2}$, these non-symmetric wormholes are present in the most part of the attractive Yukawa charge ($S_2^-<0$) and in a little area with small repulsive Yukawa charge ($S_2^->0$). The last two region are populated by the attractive naked singularities, namely $(\text{-}2,2)_{0}^1$, they represent a zone with repulsive Yukawa charge but with a mass much larger than the Yukawa charge, and a small region with an attractive mass and an attractive Yukawa charge, but where both take small values. Finally, in this space black holes populate only zero-measure regions of the diagram, and are represented by two different line for the two families, Schwarzschild with $S_2^-=0$ and non-Schwarzschild black holes. Moreover, the non-Schwarzschild are depicted with two different colors indicate whether they have negative or positive Yukawa charge. The specific
location of these lines separate different families of solution analogously at thermodynamic phase diagram they represent a sort of "phase transition". It is worth noting that astrophysical black holes should belong to the Schwarzschild family, but recently it has been proposed that during the final stages of evaporation a phase transition can occur that can lead to the other branch \cite{Bonanno:2024fcv} and the endpoint of this process is a static attractive naked singularity \cite{Bonanno:2025ffy,Bonanno:2025ohx}. Since there is no extensive analysis of this family we have concentrated on the analysis of the perturbative dynamics of this family as well as of black holes alone.

\subsection{Numerical methods and semi-analytical approximation for non linear regime} \label{NM}

Having shown in the previous subsections the approximations used to describe the metric in various regimes, we present here the main numerical tools used to investigate the non-linear regime and the global structure of the solutions. In particular, in our works we used two different methods; the first is numerical, where our analysis is divided into at least two spatial regions, large distances in the asymptotic weak-field regime that will be described by the expansion shown in \eqref{eq|weakfield}, and from a large radius $x_\infty$ down to a point close to the origin where is sufficient to integrate the equations \eqref{eom} using \eqref{eq|weakfield} as initial conditions. The integration was carried out in \textit{Fortran} using the adaptive step size Runge-Kutta integrator \texttt{DO2PDF} implemented by the N.A.G. group (see \href{https://www.nag.com}{https://www.nag.com})~\cite{Press:1992zz}. It is important to note that for some cases to study the solutions in greater detail, however, it is necessary to use a more refined method, near the origin, in particular a shooting method can be implemented, dividing in more small regions the integration (see \cite{Silveravalle:2023lnl} for details). 

The second method, instead, is based on a semi-analytical approximation and can be more efficient for successive numerical computation. Adopting the general parametrization for asymptotically flat, spherically symmetric black holes in arbitrary metric theories of gravity, known as the Rezzolla–Zhidenko parametrization~\cite{Rezzolla:2014mua}. This approach is conceptually similar to the post-Newtonian parametrization, but unlike the latter, it remains valid across the entire domain outside the event horizon, extending all the way to spatial infinity.

The Rezzolla–Zhidenko framework provides a systematic and convergent expansion that captures both near-horizon and asymptotic behavior. It was later generalized to axially symmetric \cite{Konoplya:2016jvv,Younsi:2016azx} black holes and has been widely used in astrophysically motivated applications, including the analytic approximation of numerical black hole solutions. 
This parametrization has proven especially useful for modeling deviations from general relativity in a theory-independent way.

The semi-analytical approximation for the metric function based on the Rezzolla-Zhidenko parametrization \cite{Rezzolla:2014mua} was introduced in \cite{Kokkotas:2017zwt},\\ which reproduces the correct asymptotic behavior at infinity and at the event horizon. It has been verified against numerical solutions of eq.~\eqref{eom} \cite{Kokkotas:2017zwt}. The metric functions are defined as
\begin{align}
    h(r) & \equiv x\,A(x)\,,\\
    \frac{h(r)}{f(r)} & \equiv B(x)^2\,,\label{eq:fh}
\end{align}
where $x = 1-r_h/r$, is a compact coordinate, with $r_h$ being the radius of the black hole event horizon. The functions $A(x)$ and $B(x)$ are parameterized as
\begin{align}
    A(x)=&1-\epsilon (1-x)-\epsilon(1-x)^2+\tilde{A}(x)(1-x)^3\,,\nonumber\\
    B(x)=&1+\tilde{B}(x)(1-x)^2\,,
\end{align}
and expressed in terms of the following continued fraction expansions to describe the metric near  the event horizon $x=0$. Fixing the expansion parameter $\epsilon$ as 
$$\epsilon=-\left(1-\frac{2M}{r_h}\right),$$
we obtain
\begin{equation}
\tilde{A}(x)=\frac{\tilde{a}_1}{1+\frac{\tilde{a}_2 x}{ 1+\frac{\tilde{a}_3 x}{1+\frac{\tilde{a}_4 x}{1+\ldots}}}}\,,
\;\tilde{B}(x)=\frac{\tilde{b}_1}{1+\frac{\tilde{b}_2 x}{1+\frac{\tilde{b}_3x}{1+\frac{\tilde{b}_4 x}{1+\ldots}}}}\,,
\end{equation}
where the coefficients depend on the dimensionless parameter $p=r_h m_2$ and are written explicitely in Appendix \ref{appendix}. 

We used both methods introduced, depending on the purpose of the work, since the semi-analytic method is clearly simpler and more effective to use in numerical calculations, but its validity extends only in certain parameter ranges (for example, where black holes are stable) \cite{Held:2022abx,Kokkotas:2017zwt}. Therefore, when performing a completely agnostic analysis regarding the parameter range and the stability of compact objects, it is preferable to integrate numerically.

\chapter{Dynamics of Perturbations in Quadratic Gravity}

In this chapter, we will discuss perturbations to black holes and veiled singularities in Einstein-Weyl gravity. Specifically, we will first delve into the metrics and characteristics of the solutions, then study their perturbative dynamics, focusing on stability.

\section{Black holes} \label{BH}

These solution have been studied in depth over the years, so their characteristics are well known. As showed in \ref{st_classification} they describe the behavior of the $(1,1)^1_{r_0}$ family, so has the general form
\begin{align}
h(r) &= h_1 (r - r_0) + h_2(f_1, f_2, r_0)\,(r - r_0)^2 + \mathcal{O}\!\left((r - r_0)^3\right), \\
f(r) &= f_1 (r - r_0) + f_2 (r - r_0)^2 + \mathcal{O}\!\left((r - r_0)^3\right),
\end{align}
and all the the curvature invariants are regular at $r=r_0$, which is an horizon $r_h$.

The main novelty is the discovery of a class of non-Schwarzschild black holes.
These solutions are characterized by a non-vanishing Yukawa charge $S_2^-$ and, as a consequence, by two distinct metric functions $h(r)$ and $f(r)$, with $h(r)\neq f(r)$.
Depending on the sign of the Yukawa charge, one can distinguish two branches, corresponding to attractive and repulsive Yukawa contributions.

These branches meet the Schwarzschild solution, for which the Yukawa charge vanishes, at a crossing point corresponding to a specific mass.
This mass may range from asteroid-scale values down to the Planck mass \cite{Bonanno:2024fcv}, depending only on the value of a constrained free parameter obtained from torsion-balance tests \cite{Giacchini:2016nta}.

We should note that (see \ref{Phase diagram} or \cite{Silveravalle:2023lnl,Silveravalle:2022wij} for the phase diagram of the solution) Yukawa repulsive black holes $(S_2^- >0)$ are present only for small horizon radius and small positive masses, while Yukawa attractive black holes $(S_2^- <0)$ have unbounded horizon radii and unbounded negative masses. The most interesting characteristic of this type of black hole, however, is that the mass decreases as the horizon radius increases, precisely the opposite of what happens in General Relativity. These properties already set some constraints on the presence of non-Schwarzschild black holes in our universe, as they are either large but gravitationally repulsive or small both in size and mass, and then they cannot model astrophysical black holes. It is therefore reasonable to expect astrophysical processes to produce Schwarzschild black holes, making the crossing point relevant only in the final stages of black hole evaporation, while the resolution of the singularity problem may require further extensions of the theory such as the inclusion of even higher-order derivative terms~\cite{Giacchini:2024exc,Giacchini:2025gzw}.

\begin{figure}[h]
\centering
\includegraphics[width=0.9\textwidth]{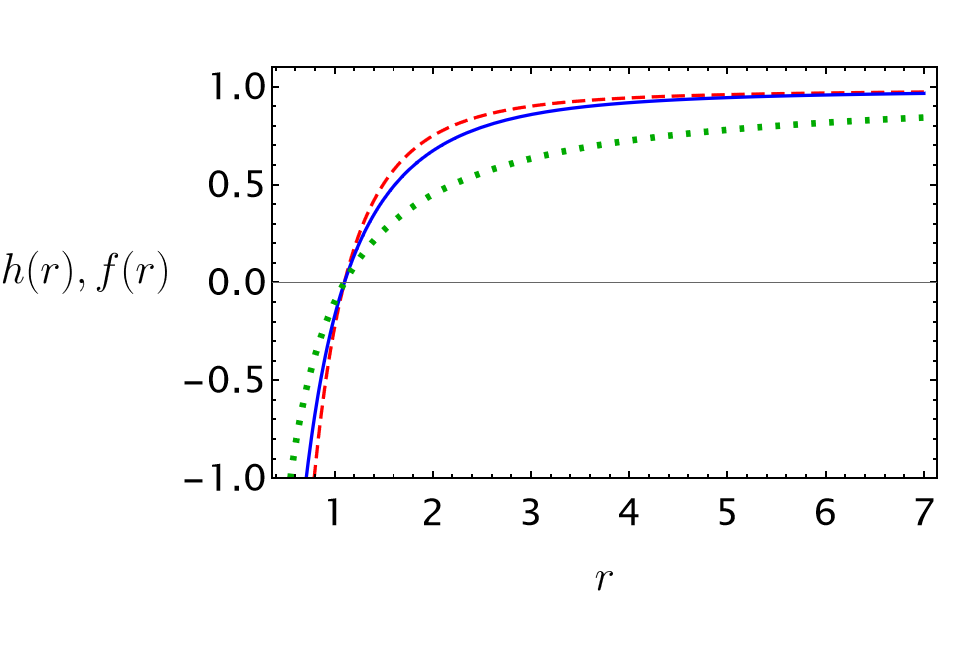}\\
\caption{We plot the metric functions $h(r)$, in red dashed line, and $f(r)$, in blue solid line, for a Yukawa attractive Non-Schwarzschild black hole with $r_h=1.1$. The green dotted line corresponds to the Schwarzschild black hole with the same horizon radius.}
\label{fig|metricBH}
\end{figure}

We plot, as example, in figure \ref{fig|metricBH} the metric functions obtained using the semi-analytical approximation illustrated in \ref{NM} with all the characteristic described before (for further example and insights see \cite{Bonanno:2019rsq,Silveravalle:2023lnl}) .

\section{Veiled singularities} \label{VS}

In our work \cite{Bonanno:2025ksz} we focus our attention on a remarkable class of solutions arising within the framework of Einstein--Weyl gravity that have no counterparts in classical General Relativity: the so-called \emph{attractive naked singularities} of the $(-2,2)$ type introduced in \ref{st_classification}. They represents an important region of the phase diagram \ref{Phase diagram} of the parameter, in fact, they are solutions of the region with an attractive mass ($M>0$) and repulsive Yukawa charge ($S_2^->0$) but with a mass much larger than the Yukawa charge, and finally a small region with an attractive mass and an attractive Yukawa charge, but where both take small values.

\subsection{From $\mathbf{(\text{-}2,2)}$ solutions to veiled singularities}

The solutions we consider are defined by the absence of an event horizon or a wormhole throat, and by the metric behaving as
\begin{equation}
    \mathrm{d}s^2\sim r^2\left(-h_{2}\mathrm{d}t^2+\frac{\mathrm{d}r^2}{f_{-2}}+\mathrm{d}\Omega^2\right)
\end{equation}
near the origin, with $h_{2}$ and $f_{-2}$ being constants. These solutions have been referred to as $(2,2)$ solutions~\cite{Lu:2015psa}, $(2,2)$-holes~\cite{Holdom:2016nek}, Bachian singularities~\cite{Podolsky:2019gro}, Holdom stars~\cite{Silveravalle:2022wij}, or attractive naked singularities~\cite{Silveravalle:2023lnl,DelPorro:2025wts}; however, we believe that a complete analysis of these solutions as viable physical objects is still lacking. In particular, the extensive work of Bob Holdom~\cite{Holdom:2002xy,Holdom:2016nek,Holdom:2019bdv,Holdom:2020onl,Holdom:2022zzo,Holdom:2022fsm} has focused on a thin-shell model, which can be tuned to mimic a black hole, rather than on the pure vacuum solutions of quadratic gravity. In this section we investigate these geometries in greater detail, showing that the local $(-2,2)_{0}^{1}$ behavior near the origin naturally extends to a global spacetime structure that can be interpreted as a \emph{veiled singularity}. In
this and following sections regarding the \emph{veiled singularities}, we made every quantity dimensionless using the mass $m_2$ as units.

\subsubsection{Characterization of the singularity}

The metric close to the origin can be expanded as
\begin{equation}\label{eq|expansion}
    \begin{split}
        h(r)=&\ h_2\,r^2\left(1-\frac{f_{-1}}{f_{-2}}r+\frac{7 f_{-1}^2+f_{-2} \left(2-6 f_0\right)}{8 f_{-2}^2}r^2+O\left(r^3\right)\right),\\
        f(r)=&\ \frac{f_{-2}}{r^2}\left(1+\frac{f_{-1}}{f_{-2}}r+\frac{f_0}{f_{-2}}r^2+O\left(r^3\right)\right),
    \end{split}
\end{equation}
where all higher-order coefficients can be expressed in terms of $h_2, f_{-2}, f_{-1}$ and $f_0$. While the Ricci scalar is identically zero due to equation~\eqref{trace}, the contracted Ricci, Riemann, and Weyl tensors diverge at the origin as
\begin{equation}\label{eq|curvature}
    \begin{split}
        R^{\mu\nu}R_{\mu\nu}=&\ 12 f_{-2}^2\ r^{-8}+O\left(r^{-7}\right),\\
        R^{\mu\nu\rho\sigma}R_{\mu\nu\rho\sigma}=&\ 24 f_{-2}^2\ r^{-8}+O\left(r^{-7}\right),\\
        C^{\mu\nu\rho\sigma}C_{\mu\nu\rho\sigma}=&\ \frac{3}{16} \left(2\left(1+f_0\right)-\frac{f_{-1}^2}{f_{-2}}\right)^2r^{-4}+O\left(r^{-3}\right),
    \end{split}
\end{equation}
which clearly signals a divergent curvature structure at the origin. Nevertheless, despite the divergence of the curvature invariants, the Weyl contribution remains integrable. Including the measure, the integral of the squared Weyl tensor reads
\begin{equation}\label{eq|intweyl}
\begin{split}
    &\int\mathrm{d}^4x\sqrt{-g}C^{\mu\nu\rho\sigma}C_{\mu\nu\rho\sigma}=\\ &\int\mathrm{d}^4x\Bigg[\sqrt{\frac{h_2}{f_{-2}}}r^4\sin(\theta)\frac{3}{16} \left(2\left(1+f_0\right)-\frac{f_{-1}^2}{f_{-2}}\right)^2r^{-4}+O(r)\Bigg]= \\
    &\,\int\mathrm{d}t\left[\frac{3}{4}\pi\sqrt{\frac{h_2}{f_{-2}}} \left(2\left(1+f_0\right)-\frac{f_{-1}^2}{f_{-2}}\right)^2 r+O\left(r^2\right)\right],
    \end{split}
\end{equation}
showing that, while the spacetime possesses a curvature singularity, its action remains finite. The approach of geodesics toward the singularity can be quantified by considering trajectories, characterized by conserved energy $E$ and angular momentum $L$, in coordinate and proper time given by~\cite{Silveravalle:2023lnl}
\begin{equation}\label{eq|time}
\begin{split}
    \Delta t=&\ \sqrt{\frac{E^2}{h_2f_{-2}\left(E^2-h_2L^2\right)}}\ r+O\left(r^2\right),\\
        \Delta\tau=&\ \sqrt{\frac{h_2}{f_{-2}\left(E^2-h_2L^2\right)}}\ \frac{r^3}{3}+O\left(r^4\right),
\end{split}
\end{equation}
whenever expansion~\eqref{eq|expansion} holds, independently of whether the geodesic is timelike or null (in the latter case, the proper time $\tau$ must be interpreted as an affine parameter). To further characterize the effect of the singularity on nearby geodesics, we consider the geodesic deviation vector $z^\mu$, which describes the infinitesimal separation between neighboring geodesics in the congruence. For a congruence of geodesics rotating in the equatorial plane, the corresponding deviation vectors for timelike and null geodesics, denoted by $z_t^\mu$ and $z_n^\mu$, respectively, behave near the origin as
\begin{equation}\label{eq|congruence}
    z_t^\mu\sim\begin{pmatrix}
        c_1\log(r)\\
        -c_1\sqrt{\frac{h_2f_{-2}\left(E^2-h_2L^2\right)}{E^2}}\log(r)\\
        -c_1\frac{h_2L}{E}\log(r)\\
        c_2r
    \end{pmatrix},\quad
    z_n^\mu\sim\begin{pmatrix}
        c_1\\
        c_2\\
        C_3\left(c_1,c_2,E,L,h_2,f_{-2}\right)L\\
        c_4r
    \end{pmatrix},
\end{equation}
in the timelike and null cases, respectively. While the congruence of null geodesics remains relatively well-behaved, timelike geodesics tend to become infinitely separated as they approach the origin; in other words, a timelike observer would be disrupted by diverging tidal forces. Taken together, Eqs.~(\ref{eq|curvature}--\ref{eq|congruence}) demonstrate that $(-2,2)_0^1$ solutions are naked singularities.

Although infalling observers encounter divergent tidal effects, the ability of signals to emerge from the singularity requires a separate analysis. In particular, recall that the redshift of a photon emitted at radius $r$ and measured at infinity is $z=h(r)^{-1/2}-1$. Using the expansion~\eqref{eq|expansion}, we obtain
\begin{equation}\label{eq|redshift}
    z=\frac{1}{\sqrt{h(r)}}-1\sim\frac{1}{\sqrt{h_2}\ r}\ \xrightarrow[r\to 0]{}\ \infty.
\end{equation}
As a consequence, radiation originating arbitrarily close to the singularity is infinitely redshifted before reaching a distant observer, or, equivalently, only a particle with infinite energy could escape. The singularity is in fact a degenerate Killing horizon for the time-like Killing vector $\xi^\mu\partial_\mu=\partial_t$, and no information can genuinely escape it. This behavior can be traced back to the attractive nature of the gravitational field, which persists all the way to the singularity, in contrast with the standard naked singularities of General Relativity, where the field becomes repulsive and leads to the emission of high-energy photons. While being naked singularities from a causal point of view, $(-2,2)_0^1$ solutions are protected by a \emph{weaker} form of the cosmic censorship conjecture, in which no particle with finite energy can leave the singularity, and are therefore \emph{veiled}.

\subsubsection{Beyond Frobenius analysis and veil radius}

\begin{figure}[H]
\centering
\includegraphics[width=\textwidth]{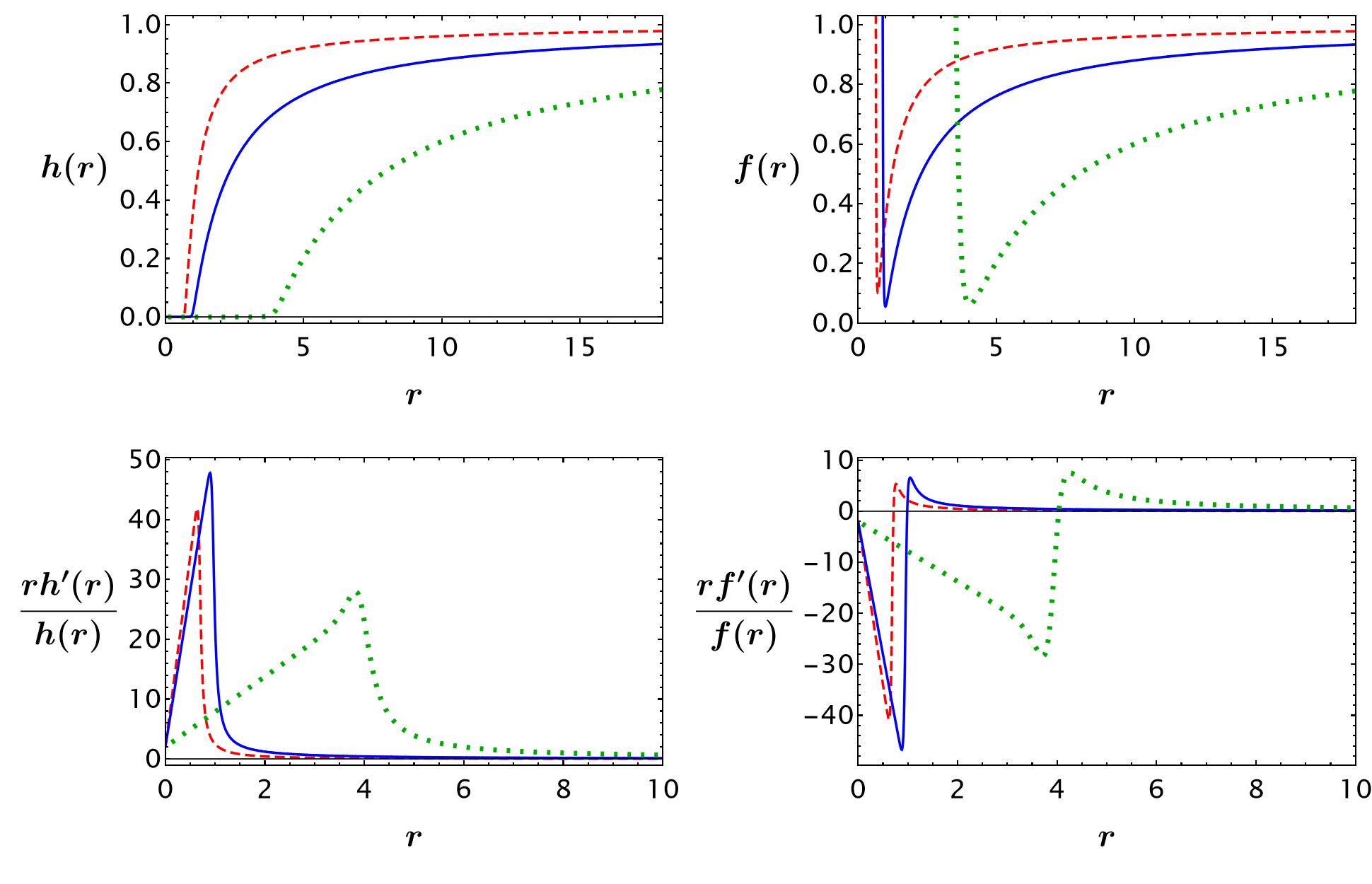}\\
\caption{In the top panels we plot the metric functions $h(r)$ and $f(r)$ for three different parameter choices, with the blue solid line corresponding to $M=0.6$ and $S^-_2=0.05$, the red dashed line to $M=0.2$ and $S^-_2=-0.2$, and the green dotted line to $M=2.0$ and $S^-_2=0.01$. The bottom panels show the logarithmic derivatives $\frac{rh'(r)}{h(r)}$ and $\frac{rf'(r)}{f(r)}$, which approach $2$ for $h(r)$ and $-2$ for $f(r)$ as $r\to 0$, in agreement with the expected near-origin scaling.}
\label{fig|metric}
\end{figure}

Although the Frobenius expansion of Eq.~\eqref{eq|expansion} captures the local structure of the singularity, it provides only a limited description of the full spacetime geometry. A comparison with the numerical solutions displayed in the upper panels of Fig.~\ref{fig|metric}, where we show the two metric functions, $h(r)$ and $f(r)$, for three different veiled singularities, highlights the limitations of the near-origin expansion, and it is clear that expansion~\eqref{eq|expansion} fails to reproduce their behavior in the region surrounding the singularity. In particular, the lapse function $h(r)$ drops to zero already at radii close to the Schwarzschild value $r\sim 2M$, and appears to remain vanishing all the way to the origin. The radial metric function $f(r)$, by contrast, exhibits a sharp spike near the same radius and continues to grow as $r$ decreases.

The behavior is not a consequence of numerical inaccuracies, indeed feature is confirmed by the logarithmic derivatives of the metric functions, $\frac{\mathrm{d}\log h(r)}{\mathrm{d}\log r}=\frac{rh'(r)}{h(r)}$ and $\frac{\mathrm{d}\log f(r)}{\mathrm{d}\log r}=\frac{rf'(r)}{f(r)}$ shown in the bottom panels of Fig.~\ref{fig|metric} confirm that the Frobenius expansion in~\eqref{eq|expansion} is valid only extremely close to the origin, and that exponential corrections must be included.


Considering the region where the log-log derivatives can be approximated linearly, the metric functions behave as
\begin{equation}
    \begin{array}{cc}
         & \frac{rh'(r)}{h(r)}\sim2+a\,r  \\[0.3cm]
         & \frac{rf'(r)}{f(r)}\sim-2-a\,r 
    \end{array}
    \qquad\implies\qquad
    \begin{array}{cc}
         & h(r)\sim h_2 r^2\,\mathrm{e}^{a\,r}  \\[0.3cm]
         & f(r)\sim \frac{f_{-2}}{r^2}\,\mathrm{e}^{-a\,r}
    \end{array}
\end{equation}
with $a,\, h_2$ and $f_{-2}$ as free parameters. This form still does not provide the full set of integration constants, it matches Eq.~\eqref{eq|expansion} only at next-to-leading order, and it does not satisfy the equations of motion, but it reproduces the qualitative behavior of the solution. An exponential dependence of this kind already appeared in the asymptotic region of non-symmetric wormholes. More generally, expanding around a vanishing metric as
\begin{equation}
\begin{split}
    h(r)=& \epsilon\, h_1(r)\left(1+\epsilon\, h_2(r)+O\left(\epsilon^2\right)\right),\\ f(r)=&\frac{1}{\epsilon\, f_1(r)\left(1+\epsilon\, f_2(r)+O\left(\epsilon^2\right)\right)},
\end{split}
\end{equation}
one finds at leading order in $\epsilon$, that is $O\left(\epsilon^{-1}\right)$, the solutions
\begin{equation}\label{eq|expo}
    h_1(r)=h_2 r^2\,\mathrm{e}^{a\,r},\qquad\qquad f_1(r)= \frac{f_{-2}}{r^2}\,\mathrm{e}^{-a\,r},
\end{equation}
while at the next order the missing parameters are recovered~\cite{Bonanno:2022ibv}. Non-symmetric wormholes appear for negative values of $a$ in the $r\to\infty$ region, while veiled singularities appear for positive values of $a$ in the $r\to 0$ region.

This motivates the introduction of a characteristic scale, which we refer as the veil radius, defined as the outermost radius for which the exponential approximation~\eqref{eq|expo} remains valid:
\begin{equation}\label{eq|effective}
    r_{\rm v}:= \max\left\{r\ |\ \left|\frac{rh'(r)}{h(r)}-(2+a\,r)\right|<\delta\ \land\ \left|\frac{rf'(r)}{f(r)}+(2+a\,r)\right|<\delta\right\},
\end{equation}
with $\delta$ small (in practice we take $\delta=10^{-3}$). Below this radius, the lapse function is essentially zero and the radial function $f(r)$ becomes very large, due to the smallness of $h_2$ (of order $10^{-15}$–$10^{-20}$) and the largeness of $f_{-2}$ (of order $10^{10}$–$10^{15}$). Within this approximation, the curvature invariants behave as
\begin{equation}\label{eq|curvaturenf}
    \begin{split}
        R^{\mu\nu}R_{\mu\nu}\sim &\ 2\,\mathrm{e}^{-2\,a\,r}\left(6+4\,a\,r+a^2r^2\right) f_{-2}^2\ r^{-8},\\
        R^{\mu\nu\rho\sigma}R_{\mu\nu\rho\sigma}\sim &\ 4\,\mathrm{e}^{-2\,a\,r}\left(6+4\,a\,r+a^2r^2\right) f_{-2}^2\ r^{-8},\\
        C^{\mu\nu\rho\sigma}C_{\mu\nu\rho\sigma}\sim &\ 0,
    \end{split}
\end{equation}
so both the Ricci and Riemann contractions are large, and the spacetime is effectively in a regime of extreme curvature. 
Geodesics falling to the origin reach it in coordinate and proper time
\begin{equation}\label{eq|timenf}
\begin{split}
        \Delta t\sim &\ \frac{1}{\sqrt{h_2f_{-2}}}\ r,\\
        \Delta\tau\sim &\ \sqrt{\frac{h_2}{f_{-2}}}\ \frac{\mathrm{e}^{a\,r}\left(2-2a\,r+a^2r^2\right)}{a^3E},
\end{split}
\end{equation}
where the angular momentum has no significant role, indicating that trajectories are essentially radial. The coordinate time is unaffected by the exponential factor, and the combination $h_2f_{-2}$ ensures it remains of moderate size; the proper time, instead, quickly tends to zero because $h_2/f_{-2}$ is very small. This mismatch already signals a large redshift for photons emitted inside $r_{\rm v}$ and received at infinity:
\begin{equation}\label{eq|redshiftnf}
    z=\frac{1}{\sqrt{h(r)}}-1\sim\frac{1}{\sqrt{h_2}\ r\ \mathrm{e}^{a\,r}}-1 \gg 1,\qquad\text{if }\ h_2 \ll 1.
\end{equation}
Radiation from inside the veil thus reaches infinity with negligible energy and is effectively hidden. Finally, although the geometry is extreme, singular behavior occurs only at the origin. The displacement vectors of geodesic congruences,
\begin{equation}\label{eq|congruence2}
\begin{split}
    z_t^\mu\sim&\begin{pmatrix}
        c_1+c_2\log(r)+\frac{aEc_2-2Lc_3}{2E}r-\frac{aLc_3}{4E}r^2\\
        \sqrt{h_2f_{-2}}\left((c_2-c_1)-c_2\log(r)-\frac{ac_2}{2}r+\frac{aLc_3}{4E}r^2\right)\\
        c_3r\\
        c_4r
    \end{pmatrix},\\
    z_n^\mu\sim&\begin{pmatrix}
        c_1-\frac{Lc_3}{2E}r\\
        c_2-\frac{\sqrt{h_2f_{-2}}Lc_3}{2E}r\\
        c_3r\\
        c_4r
    \end{pmatrix},
    \end{split}
\end{equation}
shows a divergent behavior only in the limit $r\to 0$.

Finally, an important feature emerges in the large-mass regime. In this limit the Yukawa charge $S^-_2$ becomes small \ref{Phase diagram} and the veil radius approaches the Schwarzschild radius, causing veiled singularities to closely mimic the external appearance of black holes.

\section{Perturbations} \label{Pert}

A major technical difficulty in the analysis of perturbations in quadratic gravity has long been the absence of a master equation of the form of \eqref{master_eq} for tensor modes. This issue was recently overcome by \cite{Antoniou:2024jku} for $l=2$ axial tensor perturbations and in \cite{Held:2022abx} for monopole $l=0$ perturbations, where such an equation was explicitly derived.

The starting point consists in rewriting the action \eqref{action} in the Einstein frame, following \cite{Hindawi:1995an}, which leads to
\begin{equation}
I_{\text{Stelle}} =
\int d^4x \sqrt{-g}\,\gamma \left[
R 
+ 2 f^{\mu\nu} G_{\mu\nu}
+ m_2^2 \left(f^{\mu\nu}f_{\mu\nu} - f^2\right)
\right],
\end{equation}
where $f_{\mu\nu}$ is an auxiliary tensor field encoding the massive spin-2 mode. In fact, the equations of motion for $g_{\mu\nu}$ are obtained by varying the action with respect to the auxiliary field $f_{\mu\nu}$:
\begin{equation}
\mathcal{E}^{(g)}_{\mu\nu}
\equiv
G_{\mu\nu}
+
\mu^2
\left(
f_{\mu\nu}
-
f\, g_{\mu\nu}
\right)
=0 ,
\label{eq:Eg}
\end{equation}
From Eq.~\eqref{eq:Eg} one finds
\begin{equation}
R_{\mu\nu}
=
-\mu^2
\left(
f_{\mu\nu}
+
\frac{1}{2}f\, g_{\mu\nu}
\right),
\qquad
R
=
-3\mu^2 f ,
\label{eq:Rf}
\end{equation}
or equivalently
\begin{equation}
f_{\mu\nu}
=
-\frac{1}{\mu^2}
\left(
R_{\mu\nu}
-
\frac{1}{6}R\, g_{\mu\nu}
\right).
\label{eq:fmunu}
\end{equation}
It is straightforward to confirm that substituting Eq.~\eqref{eq:fmunu} into the action reproduces the original action \cite{Antoniou:2024jku}.

Using the Bianchi identity, the equations of motion obtained by varying the action with respect to $g_{\mu\nu}$ take the form
\begin{equation}
\begin{split}
    &
    \mathcal{E}^{(f)}_{\mu\nu}\equiv
    G_{\mu\nu}+G_{(\mu}{}^\rho f_{\nu)\rho}-g_{\mu\nu}G^{\rho\sigma}f_{\rho\sigma}-fR_{\mu\nu}
    \\
    &
    +R f_{\mu\nu}+\Box f_{\mu\nu}+(\nabla_\mu\nabla_\nu-g_{\mu\nu})f
    -2\nabla_\rho\nabla_{(\mu}f_{\nu)}{}^\rho\\
    &
    +g_{\mu\nu}\nabla_\rho\nabla_\sigma f^{\rho\sigma}+m_2^2\big[\big(f^2-f_{\rho\sigma}f^{\rho\sigma}\big)/2\\
    &
    +2\big(f_\mu{}^\rho f_{\nu\rho}-f\,f_{\mu\nu}\big)\big]=0.
 \end{split}
\label{eq:Ef}
\end{equation}
General Relativity is recovered in the limit $m_2 \rightarrow \infty ,$ or equivalently $\alpha \rightarrow 0$. As follows from Eq.~\eqref{eq:fmunu}, the auxiliary tensor $f_{\mu\nu}$ vanishes in this limit, removing the higher-derivative corrections and leaving the standard vacuum Einstein equations \eqref{eq:Ef}.

We then consider linear perturbations around a background configuration,
\begin{equation}
g_{\mu\nu} = \bar{g}_{\mu\nu} + \delta g_{\mu\nu}, \quad
f_{\mu\nu} = \bar{f}_{\mu\nu} + \delta f_{\mu\nu},
\end{equation}
and restrict to Ricci-flat background solutions, such that $\bar{R}  = \bar{f} = 0$. Under these assumptions, the linearized equations of motion take the form
\begin{align} \label{eq:field}
\delta G_{\mu\nu} + m_2^2 \left(\delta f_{\mu\nu} - \bar{g}_{\mu\nu}\,\delta f\right) &= 0, \\
\bar{\Box}\,\delta f_{\mu\nu}
- \bar{\nabla}_\mu \bar{\nabla}_\nu \delta f
+ 2 \bar{R}_{\mu\rho\nu\sigma}\,\delta f^{\rho\sigma}
+ 2 \bar{f}^{\rho\sigma}\,\delta R_{\mu\rho\nu\sigma}+
\nonumber \\
\quad
- m_2^2 \Bigg[
\delta f_{\mu\nu}
- \bar{g}_{\mu\nu}\,\bar{f}^{\rho\sigma}\delta f_{\rho\sigma}
- \left(\bar{g}_{\mu\nu} + \bar{f}_{\mu\nu}\right)\delta f
- \frac{1}{2}\bar{f}^{\rho\sigma}\bar{f}_{\rho\sigma}\,\delta g_{\mu\nu}
\Bigg] &= 0,
\end{align}
where $\delta G_{\mu\nu}$ and $\delta R_{\mu\rho\nu\sigma}$ denote the first-order variations of the Einstein and Riemann tensors, respectively. The operators $\bar{\nabla}_\mu$ and $\bar{\Box}$ are the covariant derivative and the d'Alembertian constructed with respect to the background metric $\bar{g}_{\mu\nu}$.

\subsection{l=0}

We now consider a general decomposition of the perturbations into spherical harmonics, separating time and angular dependence as
\begin{equation}
\begin{split}
\delta g_{\mu\nu}(t,r,\theta,\phi) = &
\hat{g}_{\mu\nu}(r)\,Y_{l m}(\theta,\phi)\,e^{-i\omega t}, 
\\
\delta f_{\mu\nu}(t,r,\theta,\phi) = &
\hat{f}_{\mu\nu}(r)\,Y_{l m}(\theta,\phi)\,e^{-i\omega t}.
\end{split}
\end{equation}

In full generality, the radial amplitudes can be written in the Regge--Wheeler form for even-parity perturbations as
\begin{equation}
\delta g_{\mu\nu} =
\begin{pmatrix}
-h(r)H_0(r) & H_1(r) & 0 & 0 \\
H_1(r) & \dfrac{H_2(r)}{f(r)} & 0 & 0 \\
0 & 0 & r^2 K(r) & 0 \\
0 & 0 & 0 & r^2 \sin^2\theta\,K(r)
\end{pmatrix}
Y_{l m} e^{-i\omega t},
\end{equation}
\begin{equation}
\delta f_{\mu\nu} =
\begin{pmatrix}
-h(r)F_0(r) & F_1(r) & 0 & 0 \\
F_1(r) & \dfrac{F_2(r)}{f(r)} & 0 & 0 \\
0 & 0 & r^2 M(r) & 0 \\
0 & 0 & 0 & r^2 \sin^2\theta\,M(r)
\end{pmatrix}
Y_{l m} e^{-i\omega t}.
\end{equation}

Monopole and dipole perturbations do not contain independent radiative degrees of freedom in the case of massless spin-2 fluctuations. In fact, in General Relativity, Regge and Wheeler showed that the corresponding perturbations are associated with variations of the parameters of the black hole solution, such as infinitesimal changes in the black-hole mass, a displacement of the center of mass, or the introduction of a small rotation, rather than genuine gravitational-wave excitations \cite{Regge:1957td}. Subsequent analyses based on gauge-invariant formulations confirmed that the propagating gravitational degrees of freedom arise only for multipoles with $l\ge 2$ \cite{Zerilli:1970wzz,Vishveshwara:1970cc}. As emphasized in \cite{Held:2022abx}, this is no longer true in the presence of the massive spin-2 sector. We therefore focus first on the monopole sector $l=0$.

For $l = m = 0$, the angular dependence becomes trivial and one can exploit the constraints
\[
\delta f = 0, \qquad \bar{\nabla}_\mu \delta f^{\mu\nu} = 0,
\]
together with a suitable gauge choice (e.g. $K(r)=H_0(r)=0$). Using the lowest-order perturbation equations, one can eliminate $H_1(r)$ and $H_2(r)$ in favor of the remaining variables, leading to a reduced dynamical system.

At this stage it is convenient to introduce the combinations
\[
\varphi(r) = -2\omega M(r), \qquad \chi(r) = F_1(r),
\]
which allow the system to be cast into two coupled second-order equations of the form
\begin{align}
&\varphi''(r) + \left(\frac{4}{r} + \frac{3h'(r)}{2h(r)} - \frac{f'(r)}{2f(r)}\right)\varphi'(r)
+ \frac{\omega^2}{h(r)f(r)}\varphi(r)+\\ \nonumber
&+ V_{\varphi\varphi}(r)\varphi(r) + V_{\varphi\chi}(r)\chi(r) = 0, \\
&\chi''(r) + \left(\frac{2}{r} + \frac{3h'(r)}{2h(r)} + \frac{3f'(r)}{2f(r)}\right)\chi'(r)
+ \frac{\omega^2}{h(r)f(r)}\chi(r)+\\ \nonumber
&+ V_{\chi\chi}(r)\chi(r) + V_{\chi\varphi}(r)\varphi(r) = 0,
\end{align}
supplemented by a first-order constraint equation of the form
\begin{equation}
\varphi'(r) + \frac{i}{2f(r)}\chi'(r) + V_{\varphi}(r)\varphi(r) + V_{\chi}(r)\chi(r) = 0.
\end{equation}

Finally, the system can be diagonalized through the ansatz
\begin{equation}
\varphi(r) = a(r)\tilde{\psi}(r) + b(r)\tilde{\psi}'(r), 
\qquad
\chi(r) = c(r)\tilde{\psi}(r),
\end{equation}
where the functions $a(r), b(r), c(r)$ are fixed by the equations of motion, and where for simplicity one may set $b(r)=0$. Introducing the master field
\[
\psi(r,t) = r\,\tilde{\psi}(r)\,e^{-i\omega t},
\]
one obtains a Zerilli-like equation of the form
\begin{equation}
\frac{d^2\psi}{dr_*^2} + \left[\omega^2 - V_{\text{eff}}(r)\right]\psi = 0,
\end{equation}
where the effective potential reads
\begin{align}\label{eq:masterEq_monopole}
	 V(r)=& m_2^2 h + \frac{f'h + fh'}{2r} + \\
&+m_2^4\frac{288 f^3 h^3 (-2h + h'r)^2}{\left(4(1 - 3f)h^2r + (f'h + fh')(3f h'r + h(-4 + 3f'r))\right)^2} + \\
&+m_2^2 \frac{24 f h^2 (2f + f'r)(2h - h'r)}{r \left(4(1 - 3f)h^2r + (f'h + fh')(3f h'r + h(-4 + 3f'r))\right)}
\end{align}  

\subsubsection{Schwarzschild black holes}
For Schwarzschild background, as said, we have $\bar{R}_{\mu\nu}=\bar{f}_{\mu\nu}=0$, so eqs.~\eqref{eq:field} for the perturbations reduce to:
\begin{align}
    &
    \delta\mathcal{E}^{(g)}_{\mu\nu}=
    \delta G_{\mu\nu} + m_2^2 \delta f_{\mu\nu}=0
    \label{eq:linear-Ricciflat-eom-g}\\
    &
    \delta\mathcal{E}^{(f)}_{\mu\nu}=
    \bar{\Box} \delta f_{\mu\nu}+2\bar{R}_{\mu\sigma\nu\rho}\delta f^{\sigma\rho}-m_2^2\delta f_{\mu\nu}=0\; .
\label{eq:linear-Ricciflat-eom-f}
\end{align}
and for radial perturbation, potential of eq.~\eqref{eq:masterEq_monopole} simplifies to: 
\begin{equation}
    V_0 = f \left( \frac{2M}{r^3} + m_2^2 + \frac{24M(M - r)m_2^2 + 6r^3 (r - 4M) \mu^4}{(2M + r^3 m_2^2)^2} \right)
\end{equation}
as shown in \cite{Brito:2013wya}.

\begin{figure}[hbt!]
    \centering
    \includegraphics[width=0.8\columnwidth]{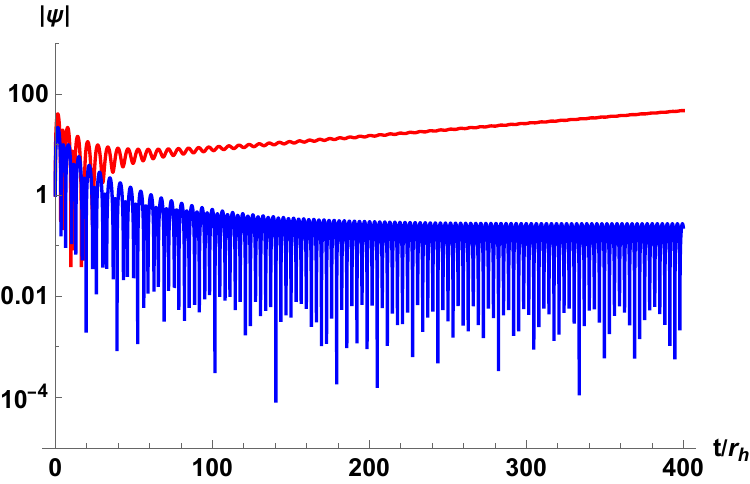}
    \caption{Semi Logarithmic plot of time evolution of massive radial peruturbation ($l=0$) for Schwarzschild background: Blue one has $p=r_h m_2=1$, red one is the unstable case with $p=r_h m_2=0.86$.}
\label{fig1}
\end{figure}

Using these expressions, we investigated the time evolution of the perturbations (the numerical method is described in ~\ref{TDI}). As the first step, we examined the stability range of the dimensionless parameter $p=r_h m_2$ for these black holes and considered the onset of the  instability of the Gregory–Laflamme type \cite{Gregory:1993vy} in light of the results reported in \cite{Held:2022abx}. Our analysis confirmed the same stability interval for $p$ ($0.876<p<1.14$), and reproduced the expected behavior associated with the Gregory–Laflamme instability, as illustrated in Fig.~\ref{fig1} for two representative values. These results are consistent with those obtained in the initial studies of the perturbation dynamics presented in \cite{Lu:2017kzi,Silveravalle:2023lnl, Bonanno:2024fcv} and with the general picture of time-domain evolution of the Gregory–Laflamme instability \cite{Konoplya:2008yy}.
We then proceeded to analyze the late-time asymptotic behavior for perturbations of black hole solutions in the  stability sector. The late-time tails are oscillatory and exhibit a decay of, at most, two orders of magnitude relative to the amplitude of the initial ringdown phase. This behavior contrasts with that of massless fields, where asymptotic tails typically become visible only after the signal has decayed by six or more orders of magnitude. Thus, we conclude that massive tails are not only oscillatory, but also decay significantly more slowly, as illustrated in Fig.~\ref{fig1}.

\begin{figure}[hbt!]
    \centering
    \includegraphics[width=0.8\columnwidth]{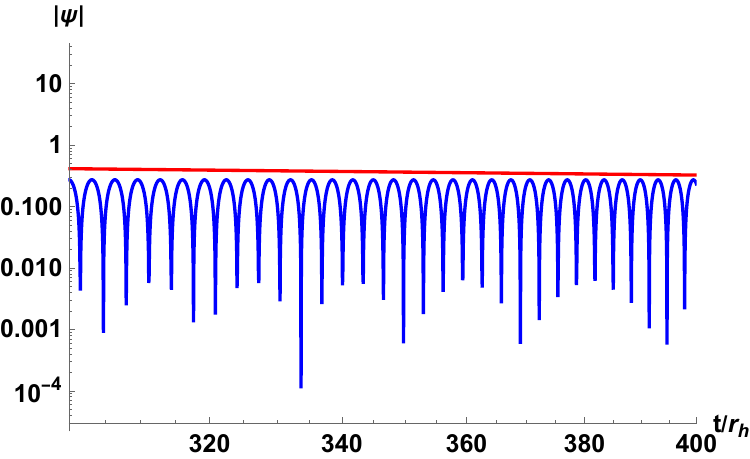}
    \caption{Logarithmic plot of the asymptotic tail part with $p=r_h m_2=1$ showed in fig.~\ref{fig1}, in red the fit of the tail with the law $\propto t^{-5/6}$}
\label{fig2}
\end{figure}

As shown by the fit in Fig.~\ref{fig2}, the late-time decay follows the power-law behavior
$$\psi\propto t^{-5/6},$$
while the oscillation amplitude is modulated by a sinusoidal term proportional to $\propto\sin{(m_2 t)}$. These findings are consistent with previous analyses of massive field perturbations \cite{Koyama:2000hj,  Konoplya:2023fmh, Konoplya:2006gq}.

\subsubsection{Non-Schwarzschild black holes}

\begin{figure}[hbt!]
    \centering
    \includegraphics[width=1.0\textwidth]{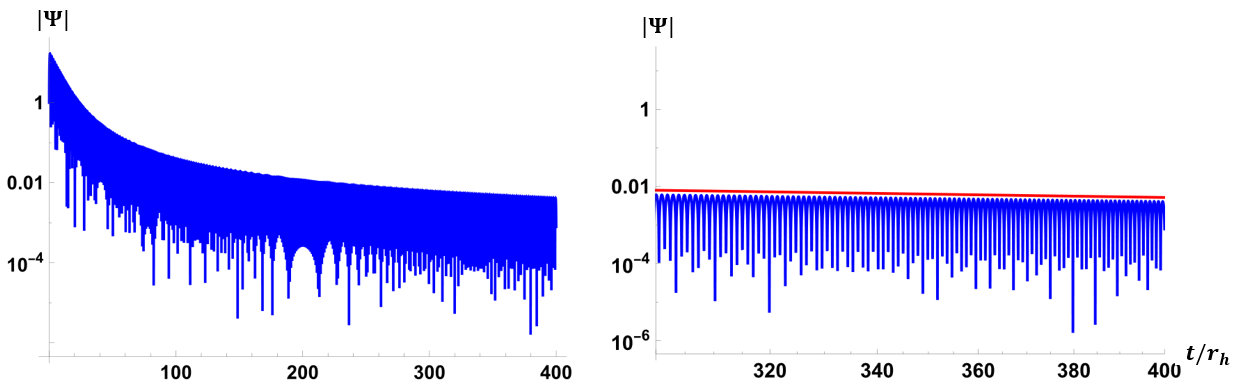}
    \caption{Left panel: Semi Logarithmic plot of time evolution of massive radial perturbation ($l=0$) for the Non-Schwarzschild black hole with $p=r_h m_2=1$.\\ Right panel: Logarithmic plot of asymptotic tail of the left panel perturbation with a fit given by the law $\propto t^{-5/6}$. }
\label{fig3}
\end{figure}

Considering the alternative branch of black hole solutions, it is necessary to use the full expression for the potential given by Eq.~\eqref{eq:masterEq_monopole}. Additionally, the metric function must be approximated using the method described in Sec.~\ref{NM}, which is meaningful only within the range of stability of the parameter $p$ \cite{Kokkotas:2017zwt}. We numerically integrated the wave-like equations to get the time evolution of perturbations of this branch of black holes, with representative results shown in Fig.~\ref{fig3}.
In this case, the decay following the ringdown phase is more pronounced, exhibiting a drop of approximately four orders of magnitude. However, the late-time tail again follows the power-law behavior
$$\psi\propto t^{-5/6},$$
confirming the universality of this decay law also for certain massive perturbations.

\subsubsection{Veiled singularities}

For Veiled singularities, initially, we found interesting results as described in \cite{Bonanno:2025ksz}. The background metric was constructed numerically using the method described in Section~\ref{NM}, and the perturbation equation with the effective potential in Eq.~\eqref{eq:masterEq_monopole}, with $m_2=1$, was integrated using the time-domain approach in light-cone coordinates introduced in Section~\ref{TDI}. Regarding the initial conditions, there is no unique prescription. A standard choice consists in imposing a Gaussian profile along the $u=0$ axis,
\begin{equation} \label{gauss}
\phi(v,0)=\exp\left[-\frac{(v-v_0)^2}{2\sigma}\right],
\end{equation}
which represents an initial localized perturbation and does not affect the late-time behavior. The second boundary condition is imposed at the singularity ($r_*=0$, corresponding to $u=v$), where the perturbation is set to zero, consistently with the assumption that this point lies outside the physical spacetime. After implementing these conditions, Eq.~\eqref{pertgrid} was integrated on the numerical grid, and the time evolution of the perturbation was extracted at a fixed radial position $x_{\rm meas}$. Several consistency tests were performed, including variations of the grid spacing near the singularity and different choices of the extraction radius. Within this setup, the solutions appeared stable over the explored parameter range and exhibited the expected universal massive late-time tail \eqref{eq:MassiveTail}.

\begin{figure}[h]
    \centering
    \includegraphics[width=0.8\columnwidth]{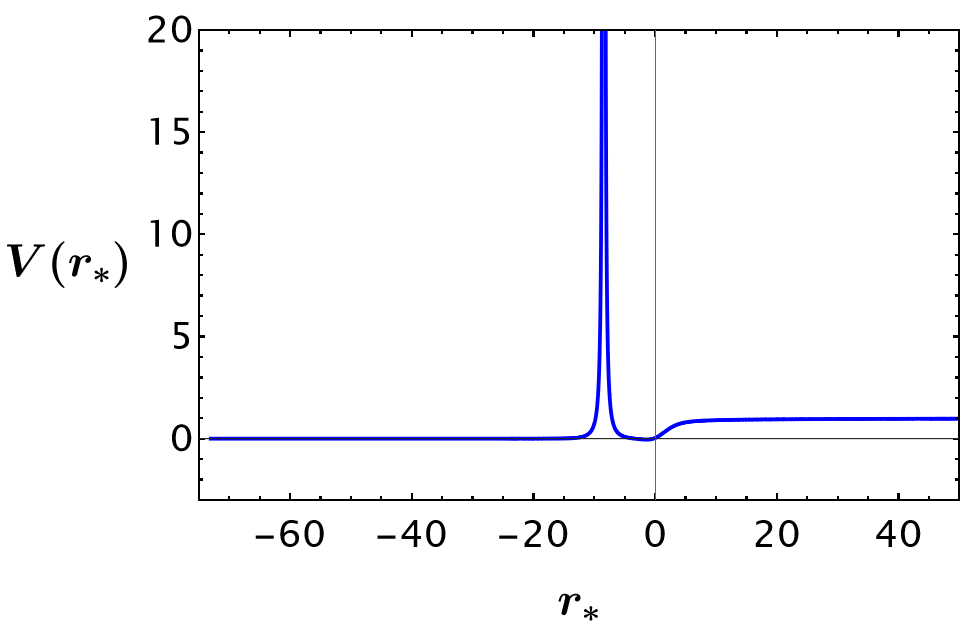}
    \caption{Effective potential of tensor perturbations with $l=0$ for veiled singularities. It is present an infinite barrier at finite radius.}
\label{VeiledPot}
\end{figure}

\begin{figure}[hbt!]
    \centering
    \includegraphics[width=0.8\columnwidth]{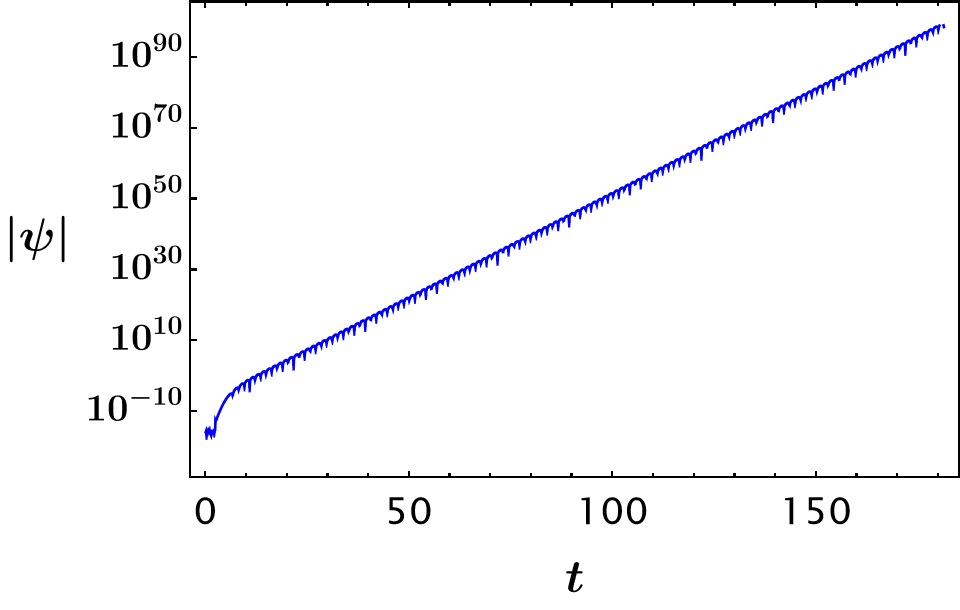}
    \caption{Plot of time evolution of the perturbation with $l=0$ for veiled singularities with effective potential \ref{VeiledPot}.}
\label{VeiledIns}
\end{figure}

A more refined analysis, however, revealed that the treatment of the tortoise coordinate in \cite{Bonanno:2025ksz} effectively suppressed the contribution from the region close to the singularity. Once this aspect is properly taken into account, the behavior of the system changes significantly. In particular, the effective potential takes the form shown in Fig.~\ref{VeiledPot}, 
and the corresponding time evolution (Fig.~\ref{VeiledIns}) displays a strong growth of the perturbations, i.e an instability of the perturbations.

At the same time, this result must be interpreted with caution. A detailed inspection of the effective potential shows that its peak diverges due to the vanishing of the denominator in Eq.~\eqref{eq:masterEq_monopole}. As a consequence, the perturbation equation in its present form becomes ill-defined in this region, and the standard time-domain integration cannot be straightforwardly applied. This suggests that the observed instability may be, at least in part, related to the breakdown of the perturbative framework rather than to a genuine physical effect. A consistent treatment likely requires a reformulation of the problem, for instance through a suitable redefinition of the perturbation variables or gauge choice \cite{Gonzalez:2008wd}, or by constructing an alternative effective potential, as proposed in related contexts \cite{Bronnikov:2012ch,Bronnikov:2011if,Kodama:2003kk}. A systematic investigation of these issues is left for future work.

\subsection{l=2}
For higher multipole numbers, the analysis becomes more involved due to the structure of the tensor harmonics. Nevertheless, the computations were carried out in \cite{Antoniou:2024jku}, where the equations governing gravitational perturbations were derived. The resulting system consists of the classical, massless component $h_1$ and the two massive components, one vectorial $F_1$ and one tensorial $F_2$, which are dynamically coupled. 

We now turn into specific derivation for axial (odd-parity) perturbations for generic multipole number $l \geq 2$. 
In this sector, it is convenient to expand both the metric and the auxiliary field in terms of axial vector and tensor spherical harmonics.

Following \cite{Antoniou:2024jku}, we can exploit the gauge freedom of the massless sector to impose the Regge--Wheeler gauge, setting the axial metric perturbations to
\begin{equation}
\delta g^{\text{ax}}_{\mu\nu} =
\begin{pmatrix}
0 & 0 & -h_0(r)\csc\theta\,\partial_\phi & h_0(r)\sin\theta\,\partial_\theta \\
0 & 0 & -h_1(r)\csc\theta\,\partial_\phi & h_1(r)\sin\theta\,\partial_\theta \\
\ast & \ast & 0 & 0 \\
\ast & \ast & 0 & 0
\end{pmatrix}
Y_{l m}(\theta,\phi)\,e^{-i\omega t}.
\end{equation}

This choice completely fixes the gauge for the massless spin-2 perturbations and does not affect the massive sector.

The axial perturbations of the auxiliary field $f_{\mu\nu}$ admit instead the most general decomposition \cite{Antoniou:2026nhh}
\begin{equation}
\delta f^{\text{ax}}_{\mu\nu} =
\begin{pmatrix}
0 & 0 & F_0(r)\csc\theta\,\partial_\phi & -F_0(r)\sin\theta\,\partial_\theta \\
0 & 0 & F_1(r)\csc\theta\,\partial_\phi & -F_1(r)\sin\theta\,\partial_\theta \\
\ast & \ast & -F_2(r)\csc\theta\,X & F_2(r)\sin\theta\,W \\
\ast & \ast & \ast & F_2(r)\sin\theta\,X
\end{pmatrix}
Y_{l m}(\theta,\phi)\,e^{-i\omega t},
\end{equation}
where $F_0(r)$ and $F_1(r)$ correspond to vector-type perturbations, while $F_2(r)$ encodes the tensorial degree of freedom.

It is important to stress that, in contrast to the massless case, the decomposition of $\delta f_{\mu\nu}$ does not reduce to the Regge--Wheeler form in the limit $m_2 \to 0$. In particular, the vector modes $F_0$ and $F_1$ become pure gauge only in this limit, whereas for finite mass they represent genuine dynamical degrees of freedom. As in the monopole case, we organize our analysis based on the choice of background solution. 

\subsubsection{Schwarzschild background.}

Let us now specialize to Ricci-flat backgrounds, such as the Schwarzschild spacetime, for which $\bar{R}_{\mu\nu}=0$ and consequently $\bar{f}_{\mu\nu}=0$. In this case, the linearized equations reduce to \eqref{eq:linear-Ricciflat-eom-g}\eqref{eq:linear-Ricciflat-eom-f}
\begin{align}
\delta G_{\mu\nu} + m_2^2 \delta f_{\mu\nu} &= 0, \\
\bar{\Box}\,\delta f_{\mu\nu} + 2\bar{R}_{\mu\rho\nu\sigma}\delta f^{\rho\sigma} - m_2^2 \delta f_{\mu\nu} &= 0.
\end{align}
The second equation governs the propagation of the massive spin-2 field and is completely decoupled from $\delta g_{\mu\nu}$. Therefore, one can first solve for $\delta f_{\mu\nu}$ and subsequently determine the metric perturbations through the first equation.

Expanding the axial perturbations as above, one obtains a system involving three radial functions $h_1(r)$, $F_1(r)$, and $F_2(r)$. The non-divergence constraint $\bar{\nabla}^\mu f_{\mu\nu}=0$ yields a first-order relation that can be used to eliminate $F_0(r)$.
Introducing the variables
$$H_1 = \frac{h(r)}{r} h_1(r), \qquad Q = h(r) F_1(r), \qquad Z = \frac{F_2(r)}{r},$$
the system can be recast in the compact form
\begin{equation} \label{system}
    -\frac{d^2}{dt^2}\boldsymbol{\Psi}+\frac{d^2}{dr_*^2}\boldsymbol{\Psi}+\boldsymbol{P}\frac{d}{dr_*}\boldsymbol{\Psi}+\boldsymbol{V}\Psi=0,
\end{equation}
where $\boldsymbol{\Psi}=\left(\psi_1,\psi_2,\psi_3\right)\equiv (H_1,Q,Z)$. The matrices $P$ and $V$ are the potential matrices of the following form:
\begin{equation} \label{matrix}
    \boldsymbol{P}=
    \begin{pmatrix}
        0 & 0 & {P}_{13}\\
        0 & 0 & 0\\
        0 & 0 & 0
    \end{pmatrix}\, , \;
    \boldsymbol{V}=
    \begin{pmatrix}
        {V}_{11} & {V}_{12} & {V}_{13}\\
        0 & {V}_{22} & {V}_{23}\\
        0 & {V}_{32} & {V}_{33}
    \end{pmatrix} \, ,
\end{equation}
whose elements are given by the following relations:
\begin{align}
    P_{13} =& -2 m_2^2 h(r) \nonumber\\
    V_{12}=& -\frac{2 m_2^2 h(r)}{r} \nonumber\\
    V_{13} =& \frac{2 m_2^2 h(r)^2}{r} - 2 m_2^2 h(r) h'(r) \nonumber\\
    V_{11} =& -\frac{h(r)}{r^2} \left( \Lambda - 3 r h'(r) \right)\nonumber\\
    V_{22}=&-h m_2^2+\big[-2 h^2 \left(-r h'+10 h+2 \left(\Lambda -1\right)\right)\nonumber\\
    &+h r^2 h'^2+h r (-2 h r h''-r h'^2+6 h h'\nonumber)\big]\\
    &(4r^2 h)^{-1}\,,\nonumber\\[2mm]
    V_{23}=&-r^{-2}\left(\Lambda -2\right) h\left(2 h-r h'\right)\,,\label{grav_pot_Schw}\\[2mm]
    V_{32}=&-2r^{-2}h\, ,\nonumber\\[2mm]
    V_{33}=&-h m_2^2+\big[r (-h h')-h (r h'\nonumber\\
    &+2 (m_2^2 r^2+\Lambda-2))(2 r^2)^{-1}\big]\,. \nonumber
\end{align}

where $\Lambda=l(l+1)$. The matrices $\boldsymbol{P}$ and $\boldsymbol{V}$ encode the coupling between the massless and massive modes. In particular, the massive sector $(Q,Z)$ forms a closed subsystem, while the metric perturbation $H_1$ is sourced by the massive field and couples all three modes.

In the limit $\delta f_{\mu\nu}=0$, the system reduces to the standard Regge--Wheeler equation of General Relativity. Conversely, when the massive modes are excited, the dynamics is governed by a coupled system describing the propagation of a massive spin-2 field on the black hole background.

\begin{figure}[h]
    \centering
    \includegraphics[width=0.8\columnwidth]{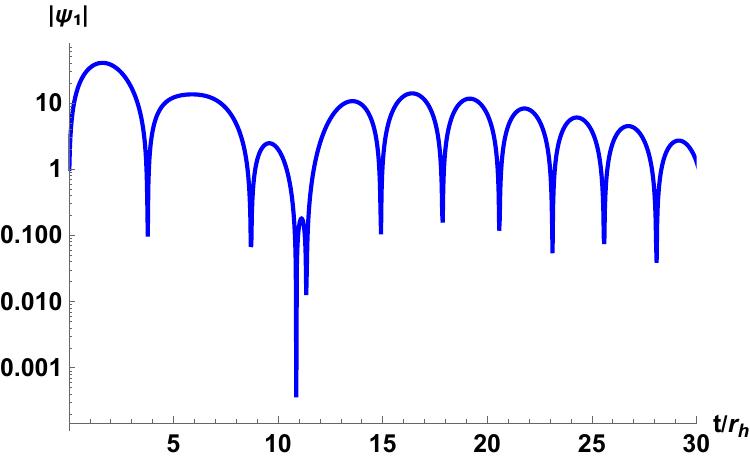}
    \caption{Semi Logarithmic plot of time evolution of perturbation ($l=2$) for Schwarzschild black holes with $p=r_h m_2=1$. We extract the behavior for the component $\psi_1\propto h_1$ which has contributions of the classical massless perturbations with the peculiar massive perturbations (see eqs.~\eqref{matrix},~\eqref{system}). We extract fundamental frequency with Prony's method and compare with massless Schwarzschild GR value.}
\label{fig4}
\end{figure}

Employing the time-domain method described in ~\ref{TDI}, we analyze the evolution of the perturbations, focusing on stability and extracting relevant physical information. As the first step, we consider, as an illustrative example,  the case $p=1$, analyzing the ringdown phase (Fig.~\ref{fig4}). Using the Prony's method (see ~\ref{Prony}), we extract the dominant frequency from the first component of $\boldsymbol{\Psi}$, whose perturbations contain the massless mode along with subdominant contributions from the massive modes. The resulting frequency is $\omega=0.732-0.177i$ deviating from the corresponding General Relativity value by less than $2\%$.

\begin{figure}[h]
    \centering
    \includegraphics[width=0.8\columnwidth]{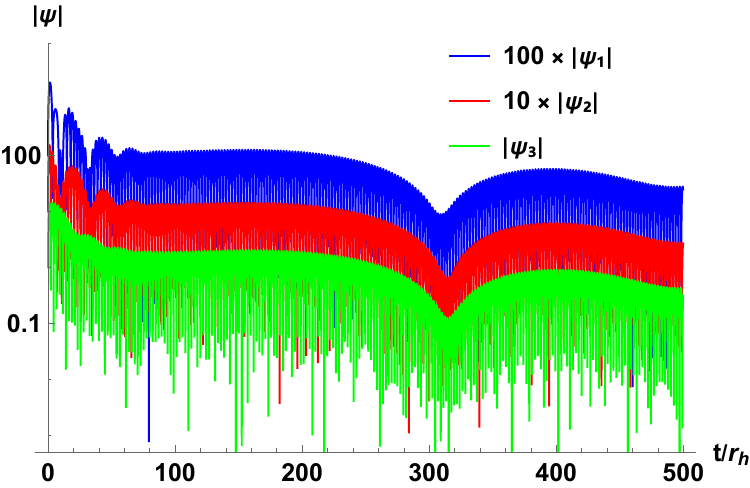}
    \caption{Semi Logarithmic plot of time evolution of perturbation ($l=2$) for Schwarzschild black holes with $p=r_h m_2=1$. We extract the behavior for the component $\psi_1$, which has contributions both from massless and massive perturbations, and for $\psi_2,\psi_3$ which have only massive perturbations. The time evolution of all components shows a similar behavior, indicating that the massive modes significantly influence the overall dynamics.}
\label{figSchw}
\end{figure}

As in the previous section, we now turn to the study of the asymptotic tail. However, for small values of \( p \), the behavior is illustrated in Fig.~\ref{figSchw}, 
where we observe the presence of echoes in the intermediate tail for all the components. These echoes persist over an extended period, preventing a clear identification of subsequent decay times. Moreover, the peculiar structure of the signal in this regime hinders the extraction of a simple analytical decay law.

\begin{figure}[h]
    \centering
    \includegraphics[width=0.8\columnwidth]{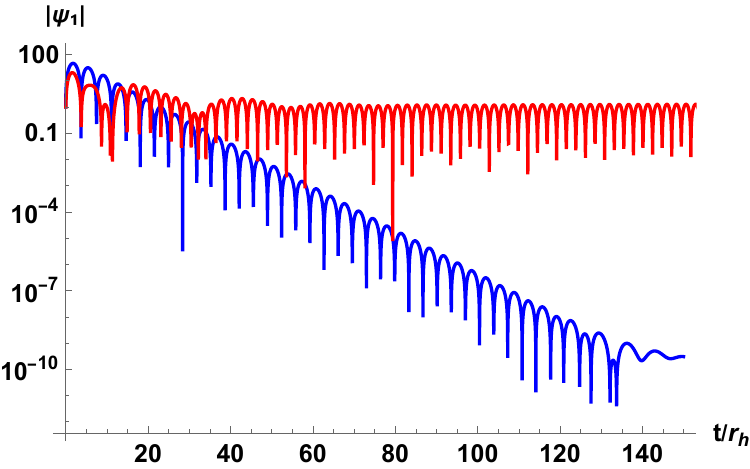}
    \caption{Semi Logarithmic plot of time evolution of perturbation ($l=2$) for Schwarzschild black holes in GR in blue and with $p=r_h m_2=1$ in EW gravity in red. For the latter we have extracted the behavior for the component $\psi_1$ which has contributions both from massless and massive perturbations.}
\label{figGR-EW}
\end{figure}

Nevertheless, the initial ringdown phase can still be meaningfully compared to that of classical massless gravitational perturbations in General Relativity, as shown in Fig.~\ref{figGR-EW}. 

The presence of massive modes significantly modifies the decay profile, resulting in a much smoother signal. The amplitude differs by several orders of magnitude compared to the massless case, underscoring the substantial influence of the massive sector on the early-time dynamics.

\begin{figure}[h]
    \centering
    \includegraphics[width=0.8\columnwidth]{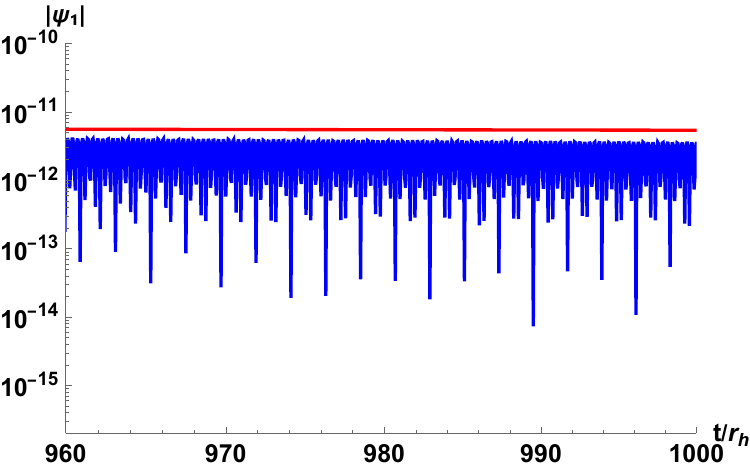}
    \caption{Logarithmic plot of the asymptotic tail of perturbation ($l=2$) for Schwarzschild black holes with $p=r_h m_2=10$ and in red the fit by the law $\propto t^{-5/6}$}
\label{fig5}
\end{figure}

Furthermore, for the Schwarzschild branch, there is no upper bound on the parameter $p$, and increasing its value leads to an earlier onset of the asymptotic tail. Characterizing this regime, we extract the late-time decay law by varying the mass parameter $\mu$ and considering higher multipole numbers $l$, in order to verify the independence of the asymptotic behavior from these parameters. The resulting trend is shown in Fig.~\ref{fig5} and is:
\begin{equation}
    \psi_1,\psi_2,\psi_3\propto t^ {-5/6}.
\end{equation}
We found that the late-time decay law observed in all other asymptotic tails studied, both in this work and in some other theoretical frameworks \cite{Koyama:2000hj,  Konoplya:2023fmh, Konoplya:2006gq}., consistently follows the same time dependence, further supporting the universality of this asymptotic behavior of massive fields in the evolution of perturbations.

\subsubsection{Non-Schwarzschild black holes}

For the Non-Schwarzschild branch of black holes, starting from the full coupled equation \eqref{eq:field}, the structure of the equations can be reduced to the previous one 
\begin{equation}
    -\frac{d^2}{dt^2}\boldsymbol{\Psi}+\frac{d^2}{dr_*^2}\boldsymbol{\Psi}+\boldsymbol{P}\frac{d}{dr_*}\boldsymbol{\Psi}+\boldsymbol{V}\Psi=0,
\end{equation}
where $\boldsymbol{\Psi}=\left(\Psi_1,\Psi_2,\Psi_3\right)$ are obtained connecting these functions to the original components as:
\[
\begin{array}{rcl}
h_1(r) &=& \dfrac{m_2^2 r^2}{\sqrt{h(r) f(r)}} \Psi_1(r),\\
F_1(r) &=& \dfrac{1}{\sqrt{h(r) f(r)}} \Biggl( \Psi_2(r) + \dfrac{\Psi_1(r)}{2 r h(r)} \dfrac{d}{dr} \left(h(r) f(r) r^2\right) \\&&- \Psi_1(r) \Biggr),\\
F_2(r) &=& \Psi_3(r) + \Psi_1(r) \dfrac{r f(r)}{\sqrt{h(r) f(r)}},
\end{array}
\]
and the matrices take the following form
\begin{equation}
    \boldsymbol{P}=
    \begin{pmatrix}
        0 & 0 & {P}_{13}\\
        0 & 0 & 0\\
        0 & 0 & 0
    \end{pmatrix}\, , \;
    \boldsymbol{V}=
    \begin{pmatrix}
        {V}_{11} & {V}_{12} & {V}_{13}\\
        {V}_{21} & {V}_{22} & {V}_{23}\\
        {V}_{31} & 0 & {V}_{33}
    \end{pmatrix} \, ,
\label{eq:axial_system}
\end{equation}
with the following matrix elements:
\begin{align} 
    V_{22}(r) =& (4 r f h' - h (4 + l + l^2 + r^2 m_2^2 \nonumber\\
    &- 4 r f') (r^{-2}) \nonumber\\
    V_{21}(r) =& (-2 + l + l^2) (3 r f h' +\nonumber\\
    & h (-4 + 3 r f'))(2 r^2)^{-1} \nonumber\\
    V_{23}(r) =& -\frac{(-2 + l + l^2) \sqrt{h f} (2 h - r h')}{r^3} \nonumber\\
    V_{31}(r) =& 3 r f^2 h'^2 + 2 h f h'(-2 + 3 r f') \nonumber\\
    & + h^2 (4 r m_2^2 - 12 r m_2^2 f- 4 f' \nonumber\\
    & + 3 r f'^2)(4 \sqrt{h f}r (2 h - r h'))^{-1} \nonumber\\
    V_{33}(r) =& \frac{r f h' - h (l + l^2 + r^2 m_2^2 - r f')}{r^2} \label{grav_pot_NonSchw} \\
    P_{13}(r) =& -\frac{2 h}{r^2} \nonumber\\
    V_{12}(r) =& -\frac{2 h}{r^2} \nonumber\\
    V_{13}(r) =& \frac{2 \sqrt{h f}r (2 h - r h')}{r^3} \nonumber\\
    V_{11}(r) =& -\frac{(-2 + l + l^2) h}{r^2}.\nonumber
\end{align} 

\begin{figure}[h]
    \centering
    \includegraphics[width=1.0\textwidth]{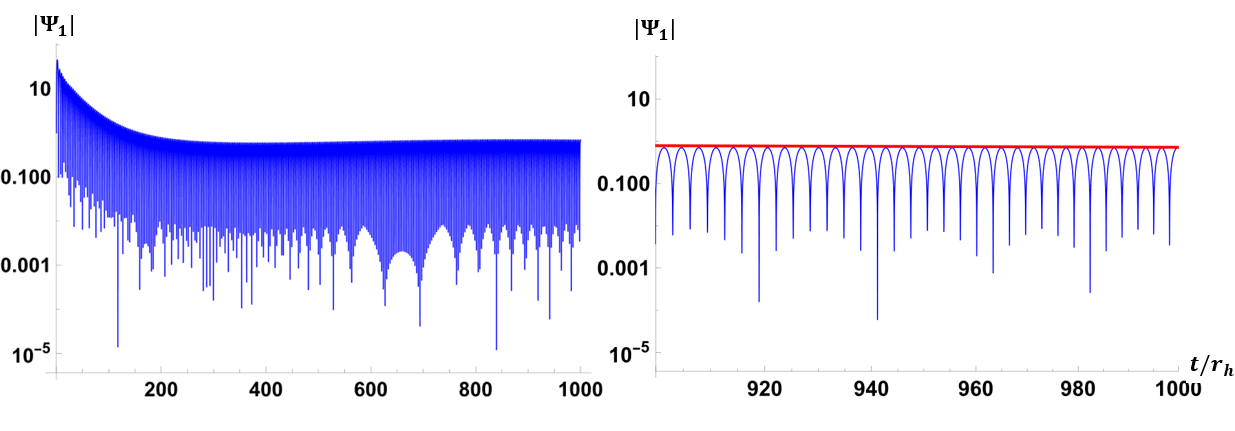}
    \caption{Left panel: Semi Logarithmic plot of time evolution of perturbation ($l=2$) for Non-Schwarzschild black holes with $p=r_h m_2=1$. We extract the behavior for the component $\psi_1$ which has contributions both from massless and massive perturbations.\\
    Right panel: Logarithmic plot of the asymptotic tail of the left panel perturbation with a fit given by the law $\propto t^{-5/6}$.}
\label{fig6}
\end{figure}

Since the system in Eq.~\eqref{eq:axial_system} is fully coupled through off-diagonal terms in the potential, perturbations in any single mode will dynamically source the others. As a result, both massless and massive components contribute to the evolution and to the spectrum of the perturbations. We proceeded with the analysis of the time evolution for this branch and observed that the characteristic tails associated with the massive modes emerge early in the signal,  and significantly affect the decay, preventing, thereby,  precise extraction of quasi-normal frequencies. At later times (see Fig.~\ref{fig6}), we confirmed the stability of the black holes within the range of the parameter $p$ where $l=0$ perturbations are stable. 

Furthermore, we analyzed the asymptotic tail (right panel of Fig.~\ref{fig6}), and in this case, we were able to study the asymptotically late-time behavior already for $p=1$. We found that, for all the components, $\Psi_1$, $\Psi_2$, and $\Psi_3$, the non-Schwarzschild branch also exhibits a power-law decay of the form
\begin{equation}
    \Psi_1, \Psi_2, \Psi_3 \propto t^{-5/6},
\end{equation}
accompanied by oscillatory behavior of the amplitude. In general, the oscillation phase is described by a nontrivial function. However, in the particular case \( p = 1 \), the signal follows, the simple form \( \propto \sin{(\mu t + \phi_i)} \), where $\phi_1$, $\phi_2$, and $\phi_3$ are the corresponding phases of the gravitational perturbations. More generally, the behavior can be expressed as
\begin{equation}
    \Psi_i \propto t^{-5/6} \sin{(F(m_2) t + \phi_i)}, \qquad i=1,2,3;
\end{equation}
where \( F \) is an unknown function that is difficult to determine via numerical fitting.

\section{Summary}

In this chapter, we have investigated both the properties of solutions and the dynamics of gravitational perturbations in Einstein--Weyl gravity. 

We focused on the  $(-2,2)^1_0 $ solutions, which describe attractive naked singularities that can be interpreted as veiled singularities. Their near-origin geometry is characterized by $h(r)\sim h_2 r^2$ and $f(r)\sim f_{-2}/r^2$, leading to divergent curvature invariants. Nevertheless, the Weyl contribution to the action remains finite. The analysis of geodesic congruences shows that timelike geodesics experience diverging tidal effects as the singularity is approached, confirming its singular nature. At the same time, the redshift diverges at the origin, so that radiation emitted arbitrarily close to the singularity cannot reach infinity with finite energy. Thus, although these solutions are naked from a causal point of view, the singularity is effectively hidden by an infinite redshift and can be regarded as a veiled singularity.

\noindent A veiled radius can be defined, where the curvature becomes extremely large and signals are strongly redshifted. In the large-mass regime, the veil radius approaches the Schwarzschild radius, making these solutions increasingly similar to black holes from the external point of view.

Finally, we investigated the dynamics of gravitational perturbations. For the monopole $l=0$ sector, the massive spin-2 degree of freedom provides a genuine propagating mode, in contrast with General Relativity. For Schwarzschild black holes, we recovered the known stability interval $0.876<p<1.14$, where $p=r_hm_2$, and confirmed the Gregory--Laflamme type instability outside this range. In the stable regime, the late-time signal is characterized by the universal massive tail $\psi\propto t^{-5/6}$, accompanied by oscillations. The same asymptotic decay was found for the non-Schwarzschild branch.

For the veiled singularities, the perturbative analysis requires particular care. An initial treatment suggested stable evolution and the standard massive tail, but a more accurate treatment of the tortoise coordinate revealed a divergent effective potential at a finite radius. The resulting growth of the perturbations cannot therefore be interpreted straightforwardly as a physical instability, since the master equation itself becomes ill-defined at the potential divergence. A consistent perturbative description of these backgrounds requires further investigation.

We also considered axial $l=2$ perturbations, for which the massless and massive sectors form a coupled system. On a Schwarzschild background, the dominant quasinormal frequency remains close to its General Relativity value, while the massive sector significantly modifies the subsequent evolution and late-time behavior. The asymptotic decay of all dynamical components again follows the $t^{-5/6}$ power law. The non-Schwarzschild branch exhibits the same qualitative behavior: massive modes contribute significantly to the evolution, making the extraction of quasinormal frequencies more difficult, while the late-time dynamics remains stable within the parameter range where the monopole sector is stable and displays the same $t^{-5/6}$ decay.

Overall, these results show that Einstein--Weyl gravity admits a rich phenomenology of both black hole and singular solutions. The massive spin-2 degree of freedom leaves a characteristic imprint on the perturbative dynamics, most notably through the modified ringdown and the universal late-time tails. At the same time, the singular backgrounds reveal important limitations of the standard perturbative framework, motivating a more detailed analysis of the perturbation equations in the presence of the singular structure.

\chapter{Regular black holes in asymptotic safety}
The typical approach to extend the domain of a classical field theory into the quantum realm is to reformulate it as a quantum field theory (QFT). Following this idea, Weinberg proposed that the effective QFT of gravity might possess an UV attractive non-Gaussian fixed point (NGFP), a condition called asymptotic safety (AS) \cite{Weinberg:1979}. An important consequence of AS is the scale dependence of the effective gravitational couplings, which can be encoded through a running Newton's constant and cosmological term. From the Wilsonian perspective, quantum field theories are understood as EFTs. With this point of view, the description of a theory is well-defined only up to a certain energy scale, beyond which the theory must be replaced by its UV completion. In this sense, the Wilsonian RG is the technique that relates an EFT to its low energy descriptions \cite{Wilson:1973jj}.\\
In fact, starting from a microscopic action $S_{\Lambda_0}[\phi_0]$, defined at a high energy cutoff scale $\Lambda_0$, the corresponding low-energy description is given by the Wilsonian effective action $S_\Lambda[\phi]$ at scale $\Lambda<\Lambda_0$ \cite{Rosten:2010vm}:
\begin{equation}
    e^{-S_\Lambda[\phi]}=\int D\phi_0\, \delta\left(\phi-\rho_\Lambda[\phi_0]\right)e^{-S_{\Lambda_0}[\phi_0]}.
\end{equation}
The blocking procedure, encoded in the smearing function $\rho_\Lambda[\phi_0]$, is at the core of the RG transformation. Its role is to zoom out the system, for example, in a spin lattice where each cell has a spin-up value of 1 and a spin-down value of -1, it consists of grouping spins into blocks and coarse-graining them into new effective spins. In this way, the details of the microscopic system fade out, but the essential information is maintained. This coarse-graining is not unique; in fact, we can take the effective spin as the average value of the spins inside the block, or alternatively, we can take the most frequent spin value. Non-universal quantities, such as the position of possible fixed points, are sensitive to different methods. On the other hand, universal quantities, including the critical exponents, remain unaffected.

In momentum space, the blocking procedure retains only the modes below the cutoff scale, while averaging out the UV modes through the regularization. More precisely, the Wilsonian effective action follows a flow from the UV scale down to the IR, providing a description of the theory at all intermediate scales.

When applied to the quantum description of Einstein gravity, renormalization group techniques predict the existence of a ultraviolet attractive non-Gaussian fixed point (NGFP), the condition known as asymptotic safety \cite{Reuter:1996cp}. 

The set of trajectories attracted towards the fixed point at high energies defines the ultraviolet critical surface. If this surface is finite-dimensional, only a finite number of parameters must be fixed experimentally, while all remaining couplings are determined by the RG flow.

Near the NGFP, the dimensionless gravitational couplings approach finite values and the Newton constant acquires a non-trivial scale dependence governed by the fixed point structure. In gravitational systems, this running typically weakens the effective gravitational interaction at short distances, generating repulsive quantum effects that can prevent the formation of spacetime singularities during gravitational collapse.

When such scale-dependent couplings are translated into position-dependent quantities via appropriate identifications of the renormalization scale with physical coordinates, one obtains quantum-corrected geometries. These have been used to construct various types of regular black hole spacetimes that avoid curvature singularities at the origin~\cite{Bonanno:2000ep, Falls:2010he}.

\section{Asymptotic Safety Framework}

We anticipated the idea that gravity might be non-perturbatively renormalizable by approaching a non-Gaussian fixed point at high energies as it was originally proposed by Weinberg under the name of asymptotic safety \cite{Weinberg:1979}. This concept was made concrete in the functional renormalization group (FRG) framework by Reuter \cite{Reuter:1996cp}. In this approach, the central object is $\Gamma_k$, a scale-dependent effective action constructed via the Wilsonian \cite{Wilson:1973jj} coarse-graining procedure described before. As first introduced  to gravity by Reuter \cite{Reuter:1996cp}, $\Gamma_k$ smoothly interpolates between the classical (bare) action $S$ in the ultraviolet ($k \to \infty$) and the full effective action $\Gamma$ in the infrared limit ($k\to 0$). The scale parameter $k$ acts as an infrared cutoff: fluctuations with momenta $p^2 > k^2$ have already been integrated out, while long-wavelength fluctuations with $p^2 < k^2$ are suppressed. In this sense, $\Gamma_k$ describes a renormalization-group trajectory in the space of effective actions, interpolating from microscopic to macroscopic physics and its evolution with the scale $k$ allows a non-perturbative investigation of the theory space. The physical identification of $k$ in curved spacetimes is not unique, but there are some general criteria which need to be respected, as will be discussed below. 

To implement the Wilsonian coarse-graining procedure at the level of the quantum effective action, one introduces a scale-dependent generating functional in which the infrared modes are suppressed through a regulator term. Starting from the Euclidean partition function,
\begin{equation}
Z_k[J]=\int \mathcal{D}\phi \,
\exp\left[
-S[\phi]-\Delta S_k[\phi]
+\int d^4x\, J\phi
\right],
\end{equation}
the regulator contribution is defined as
\begin{equation}
\Delta S_k[\phi]
=
\frac{1}{2}
\int d^4x \,
\phi \, R_k \,\phi ,
\end{equation}
where $R_k$ is an infrared cutoff function. In momentum space, modes with $p^2 \ll k^2$ acquire an effective mass term and are therefore suppressed, while modes with $p^2 \gg k^2$ remain essentially unaffected. In this way, quantum fluctuations are integrated out progressively shell by shell in momentum space.

The scale-dependent effective action $\Gamma_k$, also called the effective average action, is then defined through a modified Legendre transform of the generating functional $W_k[J]=\ln Z_k[J]$:
\begin{equation}
\Gamma_k[\phi]
=
\int d^4x\, J\Phi - W_k[J]
-\Delta S_k[\phi].
\end{equation}
where $\Phi$ is defined as the expectation value $\Phi(x)=\frac{\delta W_k[J]}{\delta J(x)}$. By construction, $\Gamma_k$ interpolates between the microscopic action and the full quantum effective action:
\begin{equation}
\Gamma_{k\to\infty}\rightarrow S,
\qquad
\Gamma_{k\to0}\rightarrow \Gamma .
\end{equation}
The most common approach is to explore the evolution of $\Gamma_k$ with an exact functional renormalization group equation, studying the variation of the effective action on the scale, which reads \cite{Wetterich:1992yh,Reuter:1996cp,Wetterich:1989xg}:
\begin{equation}
   k \partial_k\Gamma_k=\frac{1}{2}Tr\left(\frac{k \partial_{k}R_k}{\Gamma_{k}^{(2)}+R_k}\right),
\end{equation}
where $\Gamma_k^{(2)}$ denotes the second functional derivative of the effective average action with respect to the fields and plays the role of the full inverse propagator at scale $k$, while $R_k$ is the infrared regulator function introduced before. The trace includes both a sum over field components and an integration over momenta. The FRGE is exact and non-perturbative, and its solutions define trajectories in the infinite-dimensional theory space of all operators compatible with the symmetries of the theory in the space of action functionals. In practical applications, one usually projects the flow onto a truncated subspace of operators, obtaining a finite set of beta functions for the running couplings.

The scale dependence of the couplings is encoded in the corresponding beta functions,
\begin{equation}
\beta_{g_i}\equiv k\partial_k g_i(k),
\end{equation}
which determine the renormalization group trajectory in theory space. A fixed point is defined by the condition
\begin{equation}
\beta_{g_i}(g^\ast)=0
\qquad
\forall i.
\end{equation}
In particular, a Gaussian fixed point corresponds to vanishing dimensionless couplings, while a non-Gaussian fixed point (NGFP) is characterized by finite non-zero values of the couplings. A theory is said to be asymptotically safe if the renormalization group trajectory approaches an ultraviolet NGFP as $k\to\infty$, ensuring that all physical couplings remain finite at arbitrarily high energies. If the ultraviolet critical surface associated with the fixed point is finite-dimensional, the theory remains predictive despite the presence of infinitely many operators compatible with the symmetries.

An alternative implementation of the renormalization group flow is provided by the proper-time (PT) formalism \cite{Bonanno:2004sy,deAlwis:2017ysy}. In this approach, the one-loop effective action is represented through a Schwinger proper-time integral and the coarse-graining procedure is implemented by introducing a cutoff function directly in the proper-time parameter. Although the resulting flow equation is not exact in the same sense as the Wetterich equation, the PT approach captures important non-perturbative aspects of Wilsonian renormalization and has proven particularly useful in gravitational applications \cite{Bonanno:2019ukb,Bonanno:2025tfj}.

This framework has been applied to a variety of physical systems, including cosmology \cite{Zholdasbek:2024pxi,Bonanno:2024xne,Platania:2020lqb,Bonanno:2017kta,Bonanno:2016dyv} and black hole physics \cite{Koch:2014cqa,Platania:2023srt,Bonanno:2000ep,Held:2019xde,Platania:2019kyx,Bonanno:2024wvb,Bonanno:2023rzk,Bonanno:2025dry}. In the following, we focus on the application of the asymptotic safety scenario to the study of black hole solutions. In particular, after we show the main black holes solutions in this framework \cite{Spina:2025wxb}, we will focus mostly on the metric we obtained and studied with proper time approach in \cite{Bonanno:2025dry}.  

\section{Static Black Holes from RG Improvement}\label{Sec2.1}

The first black hole solution within the Asymptotic Safety framework was developed (by Bonanno and Reuter in Ref. \cite{Bonanno:2000ep}) using a procedure known as RG improvement. This method consists of substituting the classical couplings in the equations of motion or in the solutions with their scale-dependent counterparts obtained from the renormalization group flow. As outlined in the review \cite{Platania:2023srt}, the standard approach begins with a truncation of the effective average action—typically starting with the simplest case, the Einstein–Hilbert truncation—in which the couplings (such as Newton’s constant and the cosmological constant) are promoted to running quantities.
\begin{equation}
    \Gamma_k=\frac{1}{16{\pi}G(k)}\int{d^{4}x\sqrt{-g}(R-2\Lambda(k))},
\end{equation}

From this ansatz, the scale dependence of the couplings is obtained by solving the corresponding beta functions. For instance, the beta function for the dimensionless Newton's constant $g_k=k^2 G_k $ takes the form:
\begin{equation} \label{beta}
    k \frac{\partial g_{{k}}}{\partial k} = \beta_g \equiv \left(2 + \eta_N\right) g_k,
\end{equation}   
{where $\eta_N$ denotes the anomalous dimension, defined as:
$$\eta_N(g_k)=\frac{ B_1 g_k}{1- B_2 g_k},$$ 
with $B_1$ and $B_2$ that are some constants determined by the truncation given by the cutoff \cite{Bonanno:2000ep}}. By integrating the flow equation for the beta function, typically derived using the background gauge field methods, one obtains an explicit expression for the running Newton’s constant \cite{Bonanno:2000ep,Platania:2023srt,Koch:2014cqa}
\begin{equation} \label{running G}
    G(k)=\frac{G_0}{1+{\omega}G_{0}k^2},
\end{equation}
where $G_0$ is the low energy Newton constant and $\omega$ is a constant which depends on the value of the fixed point \cite{Bonanno:2000ep}.

{We now apply the RG improvement procedure to classical black-hole solutions.} In this review we considered only spherically symmetric spacetime of the form 
\begin{equation} \label{line element}
  ds^2=-f(r)dT^2+\frac{1}{f(r)}dr^2+{r^2}d\Omega^2.  
\end{equation}
with $f(r)=1-\frac{2G(r)M}{r}$, where we have to substitute the classical gravitational constant with his running counterpart, which contains the deviations from general relativity.

{A key point concerns the identification of the physical scales of the system with the cutoff scale $k$.  This identification allows $k$ to be understood as a physical IR cutoff associated with a specific length scale of the system.} There are some general criteria that any physically meaningful identification of the cutoff scale $k(r)$ is expected to fulfill \cite{Koch:2014cqa}. First, the choice of $k(r)$ should be coordinate-independent, meaning that it must be constructed from diffeomorphism-invariant quantities, such as proper distances or curvature scalars, so as to preserve the geometric nature of the theory. Lastly, the functional form of $k(r)$ should be compatible with the symmetries of the underlying spacetime geometry. {Different choices of cutoff identification therefore give rise to} different improved solutions.

\subsection{Bonanno-Reuter {Metric}}\label{Sec2.1.1}
{The first solution we consider is also the first historically developed, namely the Bonanno–Reuter model \cite{Bonanno:2000ep}}, further explored in \cite{PhysRevD.73.083005}. In analogy with QED theory, where the Wilsonian infrared cutoff scale \(k\) is typically identified with the inverse of the radial coordinate, $(k \sim 1/r)$, Bonanno and Reuter proposed in \cite{Bonanno:2000ep,Bonanno:1998ye} a scale-setting procedure based on proper distances. In their approach, the RG scale $k(r)$ is linked to the proper radial distance $d(r)$ from the center $r = 0$ to a generic point at radius $r$ along a radial geodesic, such that
\begin{equation}
    k(r) \sim \frac{\xi}{d(r)},
\end{equation}
where $\xi$ is a numerical constant that sets the scale of quantum gravity effects. $d(r)$, in particular, {is obtained by interpolating the behavior between the} small and large radii of the radial proper distance, and reads:
\begin{equation}
    d(r)=\sqrt{\frac{2r^3}{2r+9G_0 M}}.
\end{equation}

This function for small radii goes as $\sim r^{3/2}$ while for large distances reduces to the classical $\sim r$. {Using this relation between $k$ and the position, one can construct the RG-improved metric function $f(r)$}:
\begin{equation} \label{B-R}
    f(r)=1-\frac{2MG_{0}r^2}{r^3+\tilde{\omega}{G_0}(r+\frac{9}{2}{G_0}M)}.
\end{equation}
where $\tilde{\omega}=\omega \xi^2$. Considering the correction to the Newtonian potential obtained from this framework \cite{Bonanno:2000ep}, one can fix the value of $\tilde{\omega}=118/15\pi$ to be in agreement with perturbative quantization of Einstein gravity \cite{Donoghue:1993eb,Hamber:1995cq}.

The metric function given in Equation \eqref{B-R} asymptotically approach\-es the Schwarzschild solution as $ r \to \infty$ (as do all the other solutions discussed in this thesis), while in the limit $ r \to 0 $ it exhibits a regular de Sitter core, thus avoiding the curvature singularity \cite{Bonanno:2000ep,Platania:2023srt}. 

As in many other scenarios (including all the models considered in this thesis), the horizon structure depends on the value of a free parameter. In particular, one typically finds three possible configurations: two horizons, a degenerate (extremal) horizon, or no horizon at all, depending on whether the parameter lies above, at, or below a critical threshold. In the Bonanno-Reuter (B-R) solution, the only free parameter is the black hole mass. Below a certain critical mass, the geometry becomes horizonless. {In the following, adopting Planck units ($G_0 = c = \hbar = 1$), so that lengths and masses are measured in multiples of the Planck length 
$$  \ell_{ Pl} = \sqrt{\frac{G_0\hbar}{c^3}} \,,$$
and the Planck mass 
   $$M_{Pl} = \sqrt{\frac{\hbar c}{G_0}} \,,$$}
 the critical value is approximately \cite{Bonanno:2000ep,Konoplya:2022hll}:
$$ M_{\text{crit}} \simeq 3.503. $$

\subsection{Hayward {Metric}}\label{Sec2.1.2}
{As discussed earlier,} the functional form of $k(r)$ is not unique and establishes different metric solutions. A reasonable choice could come from curvature scalars, particularly, in Ref. \cite{Held:2019xde}, they have connected $k$ with the Kretschmann scalar, which for classical case is $K=48 G_0^2 M^2 /r^6$ and they choose
\begin{equation}
    k^2=\alpha K^{1/2}.
\end{equation}
with the appropriate choice of $\alpha$ and {measuring the mass in units of length,} one can rewrite the metric  as:
\begin{equation}
    f(r)=1-\frac{2r^2/M^2}{r^3/M^3 + \gamma}
\end{equation}
where {$\gamma$ encodes both the non-universal value of the fixed point and the parameter $\alpha$, thereby providing} the quantum correction to the metric.

It is straightforward to see that this is the form of the notorious Hayward metric \cite{Hayward:2005gi}, which was obtained from simpler argumentation and not correlated with the AS or FRG and describes a quite simple regular black hole.

As the previous case, we recover the Schwarzschild solution asymptotically, {while varying the parameter determines the horizon structure}. Specifically, the critical value is 
$$\gamma_{crit}\simeq \frac{32}{27}$$
{below which the solution describes a black hole with two horizons}.

\subsection{Dymnikova {Metric}}\label{Sec2.1.3}

The scale identification $k(r)$, which links the deviation from classical spacetime behavior to the running of the gravitational couplings under the renormalization group flow, present some criticalities. When computing physical quantities such as the proper distance or the Kretschmann scalar, one typically relies on the classical background metric $\bar{g}_{\mu\nu} = g^{(0)}_{\mu\nu}$. However, this classical description breaks down at very short distances ($r \ll \ell_{\text{Pl}}$), precisely where quantum gravitational corrections are expected to dominate. Instead, by evaluating these invariants on the improved geometry $g^{(1)}_{\mu\nu}$—which incorporates running couplings via the renormalization group (RG)—one obtains different results, reflecting quantum effects on spacetime structure.

Moreover, in gravity, the identification of the cutoff scale $k(x)$ is not independent of the coupling being improved. Since $k(x)$ is defined using geometric quantities, which themselves depend on the metric, and hence on $G_0$, a nontrivial feedback occurs. This implies that the improvement procedure influences the spacetime in a way that affects the very definition of $k(x)$, leading to potential backreaction effects.

To address this interdependence, one can implement the improvement recursively \cite{Platania:2019kyx}: define a sequence of metrics $g^{(n)}_{\mu\nu}$ where, at each iteration, the running coupling $G_n(r)$ is determined through a cutoff $k_n(x)$ that depends on the previous geometry, i.e., $k_n[g^{(n-1)}_{\mu\nu}]$. Through this iterative construction, a self-consistent quantum-corrected metric $g^\infty_{\mu\nu}$ may be obtained, accounting for the mutual dependence between the geometry and the RG flow.\\
We begin by considering the classical Schwarzschild solution, described by the standard spherically symmetric line element \eqref{line element}. This geometry satisfies the vacuum Einstein field equations,
\begin{equation}
    R_{\mu\nu} - \frac{1}{2} R\, g_{\mu\nu} = 0,
\end{equation}
and is characterized by the lapse function
\begin{equation}
    f^{(0)}(r) = 1 - \frac{2 m G_0}{r}, \label{eq:Schwarzschild_f}
\end{equation}
where $G_0$ denotes the classical (constant) Newton coupling. To incorporate quantum gravity effects via the renormalization group (RG), we now perturb the system by promoting the Newton constant to a scale-dependent quantity, replacing $G_0$ with its running counterpart $G(k(r))$ given by Equation \eqref{running G}. The resulting quantum-corrected geometry can be interpreted as an exact solution of the Einstein field equations sourced by an effective energy-momentum tensor. This tensor takes the form
\begin{equation}
    T^{\text{eff}}_{\mu\nu} = (\rho + p)\left( \ell_\mu n_\nu + \ell_\nu n_\mu \right) + p\, g_{\mu\nu},
\end{equation}
where $\ell^\mu$ and $n^\mu$ are null vectors. The quantities $\rho$ and $p$ represent, respectively, an effective energy density and pressure, and arise from the radial variation of the Newton coupling $G(r)$ (see Ref. \cite{Platania:2019kyx} for details).

This energy-momentum tensor can be interpreted as the manifestation of quantum gravitational vacuum polarization. In this picture, $\rho$ corresponds to an effective self-energy contribution due to quantum gravitational effects. A small variation of $G(r)$ induces a modification in the background geometry, which in turn leads to further corrections to $G(r)$ through backreaction. Consequently, the energy density, $\rho \propto \partial_r \log G(r)$, can be regarded as a proxy for the strength of quantum effects within the black hole interior. This relation may be exploited in iterative schemes aimed at constructing a self-consistent cutoff identification $k^{(n+1)}(r)$ for $n > 1$.

The iterative procedure begins with the classical Schwarzschild geometry as the zeroth-order approximation. The first step corresponds to the replacement of the constant Newton coupling $G_0$ with its scale-dependent counterpart \eqref{running G}, as described earlier. 

Subsequent iterations, for $n > 1$, are defined by the recursive update
\begin{equation}
    G^{(n+1)}(r) = \frac{G_0}{1 + g^{-1} G_0\, k^{2}_{(n+1)}(r)},
\end{equation}
where the cutoff function $k^{(n+1)}(r)$ is constructed as a functional of the energy density $\rho^{(n)}(r)$ determined from the variation of $G^{(n)}(r)$ in the previous step.
\begin{equation}
    k^{2}_{(n+1)}(r) \equiv \mathcal{K}[\rho^{(n)}(r)].
\end{equation} 

Provided a specific choice for the functional $\mathcal{K}$ and under the assumption that the iterative procedure converges, one can explore the existence of a fixed-point configuration by considering the asymptotic limit $n \to \infty$. In this regime, the gravitational coupling becomes fully scale-dependent, yielding $G(r) \equiv G^{(\infty)}(r)$.

To constrain the structure of $\mathcal{K}$, Ref.~\cite{Platania:2019kyx} examines how classical curvature invariants—such as the Ricci scalar and the Kretschmann scalar-- are affected by the spatial variation of Newton’s constant. {Requiring regularity at the center implies a balance between classical tidal forces and their quantum corrections, which guides} the construction of a self-consistent relation between the cutoff scale and the effective energy density encoded in the running of $G(r)$.

In particular, in the quantum-corrected curvature invariants, there is the appearance of the effective scale $G_0 \rho$, which arises naturally as a consequence of the spatial variation of the gravitational coupling. This quantity can be interpreted as a dynamical infrared cutoff, regulating the otherwise divergent classical invariants in the deep interior region.

Accordingly, a physically motivated identification for the RG scale is
\begin{equation}
k^2 \equiv \mathcal{K}[\rho] = \xi\, G_0 \rho,
\end{equation}
where $\xi$ is a positive, dimensionless parameter. This relation provides a concrete prescription to close the iterative improvement scheme by expressing the cutoff scale as a function of the effective energy density. Substituting the constrain, we obtain the metric function as \cite{Platania:2019kyx}:
\begin{equation}
    f(r)=1-\frac{2M}{r}\left(1-e^{-\frac{r^3}{2l^2 M}} \right).
\end{equation} 
Here $l$ is defined as 
$$l=\sqrt{\frac{3\xi}{8\pi g_*}}\ell_{Pl},$$ 
where $\ell_{Pl}$ denotes the Planck length and $g_*$ is the non-trivial fixed-point value of the dimensionless Newton coupling. This solution is a regular black hole of Dymnikova-type \cite{Dymnikova:1992ux}, with a de-Sitter core and which recover Schwarzschild, like the other solutions, for large radii. Moreover, $l$ is a characteristic length scale of the order of the Planck length. Its critical value corresponds to the maximum scale for which an event horizon still exists; for larger $l$ the horizon disappears. Approximately, one finds \cite{Konoplya:2023aph}:
\begin{equation}
    l_{\text{cr}} \approx 1.138\, M .
\end{equation}

\section{Dynamical Black Holes from Gravitational Collapse}\label{Sec2.2}

While the static RG-improved black hole solutions provide important insights into possible quantum-gravitational modifications of the spacetime geometry, they lack a dynamical process of black hole formation. To address this limitation, it is natural to consider more realistic scenarios involving gravitational collapse, where quantum effects are taken into account during the dynamical evolution of the matter content. This approach allows for a more physical modeling of black hole spacetimes, as it directly incorporates the quantum-gravity-improved dynamics leading to their formation.

In this context, one can regard these models as quantum-gravity-improved versions of the classical Oppenheimer-Snyder-Datt (OSD) \cite{Oppenheimer:1939ue,Datt:1938uwc} collapse scenario, where the RG flow is implemented dynamically. This idea, originally proposed by Markov and Mukhanov \cite{Markov1985DeSI}, in recent years, has gained a lot of interest and several works have developed such improved collapse models within the framework of asymptotic safety \cite{Bonanno:2023rzk,Harada:2025cwd,Bonanno:2025dry}.The key point is that the classical structure of the theory is preserved, while the information about quantum corrections is entirely encoded in the coupling function, which allows one to reinterpret quantum corrections in terms of an effective Newton constant and an effective cosmological term. These studies aim to understand how the running of couplings can influence the causal structure and the formation of singularities, and whether quantum-gravity effects can lead to singularity resolution in a self-consistent way. Furthermore, cosmologically the role of the inflaton
field could be replaced by the effects of the matter-gravity coupling allowing for the dark energy component to be dynamical and act as the inflaton field in the early universe \cite{Chakrabarty:2025gha}. 

We focused, following what we did in \cite{Bonanno:2025dry}, on two correlated models that share the physical approach to obtain the new metric. Starting from the action of GR with matter \cite{Markov:1985py}:
\begin{equation}
\label{actionMM}
S = \frac{1}{16 \pi G_0} \int d^4 x \sqrt{-g} \left[R + 2  \chi(\epsilon)  \mathcal{L}\right] .
\end{equation}
where, $\mathcal{L}$ represents the matter Lagrangian, $\chi$ expresses a gravity–matter coupling function, which depends on the proper energy density $\epsilon$ of the matter fluid. To obtain the total variation of the action, we take the matter Lagrangian as a barotropic perfect fluid $\mathcal{L}^{(m)}=-\epsilon,$ and consider separately the matter sector with the rest. The variation of the Einstein--Hilbert term gives, up to boundary terms,
\begin{equation}
\frac{1}{\sqrt{-g}}\delta\left(\sqrt{-g}R\right)
=
\left(R_{\mu\nu}-\frac{1}{2}g_{\mu\nu}R\right)
\delta g^{\mu\nu}.
\end{equation}
The matter contribution requires more care, since \(\epsilon\) depends implicitly on the metric. One obtains
\begin{equation}
\frac{1}{\sqrt{-g}}\delta\left(2\sqrt{-g}\chi\epsilon\right)
=
-\chi\epsilon\, g_{\mu\nu}\delta g^{\mu\nu}
+2\frac{\partial(\chi\epsilon)}{\partial\epsilon}
\delta\epsilon .
\end{equation}
For a perfect fluid with rest-mass density \(\rho\), four-velocity \(u^\mu\), and
particle-number conservation
\begin{equation}
\nabla_\mu(\rho u^\mu)=0 ,
\end{equation}
the metric variation of \(\rho\) is \cite{Minazzoli:2012md,Harko:2010zi}
\begin{equation}
\delta\rho
=
\frac{\rho}{2}
\left(g_{\mu\nu}+u_\mu u_\nu\right)\delta g^{\mu\nu}.
\end{equation}
Using the thermodynamic relation for a non dissipative fluid
\begin{equation}
\frac{\delta\rho}{\rho}
=
\frac{\delta\epsilon}{\epsilon+p},
\end{equation}
one finds
\begin{equation}
\delta\epsilon
=
-\frac{1}{2}
(\epsilon+p)
\left(g_{\mu\nu}+u_\mu u_\nu\right)\delta g^{\mu\nu}.
\end{equation}
Substituting this expression into the variation of the matter sector gives
\begin{align}
\frac{1}{\sqrt{-g}}\delta\left(2\sqrt{-g}\chi\epsilon\right)
&=-
\left[
\chi\epsilon\,g_{\mu\nu}
+
\frac{\partial(\chi\epsilon)}{\partial\epsilon}
(\epsilon+p)
\left(g_{\mu\nu}+u_\mu u_\nu\right)
\right]\delta g^{\mu\nu}
\nonumber\\
&=-
\left[
\frac{\partial(\chi\epsilon)}{\partial\epsilon}
T_{\mu\nu}
+
\epsilon^2\frac{\partial\chi}{\partial\epsilon}
g_{\mu\nu}
\right]\delta g^{\mu\nu},
\end{align}
where
\begin{equation}
T_{\mu\nu}
=
(\epsilon+p)u_\mu u_\nu+p\,g_{\mu\nu}
\end{equation}
is the usual perfect fluid energy-momentum tensor which is conserved. Requiring \(\delta S=0\) for arbitrary variations \(\delta g^{\mu\nu}\), the total variation of the action yields the following field equations \cite{Harada:2025cwd,Bonanno:2023rzk}:
\begin{equation}
\label{effectiveEQ}
R_{\mu\nu}- \frac{1}{2}g_{\mu\nu}R = \frac{\partial (\chi \epsilon)}{\partial\epsilon}
T_{\mu\nu}+\frac{\partial \chi}{\partial \epsilon} \epsilon^2 g_{\mu\nu} \equiv T_{\mu\nu}^{\rm eff},
\end{equation}
This equations establish a relation between the matter sector with the coupling function $\chi$ and the effective Newton constant and cosmological constant, respectively.
\begin{equation}
\label{efeg}
8\pi G(\epsilon)=\frac{\partial (\chi \epsilon)}{\partial \epsilon}, \quad \Lambda(\epsilon)=-\frac{\partial \chi}{\partial \epsilon} \epsilon^2.
\end{equation} 
so that the modified Einstein equations can be written as
\begin{equation}
R_{\mu\nu}-\frac{1}{2}g_{\mu\nu}R
=
8\pi G(\epsilon)T_{\mu\nu}
-\Lambda(\epsilon)g_{\mu\nu}.
\end{equation}
Inverting the equations \eqref{efeg}, holds
\begin{equation}
\chi(\epsilon)=\frac{1}{\epsilon}\int_0^{\epsilon} 8\pi G(s)\,ds,
\end{equation}
and hence we could write
\begin{equation}
\Lambda(\epsilon)=\int_0^{\epsilon} 8\pi G(s)\,ds \;-\; 8\pi\,\epsilon\,G(\epsilon) \;.
\end{equation}
Now, considering spherically homogeneous collapse, we assume a FLRW metric which is dependent on the scale factor $a(t)$
\begin{equation} \label{FLRW}
    ds^2 = -dt^2 + a^2(t) \left( \frac{dr^2}{1 - K r^2} + r^2 d\Omega^2 \right).
\end{equation}

In this framework, one can set $K = 0$ without loss of generality, as it can be shown that the final result obtained from the matching conditions does not depend on the specific value of $K$ \cite{Harada:2025cwd,Bonanno:2023rzk,Bonanno:2025dry}. Under this assumption, the modified field equations simplify to a Friedmann-like equation:
\begin{equation}
H^2 \equiv \left(\frac{\dot a}{a}\right)^2
= \frac{8\pi}{3}\,G(\epsilon)\,\epsilon + \frac{\Lambda(\epsilon)}{3}
= \frac{1}{3}\int_0^{\epsilon(a)} 8\pi G(s)\,ds.
\end{equation}
It is reduced to a relation for the scale factor \cite{Harada:2025cwd,Bonanno:2023rzk}:
\begin{equation} \label{adot-pot}
    \dot{a}^2=-V(a)
\end{equation}
where the effective potential $V(a)$ encodes the dynamics of the collapsing matter content. As discussed previously, this is related by the scale dependence of the Newton coupling, and the potential is defined as:
\begin{equation}\label{V(a)}
    V(a) = -\frac{a^2}{3} \int_0^{\epsilon(a)} G(s) \, ds,
\end{equation}
with $\epsilon(a)$ representing the energy density as a function of the scale factor. The presence of the running coupling $G(s)$ captures the leading quantum gravitational corrections to the classical collapse dynamics. Once the functional form of $G(\epsilon)$ is specified, the full collapse dynamics can be derived. 

\subsection{Matching conditions}

As in the standard Oppenheimer-Snyder (OSD) collapse model, we assume that the interior homogeneous dust spacetime can be smoothly matched to an exterior, static and spherically symmetric geometry, whose line element is given by
\begin{equation}
  ds^2=-f(R)dT^2+\frac{1}{f(R)}dR^2+R^2 d\Omega^2.  
\end{equation}

The matching between the interior and exterior spacetimes is performed across a timelike hypersurface $\Sigma$ located at a fixed comoving radius $r=r_b$. This hypersurface represents the boundary of the collapsing cloud.

From a geometrical point of view, the Israel junction conditions \cite{Israel:1966rt} require the continuity across $\Sigma$ of the first and second fundamental forms, namely the induced metric $\gamma_{ab}$ and the extrinsic curvature $K_{ab}$. Explicitly, the junction conditions can be written as
\begin{equation}
[\gamma_{ab}] \equiv \gamma^{+}_{ab} - \gamma^{-}_{ab} = 0,
\qquad
[K_{ab}] \equiv K^{+}_{ab} - K^{-}_{ab} = 0.
\end{equation}

Mathematically, the first condition ensures the continuity of the metric across the boundary, while the second enforces the continuity of its first derivatives normal to $\Sigma$. Physically, these conditions guarantee that no singular surface layer of matter (such as a thin shell with distributional stress-energy tensor) is present at the boundary.

The induced metric on the three-dimensional hypersurface $\Sigma$ is defined as
\begin{equation}
ds^2_\Sigma = \gamma_{ab} dy^a dy^b,
\end{equation}
with
\begin{equation}
\gamma^{\pm}_{ab} = 
\frac{\partial x^\mu_{\pm}}{\partial y^a}
\frac{\partial x^\nu_{\pm}}{\partial y^b}
g^{\pm}_{\mu\nu},
\end{equation}
where $y^a$ are intrinsic coordinates on $\Sigma$, and $x^\mu_{\pm}$ denote the spacetime coordinates in the interior ($-$) and exterior ($+$) regions.

In our case, choosing $y^a = (t,\theta,\phi)$, the induced metric from the interior side is
\begin{equation}
ds^2_\Sigma = -dt^2 + r_b^2 a^2(t)\, d\Omega^2,
\end{equation}
while from the exterior side we obtain
\begin{equation}
ds^2_\Sigma = -\left[f(R_b)+\frac{1}{f(R_b)}\left(\frac{dR_b}{dT}\right)^2\right] dT^2 + R_b^2 d\Omega^2.
\end{equation}

The continuity of the induced metric immediately provides two conditions. The angular part implies
\begin{equation}
    R_b(T(t)) = r_b a(t),
\end{equation}
which identifies the areal radius of the boundary in the exterior spacetime with the physical radius of the collapsing shell. The temporal part instead relates the time coordinates $t$ and $T$ through
\begin{equation}
    \left(\frac{d t}{d T}\right)^2 = f(R_b)+\frac{1}{f(R_b)}\left(\frac{d R_b}{d T}\right)^2.
\end{equation}

We now turn to the extrinsic curvature, defined as
\begin{equation}
  K_{ab}^{\pm}=  \frac{\partial x^\alpha_{\pm}}{\partial y^a}
\frac{\partial x^\beta_{\pm}}{\partial y^b}\nabla_\alpha n_\beta
\end{equation}
where $n^\beta$ is the unit normal vector to $\Sigma$.

Physically, $K_{ab}$ encodes how the hypersurface $\Sigma$ is embedded in the spacetime, i.e. how it is curved with respect to the surrounding geometry. Its continuity ensures that the embedding is smooth and that no distributional sources are present at the boundary.

Due to spherical symmetry, the only non-trivial conditions arises from the angular and the temporal components. For the angular part, on the exterior side one finds \cite{Malafarina:2022wmx,Panassiti:2025diw}
\begin{equation}
K^{+}_{\theta\theta} = R_b \, \Delta(R_b),
\qquad
\Delta(R_b) \equiv \sqrt{f(R_b) + \dot{R}_b^2},
\end{equation}
while on the interior side
\begin{equation}
K^{-}_{\theta\theta} = r_b a(t).
\end{equation}
One can also compute the temporal component of the extrinsic curvature. On the exterior side it is given by
\begin{equation}
\frac{1}{\sqrt{f(R_b)+\dot{R}_b^2}}
\left(
\ddot{R}_b+\frac{f_{,R}}{2}
\right),
\end{equation}
while from the interior spacetime one finds
\begin{equation}
K^{-}_{tt}=0.
\end{equation}
However, this condition is not independent from the angular one. Indeed,  differentiating the relation we obtain from the continuity of $K_{\theta\theta}$ together with the matching condition $R_b=r_ba(t)$ automatically yields the continuity of the temporal component. For this reason, imposing $[K_{\theta\theta}] = 0$ and using $R_b = r_b a(t)$, one finally obtains the matching condition
\begin{equation} \label{matching}
1 = f(R_b) + \dot{R}_b^2,
\end{equation}
which governs the dynamics of the boundary surface.

This equation admits a clear physical interpretation: it describes the motion of the boundary as that of a particle with unit energy moving in the effective potential determined by the exterior metric function $f(R)$, thus providing the bridge between the interior collapse dynamics and the exterior spacetime.

\subsection{Proper Time Approach}\label{Sec2.2.1}
To obtain the form of the running gravitational constant we will consider the asymptotic safety framework, hypothesizing that the collapse occurs within that scenario. In this work we are interested in the RG equation regularized via the proper-time integral \cite{Bonanno:2019ukb,Bonanno:2004sy}, and more specifically to \cite{Bonanno:2025tfj}, where the authors apply it on the Einstein-Hilbert truncation:
\begin{equation}
    \Lambda\partial_\Lambda S_\Lambda=\frac{1}{2}\int_0^\infty \frac{ds}{s} r_\Lambda(s)\mathrm{STr}\left( e^{-s \tilde{S}_\Lambda^{(2)}} \right).
\end{equation}

Here, $\tilde{S}_\Lambda^{(2)}$ is the hessian of the gauge fixed Wilsonian effective action and of the ghost action, and the regulating function is contained in
\begin{equation} 
r_\Lambda(s)=\Lambda\partial_\Lambda\rho_{k,\Lambda}(s).
\label{r_function} 
\end{equation}
The flow equation can be treated with different approaches.\\ 
The coarse-graining procedure is not unique even in momentum space, due to the different regularizing functions we can choose, which encode different ways of integrating the UV modes. To analyze how the regularization may affect the observables, it is important to test the theory using different choices. For this purpose, we utilize a family of schemes, parametrized by $m$
\begin{equation} \rho_{k,\Lambda}(s;m)=\frac{\Gamma\left(m,msk^2\right)-\Gamma\left(m,ms\Lambda^2\right)}{\Gamma\left(m\right)} \end{equation}
that implies
\begin{equation} r_\Lambda(s;m)=\frac{2}{\Gamma\left( m\right)}\left(ms\Lambda^2\right)^m e^{-ms\Lambda^2}. \end{equation}
To guarantee the correct suppression of the modes, this regulator must be adapted to hessians where the laplacians have coefficients $A_{\Lambda,i}$. In this case we simply rescale the regulating functions by $A_{\Lambda,i}$. This rescaling can be applied in two different ways.
First, if we rescale directly  $r_\Lambda$, it results
\begin{equation} r_\Lambda(A_{\Lambda,i} s;m)=\frac{2}{\Gamma(m)}\left(mA_{\Lambda,i} s\Lambda^2\right)^m e^{-mA_{\Lambda,i} s\Lambda^2}. \end{equation}
Alternatively, if we rescale $\rho_{k,\Lambda}$ as in
\begin{equation} \rho_{k,\Lambda}(A_{\Lambda,i}s;m)=\frac{\Gamma\left(m,mA_{\Lambda,i}sk^2\right)-\Gamma\left(m,mA_{\Lambda,i}s\Lambda^2\right)}{\Gamma\left(m\right)}, \end{equation}
the eq. \eqref{r_function} implies
\begin{equation}
	r_\Lambda(A_{\Lambda,i} s;m)=\frac{2}{\Gamma(m)}\left(1+\frac{\Lambda}{2}\partial_\Lambda \log A_{\Lambda,i}\right)\left(mA_{\Lambda,i} s\Lambda^2\right)^m e^{-mA_{\Lambda,i} s\Lambda^2}.
\end{equation}
We combine these two families into one, introducing a second parameter $\varepsilon$
\begin{equation}
	r_\Lambda(A_{\Lambda,i} s;m,\varepsilon)=\frac{2}{\Gamma(m)}\left(1+\varepsilon\frac{\Lambda}{2}\partial_\Lambda \log A_{\Lambda,i}\right)\left(mA_{\Lambda,i} s\Lambda^2\right)^m e^{-mA_{\Lambda,i} s\Lambda^2}.
	\label{epsilon scheme}
\end{equation} 
In the following, we use the same nomenclature as in \cite{Bonanno:2025tfj}. For $\varepsilon=0$, the derivative of $\log A_{\Lambda,i}$ is absent, leading to the ``C  scheme'', while for $\varepsilon=1$ the derivative is present, corresponding to the ``B  scheme''.

Other choices in the construction of the effective action include the field parametrization, which can be either linear or exponential, as well as multiple gauge-fixing options, such as the background field gauge and two different physical gauges.

By analyzing all these configurations, it is possible to obtain an explicit form of the beta functions, which reveal a residual dependence on these unphysical choices. More specifically, we are interested in the equation for the dimensionless running gravitational constant $g_\Lambda$, evaluated in spacetime dimension $d=4$, and for a vanishing cosmological constant. For all the configurations it can be expressed as:
\begin{equation} \label{betapt}
    \Lambda \partial_\Lambda g_\Lambda = \beta_g \equiv \left(2 + \eta_N\right) g_\Lambda,
\end{equation}
where \cite{Bonanno:2025tfj}
\begin{equation}
    \eta_N=-\frac{\Omega g_\Lambda}{1-\varepsilon \frac{\Omega}{2}g_\Lambda}.
\end{equation}
$\Omega$ is a dimensionless constant that encodes the dependencies on the gauge, the parametrization, and the regulating parameter $m$.
The beta function posses a UV attractive non-Gaussian fixed point
\begin{equation}\label{NGFP}
    g_*=\frac{2}{\left(1+\varepsilon\right)\Omega}
\end{equation}
with the critical exponent
\begin{equation}
    \left.\frac{\partial \beta_g}{\partial g_\Lambda}\right|_{g_\Lambda=g_*}=-2\left(1+\varepsilon\right)
\end{equation}
that shows, independently of the unphysical choices, the property of AS.\\

We are interested in the flow of the dimensionful coupling constant $G(\Lambda)=g_\Lambda \Lambda^{-2}$, obtained by integrating equation \eqref{betapt}. Remarkably, we find that the solution, for both schemes, is not explicitly dependent on the unphysical value of $\Omega$. This dependence is only implicit, through the value of the fixed point \eqref{NGFP}, a well-known non-universal quantity. However, as we will show later, this dependence can be reabsorbed through a redefinition of the free parameter, which relates the scale $\Lambda$ to the energy density, and is left undetermined.\\
The only variable that affects the functional form of the running coupling is the choice of the regularization scheme. Setting the boundary condition $G_0=G(\Lambda=\Lambda_0)$, with the C scheme, we obtain
\begin{equation}\label{G running C scheme}
    G^{(C)}(\Lambda)=\frac{G_0}{1+\left(\frac{G_0}{g_*}\right)(\Lambda^2-\Lambda_0^2)},
\end{equation}
while with the B scheme
\begin{equation}\label{G running B scheme}
    G^{(B)}(\Lambda)=\frac{G_0}{\frac{\Lambda^2}{2}\left(\frac{G_0}{g_*}\right)+\sqrt{1-\Lambda_0^2\left(\frac{G_0}{g_*}\right)+\frac{1}{4}\Lambda^4\left(\frac{G_0}{g_*}\right)^2}}.
\end{equation}
The C scheme reproduces the standard form previously derived in earlier works, such as \cite{Bonanno:2021squ,Bonanno:2004sy}, and subsequently adopted in the recent related study \cite{Bonanno:2023rzk} and whose test fields perturbations have been studied in \cite{Stashko:2024wuq,Spina:2024npx}, while the B scheme is the one that we use in this work \cite{Bonanno:2025dry}. 
In order to connect the energy scale $\Lambda$ with the proper energy density $\epsilon$, i.e. the physical properties of the system, we follow the prescription adopted in earlier applications of the AS \cite{Bonanno:2023rzk,Bonanno:2000ep,Bonanno:2020bil,Platania:2019kyx} where the cutoff scale is assumed to be inversely proportional to the proper distance $\Lambda \sim 1/d.$ So we can define $\Lambda^2=q\epsilon$, where $q$ is a free parameter that absorbs dependence on the precise choice of cutoff identification and obtain, for $G_0=1$ and $\Lambda_0=0$:
\begin{equation}
    G^{(B)}(\epsilon)=\frac{2g_*}{q \epsilon+\sqrt{4g_*^2+q^2\epsilon^2}},
\end{equation}
and
\begin{equation} \label{G_schemeC}
G^{(C)}(\epsilon)=\frac{1}{1+\frac{q}{g_*} \epsilon}.
\end{equation}
that is the form used in \cite{Bonanno:2023rzk}.\\
Thanks to the freedom to redefine the parameter $q$, $g_*$ can be absorbed via the substitution $q \rightarrow g_* / q$. Consequently, the running Newton constant is independent of gauge and parametrization choices in both schemes:
\begin{itemize}
    \item [-] Scheme C
\begin{equation}
    G^{(C)}(\epsilon)=\dfrac{1}{1+\dfrac{\epsilon}{q}}.
\end{equation}
From this functional form of the running gravitational constant and imposing the matching conditions, in \cite{Bonanno:2023rzk} obtained the solution:
\begin{equation}
     f(r)=1-\frac{r^2}{3\xi}\log{(1+\frac{6M\xi}{r^3})},
\end{equation}
with $\xi=1/q$. In this solution, the scale factor never goes to zero, Schwarzschild is recovered at large radii, and a critical parameter value determines the horizon structure. Specifically, in units of mass ($G_0=1$), the critical value is $\xi\simeq0.46$, which sets the maximum parameter value for which horizons still exist.

Due to the extensive existing literature on the C scheme, as in \cite{Bonanno:2025dry}, we focused our analysis on the B scheme unless otherwise specified.
    \item [-]Scheme B
\begin{equation} \label{G_running}
    G^{(B)}(\epsilon)=\dfrac{1}{\dfrac{\epsilon}{2q}+\sqrt{1+\left(\dfrac{\epsilon}{2q}\right)^2}},
\end{equation}

Having the functional form of the running Newton constant in function of $\epsilon$, first we can integrate $G$ of \eqref{G_running} over the energy density  and from \eqref{V(a)} obtain the potential. Then, assuming a dust collapse, so that $\epsilon\propto m_0 a^{-3}$ \cite{Malafarina:2022wmx,Bonanno:2023rzk}, we find the behavior of the scale factor $a(t)$ from eq.\eqref{adot-pot}, obtaining:

\begin{equation} \begin{split}
\dot{a}=&-\sqrt{- \frac{m_0 \left(-3 m_0 + \sqrt{9 m_0^2 + 4 a^6 q^2} \right)}{4 a^4 q} +}\\ &
\overline{+\frac{2}{3} a^2 q \, \text{ArcTanh}\left( \frac{2 a^3 q - \sqrt{9 m_0^2 + 4 a^6 q^2}}{3 m_0} \right)}. 
\end{split} \end{equation}
Solving the equation, we find the behavior of the scale factor in our solution, as shown in fig.\ref{fig_scale} \cite{Bonanno:2025dry}. At large time, the scale factor decays exponentially as
\begin{equation}
   a(t) \sim e^{-\frac{q t^2}{4}}, \quad t\to \infty
\end{equation}
never reaching zero, thereby, preventing the formation of a singularity within a finite time.
\begin{figure}[h!]
    \centering
    \includegraphics[width=0.7\columnwidth]{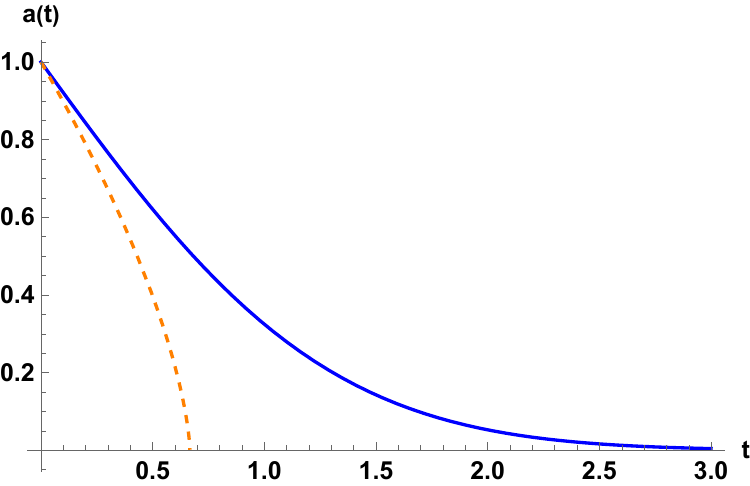}
    \caption{The thick blue line shows the scale factor of the new regular collapse mode with $q=1.40$, while the dashed orange line is the classical OSD collapse model. }
\label{fig_scale}
\end{figure}

At this stage, imposing the junction conditions, whose implies also the relation  $m_0 r_b^3 = 2 M_0$, from \eqref{matching} we derive the modified metric describing the dust collapse solution within the AS framework, as obtained through the PT formulation of the Wilsonian RG.
\begin{equation} 
\begin{split}
\label{metric}
    f(R)=&\frac{3M_0^2+qR^4-M_0\sqrt{9M_0^2+q^2 R^6}}{qR^4}+\\& +\frac{2}{3}qR^2\text{Arctanh}\left(\frac{\left(q-\sqrt{q^2+\frac{9M_0^2}{R^6}}\right)R^3}{3M_0}\right),
\end{split}
\end{equation} 
where $M_0$ represents the asymptotic mass and $q$ is the quantum scale parameter from AS. We should note that we will recover\\ Schwarzschild when $q\to\infty$ (i.e. $R\to \infty)$, in fact expanding we get 
\begin{equation}
    f(R) = 1 - \frac{2 M_0}{R} + \frac{3 M_0^2}{q R^4} - \frac{3 M_0^3}{q^2 R^7} + \mathcal{O}\left( \frac{1}{q^4} \right),
\end{equation}
where we have the Schwarzschild function plus terms of successive orders. Making the expansion around the origin, $R\to 0$ we obtain
\begin{equation}
    f(R) = 1 
+ q\, R^{2} \ln R
- \frac{q}{6} \left[ 1 + \ln 36 - 2\ln q + 2\ln M_0 \right] R^{2} 
+ \mathcal{O}\!\left(R^{8}\right).
\end{equation}
The presence of the term $q\, R^2 \ln R$ makes the curvature scalars singular but, since $R\geq R_b= r_b a(t) $ and the scale factor satisfies $a(t)>0$ at all times as shown before in fig.\ref{fig_scale}, the solution remains regular throughout the spacetime, thereby circumventing the formation of any singularities. Analyzing \eqref{metric}, we want to study the horizons obtained from the equation $f(R)=0$. We note that, as mostly for regular black holes in literature \cite{Hayward:2005gi,Bonanno:2000ep,Dymnikova:1992ux,Bonanno:2023rzk}, as shown before, we have the possibility of the formation from zero to two horizons, depending on the specific values of the parameter. In this case the critical value for the formation of the horizon is $q_{\text{cr}}\simeq1.37$ in mass unit, and for $q>q_{\text{cr}}$ we have the formation of the second one. We summarize  the three cases in fig.\ref{fig_metric} \cite{Bonanno:2025dry}.  
\begin{figure}[H]
    \centering
    \includegraphics[width=0.8\columnwidth]{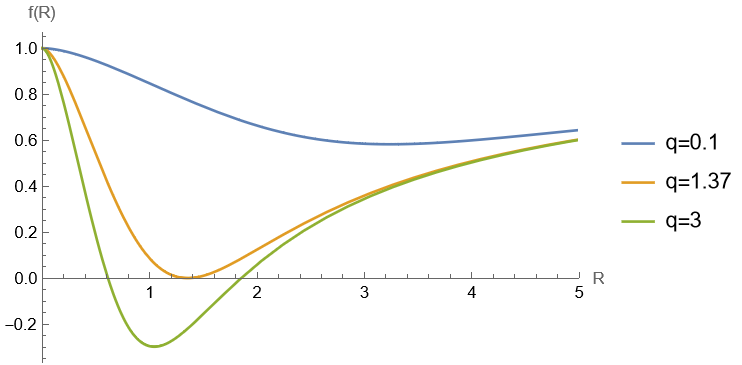}
    \caption{The behavior of the metric function $f(R)$ for different values of the parameter $q$ for $M_0=1$. There are an inner and an outer horizon for $q>1.37$, one horizon  for the critical value $q=1.37$ and no horizon for minor values of this. }
\label{fig_metric}
\end{figure}
An important aspect of the collapse dynamics is the formation of trapped surfaces and the corresponding apparent horizon. In a dynamical spacetime, the apparent horizon is defined locally as the hypersurface on which the expansion of outgoing null geodesics vanishes. Physically, it represents the boundary of the trapped region, namely the region where even outward-directed light rays are forced to move toward smaller areal radii due to the strong gravitational field generated during the collapse.

For a spherically symmetric spacetime, the apparent horizon is determined by the condition
\begin{equation}
g^{\mu\nu}\partial_\mu R \partial_\nu R =0,
\end{equation}
where $(R(t,r))$ is the areal radius. In the interior FLRW geometry considered in this work,
\begin{equation}
R(t,r)=a(t) r,
\end{equation}
and therefore
\begin{equation}
\partial_t R = r\dot a,
\qquad
\partial_r R = a.
\end{equation}
Using the interior metric \eqref{FLRW}, the apparent horizon condition becomes
\begin{equation}
1-r^2\dot a^2=0.
\end{equation}
This yields the comoving radius of the apparent horizon.
\begin{equation}
r_{\rm ah}(t)=\frac{1}{|\dot a(t)|},
\end{equation}
The apparent horizon separates the trapped region from the untrapped region formed during the collapse. Outside, outgoing null geodesics still increase their areal radius, while inside the trapped region both ingoing and outgoing null congruences converge toward smaller radii. In this sense, the apparent horizon marks the onset of local light trapping during the dynamical evolution. As shown in Fig.~\ref{rapphor}, the evolution of the apparent horizon in the regular collapse model significantly differs from the classical Oppenhei\-mer--Snyder--Datt case. Initially, the apparent horizon decreases similarly to the classical behavior, reflecting the accelerating phase of the collapse. However, once quantum-gravity effects become relevant at high densities, the running Newton coupling weakens the effective gravitational attraction, slowing down the collapse dynamics. As a consequence, the comoving apparent horizon reaches a minimum value and subsequently increases, reflecting the asymptotic freezing of the collapse. In contrast, in the classical OSD model the apparent horizon monotonically shrinks to zero at finite time, corresponding to the formation of the spacelike singularity.
\begin{figure}[H]
    \centering
    \includegraphics[width=0.8\columnwidth]{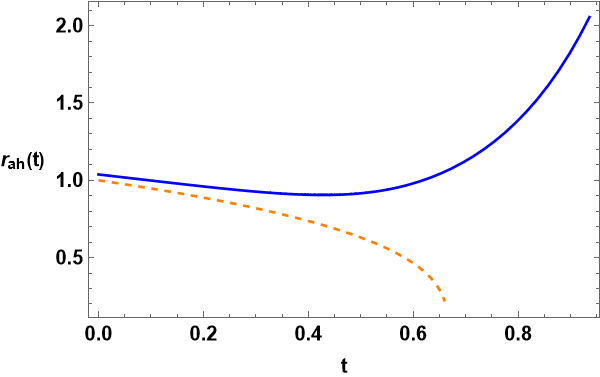}
    \caption{Evolution of the comoving apparent horizon $r_{\rm ah}(t)$ for the regular collapse model (solid blue curve) compared to the classical Oppenheimer--Snyder--Datt collapse (dashed orange curve). While in the classical case the apparent horizon monotonically shrinks to zero at finite time, signaling singularity formation, the regular model exhibits a minimum followed by an asymptotic growth due to the quantum-gravity-induced slowing down of the collapse dynamics.}
\label{rapphor}
\end{figure}
The apparent horizon intersects the boundary surface of the collapsing cloud when
\begin{equation}
r_{\rm ah}(t)=r_b,
\end{equation}
which implies
\begin{equation}
\dot R_b^2=1.
\end{equation}
Using the matching condition
\begin{equation}
1=f(R_b)+\dot R_b^2,
\end{equation}
one obtains
\begin{equation}
f(R_b)=0.
\end{equation}
Therefore, the intersections of the apparent horizon with the boundary surface occur precisely at the roots of the exterior metric function,
$$
f(R_\pm)=0,
$$
corresponding respectively to the outer event horizon $(R_+)$ and the inner Cauchy horizon $(R_-)$. The qualitative causal structure of the collapse can then be summarized by the Penrose diagram shown in Fig.~\ref{Penrose}.
\begin{figure}[H]
    \centering
    \includegraphics[width=0.4\columnwidth,height=0.4\textheight]{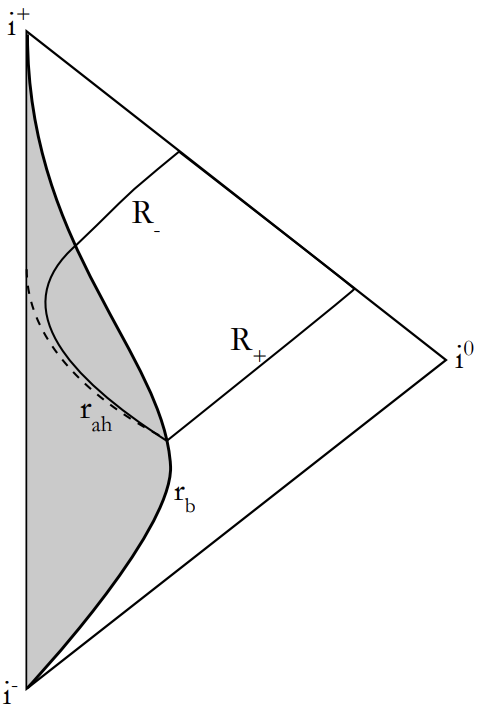}
    \caption{Qualitative Penrose diagram of the regular gravitational collapse model within the AS framework. The gray region represents the interior collapsing cloud, whose boundary surface is described by the timelike curve $r_b$. An apparent horizon $r_{ah}$ develops inside the cloud, connecting the outer event horizon $R_+$ with the inner Cauchy horizon $R_-$. Unlike the classical O-S-D collapse (dashed curve), the evolution does not terminate in a spacelike singularity.}
\label{Penrose}
\end{figure}
During the collapse, a trapped region dynamically forms and an apparent horizon develops inside the collapsing cloud, connecting the outer and inner horizons. Unlike the classical Oppenheimer-Snyder collapse, however, the quantum-gravity corrections prevent the formation of a spacelike singularity. Although horizons and trapped regions are present, the spacetime remains regular throughout the evolution.

\end{itemize}
\subsubsection{Metric Solutions}
\begin{figure}[H]
    \centering
    \includegraphics[width=0.8\columnwidth]{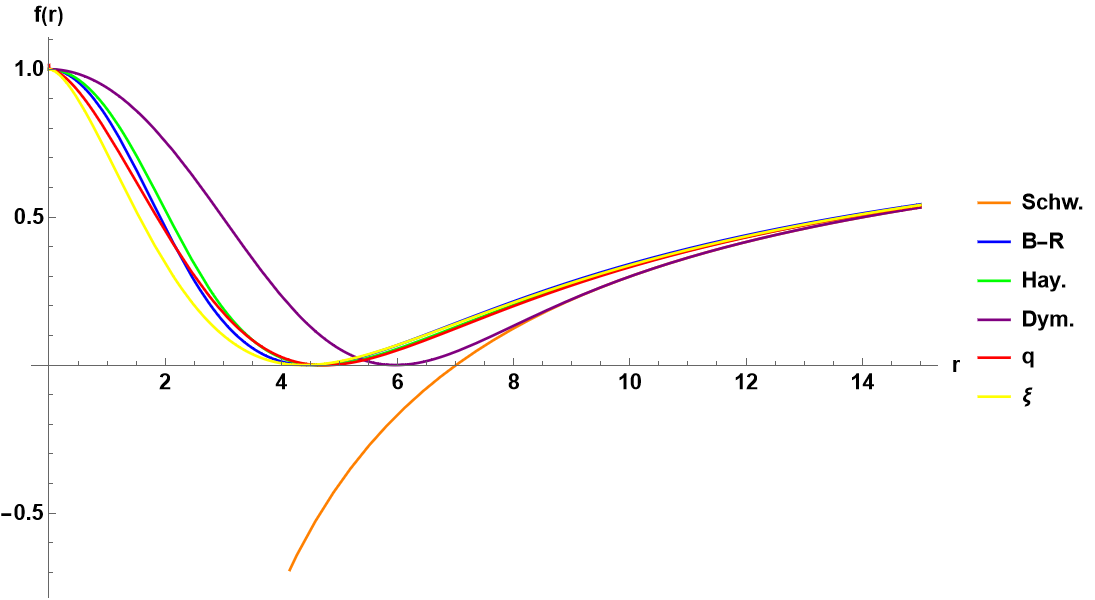}
    \caption{Metric functions $f(R)$ for all the regular black hole solutions at their critical values with $M=3.503$ (critical mass for B-R). }
\label{Metrics}
\end{figure}

After introducing all the solutions of interest, we proceeded to plot the corresponding metric functions $f(r)$ in order to analyze their behavior and perform a direct comparison \cite{Spina:2025wxb}, as shown in Figure~\ref{Metrics}. In particular, we focused on the cases close to the critical values for each model since it is the situation that differs most from the classical case, fixing the mass to $3.503$, which corresponds to the critical value for the Bonanno-Reuter (B-R) solution.
As previously mentioned, we can observe that all the metric profiles converge to the Schwarzschild solution at large distances. However, a distinctive feature of these models inspired by Asymptotic Safety is that this convergence occurs already within a few horizon radii ($2-3 r_H$), as can be seen from the figure \ref{Metrics}.  {Thus, quantum corrections remain negligible at large scales} and become significant only in the near-horizon region, where high-energy effects and strong gravitational fields are expected to play a relevant role, precisely as one would anticipate in a quantum gravity framework.

\subsection{Gravitational perturbations equation}
\label{sec:perturbations}
\
In this section, we focus on axial gravitational perturbations and their associated quasinormal modes (QNMs). The study of gravitational perturbations is intricate in this context, in fact, in general to derive the master equation governing the perturbations, one typically perturbs both the gravitational field equations and the energy-momentum tensor (if present). However, the background metric is not an exact solution of the Einstein equations with quantum corrections, but rather arises from an effective approach. As a result, performing a fully rigorous analysis of gravitational perturbations is challenging. Nonetheless, rather than working directly with these effective equations, we adopt the same approach showed by Ashtekar et al. \cite{Ashtekar:2018cay}, assuming that quantum corrections can be effectively modeled as arising from the energy-momentum tensor of an anisotropic fluid within classical Einstein gravity. This framework makes it possible to investigate perturbations.

Following the works \cite{Bonanno:2025dry,Bouhmadi-Lopez:2020oia,Konoplya:2024lch,Konoplya:2025hgp}, one can study axial gravitational perturbations under the assumption that fluctuations in the anisotropy direction are negligible in the axial sector. Similar simplifications have been adopted in other contexts where certain perturbative contributions are considered subdominant and are thus neglected. Although this approximation may overlook some features of the full gravitational spectrum, it is reasonable when the black hole geometry deviates only slightly from the classical Schwarzschild case, as in our context. This approach is consistent with the notion of perturbative quantum corrections, which are expected to be small.

In the Regge–Wheeler gauge~\cite{Regge:1957td}, the axial gravitational perturbation tensor $h_{\mu\nu}$ takes the form
\begin{equation}
   h_{\mu\nu}^a= \begin{bmatrix}
        0&0&0&h_0(r)\\
        0&0&0&h_1(r)\\
        0&0&0&0\\
        h_0(r)&h_1(r)&0&0
    \end{bmatrix}\left(\sin{\theta\frac{\partial}{\partial\theta}}\right)e^{-i\omega t}P_l(\cos{\theta})
\end{equation}
where $h_0(r)$ and $h_1(r)$ are two unknown functions, and $P_l(x)$ is the Legendre polynomial of degree $l$.

These perturbations can be interpreted as solutions to the Einstein equations sourced by an anisotropic fluid with energy-momentum tensor
\begin{equation}
    T_{\mu\nu}=(\rho+p_t)u_{\mu\nu}+g_{\mu\nu}p_t+(p_r-p_t)s_{\mu}s_{\nu},
\end{equation}
where $\rho$ is the energy density, $p_r$ and $p_t$ are the radial and tangential pressures.

The quantities $\rho$, $p_r$, and $p_t$ transform as scalars under rotations and thus do not contribute to axial perturbations. However, the perturbations of the vectors $\delta s_\mu$ and $\delta u_\mu$ may include non-zero components. We assume that there are no perturbations in the anisotropy direction, i.e., $\delta s_\mu = 0$. Then, from the conservation equation $\nabla_\mu T^{\mu \nu} = 0$, it follows that $\delta u = 0$.

Considering what has just been said, by introducing the tortoise coordinate and a new variable $\Psi$,
\begin{equation}
h_1(r) = \frac{r}{f(r)} \Psi(r), \quad \frac{dr_*}{dr} = \frac{1}{f(r)},
\end{equation}
we obtain that the perturbation equation is reduced to a Schrödinger-like wave equation
\begin{equation}
\frac{d^2\Psi}{dr_*^2} + \left[ \omega^2 - V(r) \right] \Psi = 0,
\end{equation}
with effective potential
\begin{equation}
     V(r)=f(r)\left(\frac{2f(r)}{r^2}+\frac{(l+2)(l-1)}{r^2}-\frac{f'(r)}{r}\right)
\end{equation}

A similar approach was also employed in \cite{Bolokhov:2025lnt} in the context of AS and in~\cite{Bouhmadi-Lopez:2020oia} to study perturbations of black holes in the context of loop quantum gravity.

\begin{figure}[h]
    \centering
    \includegraphics[width=0.7\columnwidth]{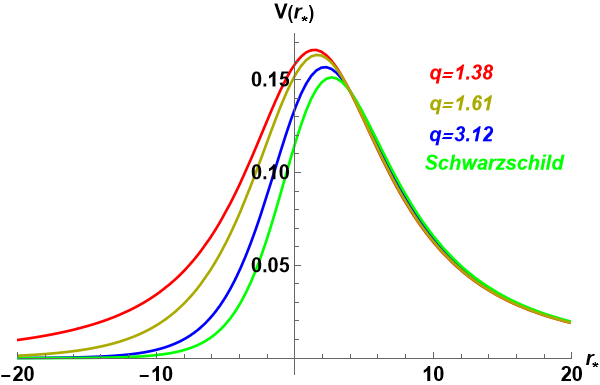}
    \caption{ Effective potential as a function of the tortoise coordinate of the $l=2$ axial gravitational perturbations for different valus of $q$ and Schwarzschild. }
\label{fig_poteff}
\end{figure}

The resulting effective potential $V$ for our solution are shown in fig.\ref{fig_poteff} \cite{Bonanno:2025dry}. One can see that it remains positive definite, in the considered class of axial perturbations within the effective anisotropic fluid framework and we could notice with the decreasing of $q$ the peak of the potential barrier increase, deviating further from Schwarzschild case, as expected.

In this work, we employ the three different methods introduced in the section \ref{num_met} to compute the quasinormal modes: WKB method with Padé approximants, Leaver method with infinite continued fraction and time-domain integration with Prony's method to extract the frequencies.

\subsection{Numerical results for quasinormal modes}

Using the methods described above, we performed an analysis of the gravitational perturbations of our regular black hole solution \cite{Bonanno:2025dry}, that we reported. To this end, we considered a range of values for the parameter $q$ (expressed in mass units), starting from the critical value $q \approx 1.37$ and increasing it gradually—initially with small steps, then with larger increments—as we approached the classical Schwarzschild limit (which is recovered in the limit $q \to \infty$). 

We began by computing the fundamental mode frequencies (in mass units) using both the Leaver and WKB methods for $l=2$ and $l=3$, analyzing the relative percentage error between the two methods and the deviation from the Schwarzschild case as a function of $q$. The resulting values are reported in Tables \ref{table1} and \ref{table2} for $l=2$, and in Tables~\ref{table3} and \ref{table4} for $l=3$. 
\begin{table}[H] 
\centering
\begin{tabular}{ccccc} 
\hline
$q$ & Leaver Re[$\omega M$] & WKB Re[$\omega M$] & WKB Error [\%] & Dev. Schw. [\%] \\
\hline
10.57 & 0.37602 & 0.37602 & 0.00135 & 0.6282 \\
3.12 & 0.38208 & 0.38213 & 0.01386 & 2.2547 \\
1.93 & 0.38787 & 0.38835 & 0.1242 & 3.8004 \\
1.52 & 0.39216 & 0.39209 & 0.0172 & 4.9469 \\
1.42 & 0.39349 & 0.39345 & 0.0103 & 5.3142 \\
1.40 & 0.39372 & 0.39375 & 0.0078 & 5.3791 \\
1.39 & 0.39391 & 0.39390 & 0.0034 & 5.4307 \\
1.38 & 0.39407 & 0.39405 & 0.0053 & 5.4717 \\
1.37 & 0.39419 & 0.39420 & 0.0026 & 5.4996 \\
\hline
\end{tabular} 
\caption{Real part of frequencies with $l=2$ for various values of the parameter: comparison between Leaver and WKB methods, with relative percentage error and deviation from the Schwarzschild value $\text{Re}[\omega_{\text{Schw}}] = 0.3736715$.}
\label{table1}
\end{table} 

\begin{table}[H] 
\centering
\begin{tabular}{ccccc}
\hline
$q$ & Leaver Im[$\omega M$] & WKB Im[$\omega M$] & WKB Error [\%] & Dev. Schw. [\%] \\
\hline
10.57 & -0.08789 & -0.08787 & 0.0241 & 1.2063 \\
3.12 & -0.08468 & -0.08457 & 0.1360 & 4.8284 \\
1.93 & -0.08079 & -0.08106 & 0.3306 & 9.1987 \\
1.52 & -0.07704 & -0.07727 & 0.3030 & 13.363 \\
1.42 & -0.07563 & -0.07574 & 0.1487 & 15.001 \\
1.40 & -0.07539 & -0.07540 & 0.0126 & 15.274 \\
1.39 & -0.07516 & -0.07523 & 0.0897 & 15.531 \\
1.38 & -0.07498 & -0.07505 & 0.0900 & 15.757 \\
1.37 & -0.07484 & -0.07486 & 0.0330 & 15.882 \\
\hline
\end{tabular}
\caption{Imaginary part of frequencies with $l=2$ for various values of the parameter: comparison between Leaver and WKB methods, with relative percentage error and deviation from the Schwarzschild value $\text{Im}[\omega_{\text{Schw}}] = -0.0889625$.}
\label{table2}
\end{table}

\begin{table}[H] 
\centering
\begin{tabular}{cccccc}
\hline
$q$ & Leaver Re[$\omega M$] & WKB Re[$\omega M$] & WKB Error [\%] & Dev. Schw. [\%] \\
\hline
1.376  & 0.631604 & 0.631790 & 0.0296 & 5.366 \\
1.384  & 0.631381 & 0.631505 & 0.0197 & 5.331 \\
1.395  & 0.631083 & 0.631224 & 0.0224 & 5.276 \\
1.408  & 0.630727 & 0.630948 & 0.0352 & 5.216 \\
1.423  & 0.630324 & 0.630410 & 0.0137 & 5.155 \\
1.519  & 0.627965 & 0.627960 & 0.0008 & 4.769 \\
1.932  & 0.620900 & 0.620934 & 0.0054 & 3.582 \\
3.120  & 0.611960 & 0.611964 & 0.0006 & 2.089 \\
10.576 & 0.602912 & 0.602915 & 0.0005 & 0.578 \\
\hline
\end{tabular}
\caption{Real part of frequencies with $l=3$ for various values of the parameter: comparison between Leaver and WKB methods, with relative percentage error and deviation from the Schwarzschild value $\text{Re}[\omega_\text{Schw}] \approx 0.599443$.}
\label{table3}
\end{table}

\begin{figure}[H]
    \centering
    \includegraphics[width=\columnwidth]{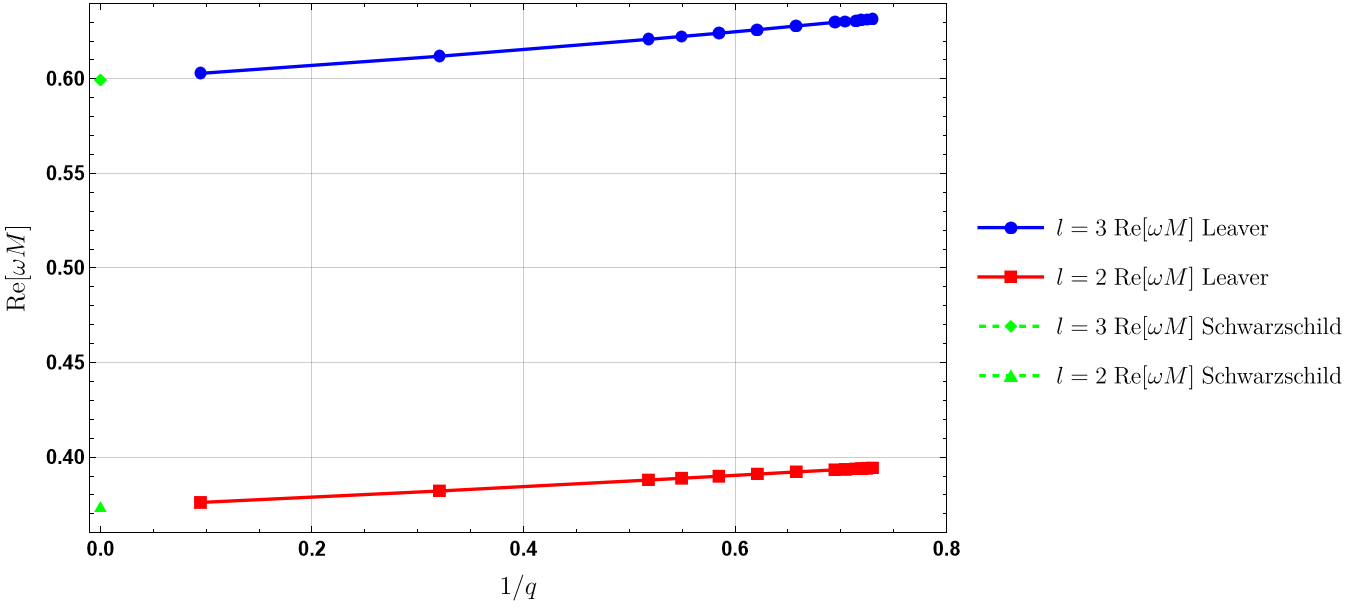}
    \caption{Real part of frequencies for $l=2,3$ for various values of the parameter with Leaver method. Schwarzschild values are also present for comparison  }
\label{figRefreq}
\end{figure}

\begin{figure}[H]
    \centering
    \includegraphics[width=\columnwidth]{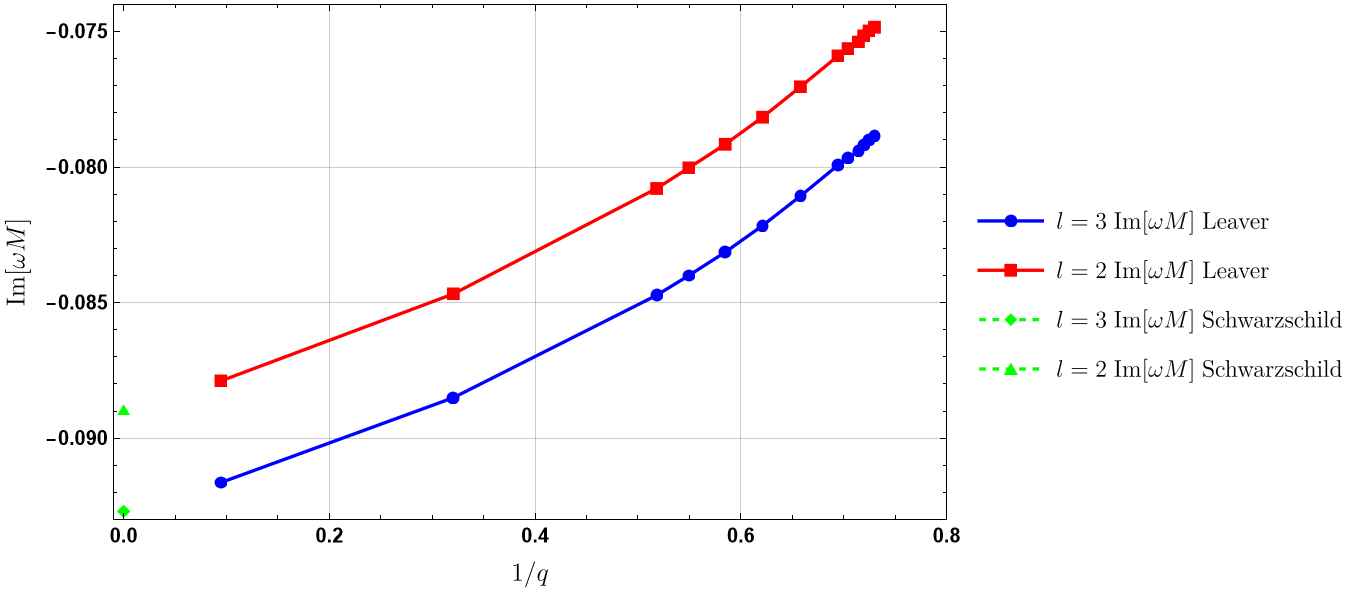}
    \caption{Imaginary part of frequencies for $l=2,3$ for various values of the parameter with Leaver method. Schwarzschild values are also present for comparison }
\label{figImfreq}
\end{figure}

\begin{table}[H] 
\centering
\begin{tabular}{cccccc}
\hline
$q$ & Leaver Im[$\omega M$] & WKB Im[$\omega M$] & WKB Error [\%] & Dev. Schw. [\%] \\
\hline
1.376  & -0.078847 & -0.078724 & 0.1559 & 15.008 \\
1.384  & -0.078991 & -0.078909 & 0.1038 & 14.872 \\
1.395  & -0.079181 & -0.079089 & 0.1165 & 14.593 \\
1.408  & -0.079407 & -0.079265 & 0.1790 & 14.406 \\
1.423  & -0.079658 & -0.079603 & 0.0691 & 14.088 \\
1.519  & -0.081065 & -0.081067 & 0.0022 & 12.552 \\
1.932  & -0.084732 & -0.084716 & 0.0190 & 8.608 \\
3.120  & -0.088509 & -0.088508 & 0.0012 & 4.526 \\
10.576 & -0.091644 & -0.091644 & 0.0002 & 1.141 \\
\hline
\end{tabular}
\caption{Imaginary part of frequencies with $l=3$ for various values of the parameter: comparison between Leaver and WKB methods, with relative percentage error and deviation from the Schwarzschild value  $\text{Im}[\omega_\text{Schw}] \approx -0.092703$.}
\label{table4}
\end{table}
We can immediately observe that the difference between the two methods is negligible, as it is almost always below $0.1\%$. As expected, lowering the value of $q$ and approaching the critical value leads to an increasing deviation from the classical case, with the largest deviations—exceeding $15\%$—found in the imaginary part of the frequencies. We have also produced two plots showing the real and imaginary parts of the fundamental mode frequencies (computed with the accurate Leaver method) as functions of $1/q$ (see Figures~\ref{figRefreq}, \ref{figImfreq}), in order to provide a clearer graphical representation of the behavior of the fundamental mode and its departure from the Schwarzschild limit.

\begin{figure}[H]
    \centering
    \includegraphics[width=0.8\columnwidth]{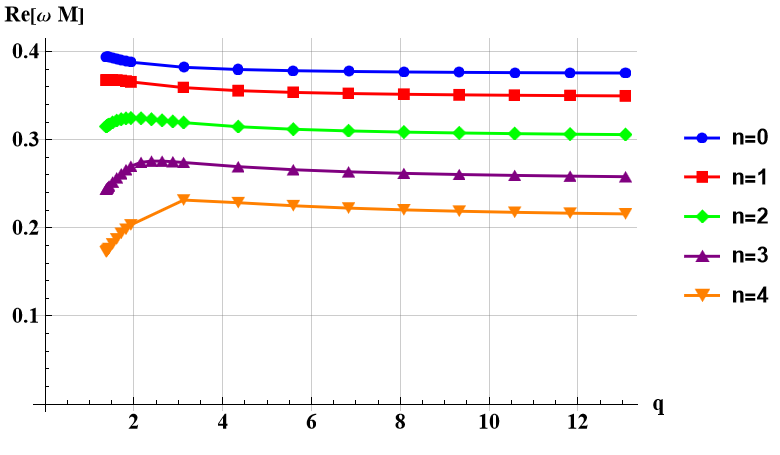}
    \caption{Real part of frequencies for $l=2$  of the overtones for various values of the parameter with Leaver method.}
\label{figReover}
\end{figure}

\begin{figure}[h]
    \centering
    \includegraphics[width=0.8\columnwidth]{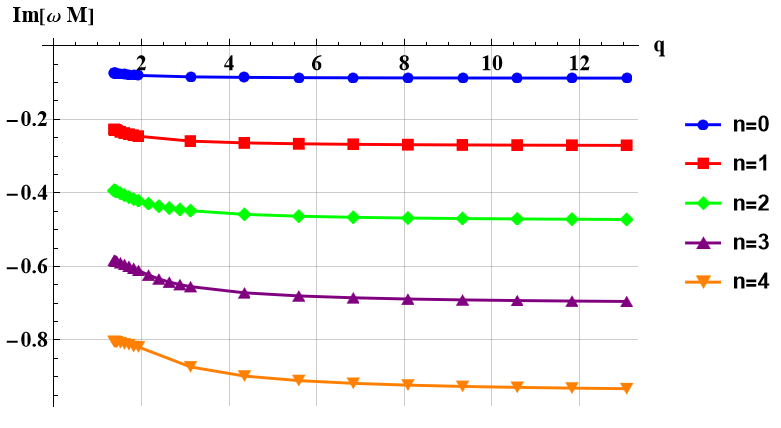}
    \caption{Imaginary part of frequencies for $l=2$ of the overtones for various values of the parameter with Leaver method.}
\label{figImover}
\end{figure}

Subsequently, we computed the first few overtones using the Leaver method for $l=2$, obtaining the results shown in Figures~\ref{figReover} and \ref{figImover}. As is often the case in similar scenarios~\cite{Konoplya:2023aph,Konoplya:2022hll,Spina:2024npx}, the variation with respect to the parameter becomes more pronounced with increasing overtone number, although not so marked for large q, since higher overtones are generally more sensitive to changes near the event horizon~\cite{Konoplya:2022pbc}.

In order to obtain accurate values of the quasinormal frequencies, including higher overtones, we employed the Leaver method, which requires that the coefficients in the corresponding differential equation be given in a rational form. To meet this requirement, we expressed the metric functions as a series expansion in powers of the small parameter $1/q$. The expansion was carried out up to order $q^{-10}$, which proved sufficient for computing the first five overtones, as demonstrated in Table~\ref{TableConvergence}. As seen in the table, for a near-extremal black hole, the frequency of the fourth overtone varies by less than $0.1\%$ when higher-order corrections are included, while the fundamental mode remains unchanged to within six decimal places.

In the above plots and tables, we observe that in the regime where the parameter $q$ approaches its critical values—corresponding to the maximal deviation from the Schwarzschild geometry—the overtones exhibit increasing sensitivity. This phenomenon was referred to as the ``outburst of overtones'' in~\cite{Konoplya:2022pbc}.  As illustrated in Fig.~\ref{fig_metric}, the corrected metric with non-zero $q$ shows significant deviation from the Schwarzschild solution in the near-horizon region, including a notable shift in the location of the event horizon. However, it rapidly converges to the Schwarzschild metric at some distance from the horizon.  As a result, in this regime, each successive overtone deviates from its Schwarzschild counterpart at an accelerating rate, producing what may be interpreted as a distinctive ``sound of the event horizon''~\cite{Konoplya:2023hqb}.

We then proceeded to compare the two solutions obtained using different regularization schemes (Scheme B and Scheme C). In particular, we computed the gravitational perturbation frequencies starting from the critical value for Scheme C ($q_{\text{cr}}\simeq 2.22$) and approaching the Schwarzschild limit, comparing, for the same parameter values, their deviation from the classical case, as reported in Tables \ref{tab 5} and \ref{tab 6}. It should first be noted that, for the model with Scheme B presented in this work, the allowed range of the parameter $q$ is broader, with the extreme values being those already shown and discussed earlier. Regarding the specific values compared in these tables, one can immediately see that the solution with Scheme C, whose frequencies were already computed in \cite{Lutfuoglu:2025ohb}, exhibits a significantly larger deviation from the classical result already at the fundamental mode, compared to the new model. This can be understood from the different critical values of the parameter, in Scheme B $q_{\text{cr}}\simeq 1.37$, while in Scheme~C $q_{\text{cr}}\simeq 2.22$. Hence, for $q\gtrsim 2.2$ we are already far from the critical regime in Scheme B, so its QNM frequencies display a weaker dependence on $q$ than in Scheme C.

\begin{table}[H] 
\centering
\begin{tabular}{ccccc} 
\hline
$q$ & Scheme B & Dev. Schw. B [\%] & Scheme C  & Dev. Schw. C [\%] \\
\hline
2.22 & 0.38619  & 3.35  & 0.40028 & 7.12 \\
2.5  & 0.38452  & 2.90  & 0.39689 & 6.21 \\
5    & 0.37877  & 1.36  & 0.38442 & 2.88 \\
10   & 0.37615  & 0.66  & 0.37874 & 1.36 \\
100  & 0.37387  & 0.053 & 0.37413 & 0.123 \\
\hline
\end{tabular} 
\caption{Real part of frequencies with $l=2$ and the deviation from Schwarzschild value $\text{Re}[\omega_{\text{Schw}}] = 0.3736715$.}
\label{tab 5}
\end{table}

\begin{table}[h] 
\centering
\begin{tabular}{ccccc} 
\hline
$q$ & Scheme B  & Dev. Schw. B [\%] & Scheme C & Dev. Schw. C [\%] \\
\hline
2.22 & -0.08209 & 7.73  & -0.07226 & 18.77 \\
2.5  & -0.08309 & 6.60  & -0.07534 & 15.31 \\
5    & -0.08646 & 2.81  & -0.08385 & 5.75  \\
10   & -0.08780 & 1.31  & -0.08657 & 2.69  \\
100  & -0.08882 & 0.160 & -0.08871 & 0.284 \\
\hline
\end{tabular} 
\caption{Imaginary part of frequencies with $l=2$ and the deviation from Schwarzschild value $\text{Im}[\omega_{\text{Schw}}] = -0.0889625$.}
\label{tab 6}
\end{table}

\begin{figure}[H]
    \centering
    \includegraphics[width=\columnwidth]{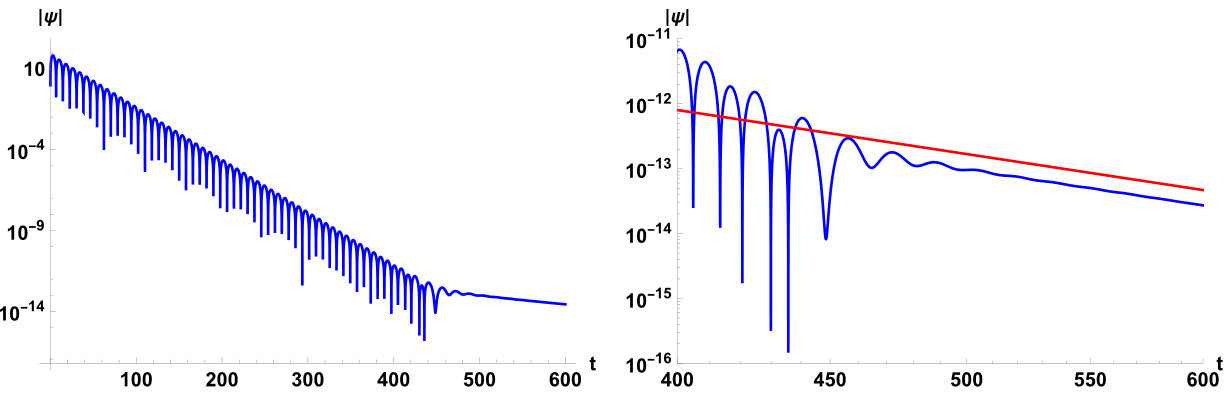}
    \caption{Left panel: Semi Logarithmic plot of time evolution of gravitational perturbation ($l=2$) with $q=1.38$. We also checked frequencies for this $q$ with Prony's method, obtaining the value of $\omega M=0.39401-0.07502i$, and for some other random values, always with errors of the same orders of WKB. \\
    Right panel: Logarithmic plot of the asymptotic tail of the left panel perturbation with a fit given by the law $\propto t^{-7}$.}
\label{fig8}
\end{figure}

As a final analysis, we studied the time evolution of the perturbations. In general, this provides information about the stability of the system — which is already known in our case—and about the asymptotic late-time behavior. As shown in Figure~\ref{fig8} for the example case of $l=2$ and a fixed value of $q$, we find that the late-time behavior follows the universal decay law for massless perturbations, known as Price's law, given by ~\cite{Price:1971fb,Price:1972pw} , 
$$\psi\sim t^{-(2l+3)}, \quad t \rightarrow \infty.$$
In addition to the confirming stability of the system, as clearly visible in the figure, the time-domain analysis also allowed us to perform an additional consistency check of the frequency values for selected representative values of $q$, using the Prony's method. The results confirmed, with high accuracy, the frequencies previously obtained through frequency-domain methods.

\begin{table}[ht]\centering
\begin{tabular}{ccc}
    \hline
  Order &  $\omega$ ($n=0$) & $\omega$ ($n=4$) \\
    \hline
  4 & 0.392968 - 0.0748433 i & 0.170112 - 0.801219 i \\
  6 & 0.394493 - 0.0748375 i & 0.172974 - 0.803734 i \\
  8 &  0.394019 - 0.0748368 i & 0.173142 - 0.804146 i \\
  10 & 0.394193 - 0.0748368  i & 0.172547 - 0.804454 i \\
  12 & 0.394193 - 0.0748368 i & 0.172603 - 0.804042 i \\
  14 & 0.394193 - 0.0748368 i & 0.172749 - 0.804155 i \\
  \hline
\end{tabular}
\caption{Stress test of the convergence  of the expansion of metric coefficient in powers of $1/q$  at $\ell=2$, $q=1.3766$.}
\label{TableConvergence}
\end{table}

After the detailed analysis for the model we built within the proper-time approach, we also performed a brief analysis and review of the QNMs for scalar perturbations of the solutions and models presented previously throughout the chapter on asymptotic safety. Indeed, for the Bonanno–Reuter solution, test-field perturbations were first studied (with limited accuracy) in \cite{Rincon:2020iwy} and later refined in \cite{Konoplya:2022hll}, which also analyzed Hayward-type solutions. The QNM of the Dymnikova solution were computed in \cite{Konoplya:2023aph}, while for the collapse-based solutions the spectra were obtained in \cite{Stashko:2024wuq,Spina:2024npx,Bonanno:2025dry} for test fields. Gravitational perturbations were investigated in \cite{Bolokhov:2025lnt} for B-R, in \cite{Bolokhov:2025egl,Malik:2025dxn} for Hayward, and in \cite{Bonanno:2025dry,Lutfuoglu:2025ohb} for the two proper-time solutions. For Dymnikova, only greybody factors were computed for gravitational perturbations \cite{Dubinsky:2025nxv}.

{Table \ref{QNM} summarizes the frequencies of the solutions close to their critical parameter values and compares them with the Schwarzschild case.} All these values were calculated in units of mass so {it cannot include the Bonanno-Reuter model.}. We report both the fundamental mode and the third overtone  {in order to highlight the peculiar behavior of the QNM spectrum.}
We can observe that, already for the fundamental mode, the imaginary part departs more significantly from the classical Schwarzschild case. In fact, all the models considered display both a larger imaginary part and a larger real part (with the exception of the Dymnikova solution, which instead shows the opposite behavior and, as can already be anticipated from the form of its metric, exhibits smaller deviations from Schwarzschild). As is well known, the fundamental mode is mostly determined by the shape of the effective potential, which in these modified metrics differs only slightly from the classical one, resulting in correspondingly small percent-level deviations.  

The situation changes for the overtones. It has been shown \cite{Konoplya:2022hll,Konoplya:2022pbc} that these modes are closely related to the structure of the event horizon, as said before, which, as illustrated in Figure \ref{Metrics}, departs significantly from the classical case. For this reason, the overtone spectrum is sometimes referred to as the distinctive ``sound of the event horizon'' \cite{Konoplya:2023hqb}. Indeed, as shown in Table \ref{QNM}, already in the scalar case the third overtone displays a much more pronounced deviation in its real part.

\begin{table}[h] 
\centering
\begin{tabular}{ccccc} 
\hline
$n=0$ & Re$(\omega)$ & Dev. Schw. [\%] & Im$(\omega)$ & Dev. Schw. [\%] \\
\hline
Hayward ($\gamma=1$)   & 0.30562 & 4.32  & 0.08559 & 12.37 \\
Dymnikova ($l=1.1$)    & 0.28874 & 1.43  & 0.09468 & 3.05  \\
Scheme B ($q=1.38$)    & 0.30598 & 4.46  & 0.08300 & 15.06 \\
Scheme C ($\xi=0.455$) & 0.31052 & 6.01  & 0.07986 & 18.22 \\
\hline
$n=3$ & Re$(\omega)$ & Dev. Schw. [\%] & Im$(\omega)$ & Dev. Schw. [\%] \\
\hline
Hayward   & 0.158106 & 22.21 & 0.70061 & 11.12 \\
Dymnikova & 0.159    & 21.75 & 0.964    & 22.33 \\
Scheme B  & 0.13019  & 35.93 & 0.70745  & 10.24 \\
Scheme C  & 0.14249  & 29.91 & 0.67509  & 14.34 \\
\hline
\end{tabular} 
\caption{Complex frequencies in units of mass of scalar QNMs ($l=1$) for different regular black hole models, with parameters close to their critical values (indicated in the table). Results are compared to the Schwarzschild case: $\omega_{\text{Schw}} = 0.29293 - 0.09766i$ for $n=0$ and $\omega_{\text{Schw}} = 0.203259 - 0.788298i$ for $n=3$. The table reports real and imaginary parts along with the corresponding relative percentage deviations.}
\label{QNM}
\end{table}

\section{Summary}

In this chapter, we have investigated the formation, geometry, and perturbative dynamics of regular black holes arising within the Asymptotic Safety framework.

\noindent We first analyzed several famous static regular black hole solutions obtained in Asymptotic Safety. All the considered solutions recover the Schwarzschild geometry at large distances, while the deviations from the classical metric are concentrated in the near-horizon region. This behavior can provide observable signatures in the black hole perturbation spectrum, in fact, comparing scalar quasinormal modes, we found that while the fundamental frequencies generally remain relatively close to their Schwarzschild counterparts, the deviations become considerably more pronounced for higher overtones, particularly when the solutions approach their critical parameter values, confirming the well-knokn "outburst of overtones". 

We then construct a dynamical collapse black hole solution using the proper-time flow within asymptotic safety.

\noindent We found that the metric obtained is mostly independent from non-physical choices, such as gauge fixing or parametrization ambiguities which appear from asymptotic safety construction, depending only on the regularization scheme adopted in the renormalization group improvement. We showed how the running of the gravitational coupling modifies the classical collapse at high densities. In contrast to the Oppenheimer--Snyder--Datt scenario, where the apparent horizon and the scale factor shrink monotonically and a spacelike singularity forms, the quantum-corrected collapse slows down when quantum-gravity effects become relevant. A trapped region and an apparent horizon still develop during the collapse, but the latter reaches a minimum and subsequently grows, while the scale factor decays exponentially and the spacetime remains regular throughout the evolution.

We subsequently investigated axial gravitational perturbations of the collapsing regular black-hole solutions. The resulting effective potential remains positive in the parameter range considered, and the time-domain evolution confirms the stability of the backgrounds in this sector. The quasinormal frequencies were computed using complementary frequency and time-domain techniques, namely the WKB method with Pad\'e approximants, the Leaver method, and Prony's method. The agreement between these approaches provides a consistency check of the numerical results. We also found that the deviations from Schwarzschild increase with overtones, as the static solutions, and as the parameter approaches its critical value, where the departure of the near-horizon geometry from the classical solution is maximal.

Overall, the results presented in this chapter show that Asymptotic Safety can modify the near-horizon geometry while preserving the classical Schwarzschild behavior at large distances and avoiding the formation of spacelike singularities in static and dynamical collapse scenario considered here. The perturbation analysis further demonstrates that these modifications are encoded more clearly in higher quasinormal overtones, providing potentially useful signatures of quantum gravity corrections to black hole spacetimes.

\chapter{Phenomenology and observational implications}
\section{Shadow}

The direct imaging of the supermassive compact objects at the center of M87 and Sagittarius A* by the Event Horizon Telescope (EHT) collaboration~\cite{EventHorizonTelescope:2019dse,EventHorizonTelescope:2022wkp} has opened a new observational window into the strong-field regime of gravity. In particular, the observation of black hole shadows provides a powerful probe of spacetime geometry near compact objects and therefore represents an important tool for testing General Relativity and alternative theories of gravity.

The shadow of a compact object is the dark region appearing in the observer sky due to the capture of photons by the gravitational field of the object. Physically, it corresponds to the set of photon trajectories that do not reach the observer because they either cross the event horizon or terminate at a singularity. The boundary of the shadow is determined by critical null geodesics orbiting the compact object on unstable circular trajectories, known as photon spheres or photon rings. Photons approaching these critical orbits can perform multiple revolutions around the compact object before either escaping to infinity or being captured, generating the bright ring structure surrounding the shadow.

The precise shape and size of the shadow depend sensitively on the underlying spacetime geometry. For this reason, shadow observations have become an important phenomenological tool to distinguish between black holes predicted by General Relativity and compact objects arising in modified theories of gravity, including regular black holes, wormholes, naked singularities, and quantum-corrected geometries, thus we decided to study them within the Asymptotic Safety framework and quadratic gravity.

A realistic determination of the shadow is typically obtained through ray-tracing simulations. In this approach, null geodesics are integrated numerically backward from the observer screen into the spacetime geometry. By determining whether each photon trajectory escapes to a distant light source or is instead captured by the compact object, one reconstructs the intensity map observed by a distant observer. Ray tracing therefore provides a direct connection between the spacetime metric and observable images, making it a fundamental tool in modern black hole phenomenology.

In order to determine the shadow of a spherically symmetric spacetime of the form~\eqref{line element}, one first needs to compute the radius of the photon sphere. This corresponds to the existence of unstable circular null geodesics, which play a central role in defining the shadow boundary.

For null geodesics, the radial motion can be written in terms of an effective potential as~\cite{alma9917223253502466,Chandrasekhar:1985kt}
\begin{equation}
\dot{r}^2 + V_{\text{eff}}(r) = E^2 , \qquad
V_{\text{eff}}(r)= f(r)\left(\frac{L^2}{r^2}\right),
\end{equation}
where $E$ and $L$ are the conserved energy and angular momentum of the photon, respectively. Circular null orbits are obtained by imposing
$$ \dot r =0, \qquad \ddot r =0, $$
together with the extremality condition
\begin{equation}
\frac{d}{dr}V_{\text{eff}}(r)=0.
\end{equation}
This condition is equivalent to~\cite{Cardoso:2008bp,Konoplya:2019sns}
\begin{equation} \label{circular}
\frac{d}{dr}\left(\frac{r^2}{f(r)}\right)=0
\iff
\left.r f'(r)-2f(r)\right|_{r=r_p}=0.
\end{equation}

Solving Eq.~\eqref{circular} for the specific metric function $f(r)$ associated with a given compact object determines the photon-ring radius $r_p$. Once the photon sphere is known, the apparent shadow radius for a distant observer is obtained from the critical impact parameter
\begin{equation} \label{Rshadow}
R_{\text{Shad}}=\frac{r_p}{\sqrt{f(r_p)}},
\end{equation}
which determines the angular size of the shadow observed at infinity.

\begin{figure}[h]
    \centering
    \includegraphics[width=\columnwidth]{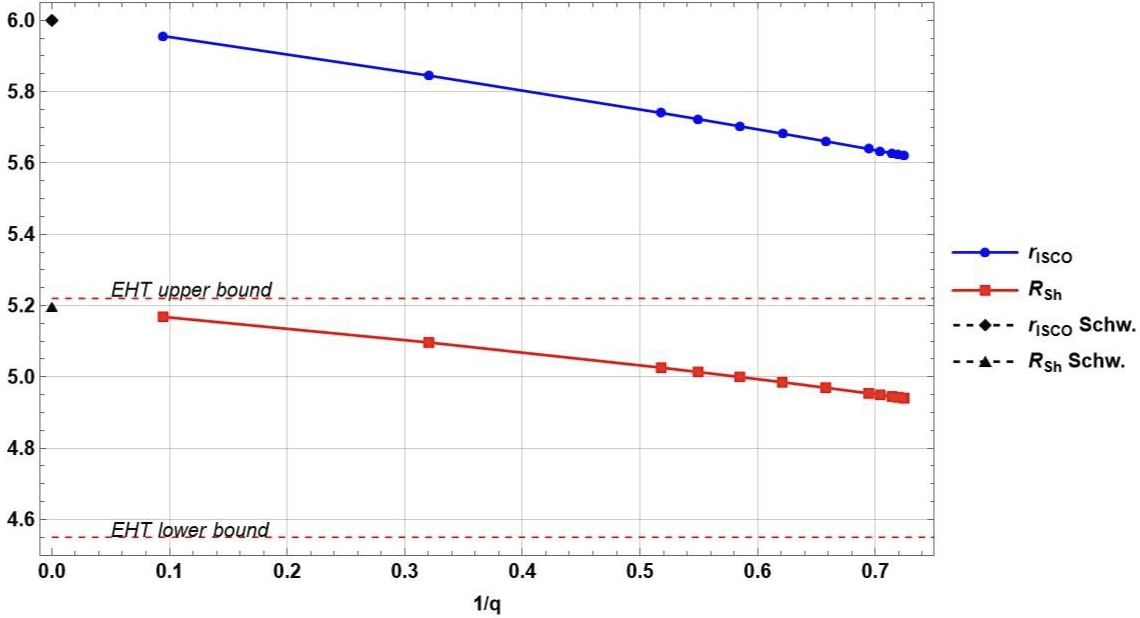}
    \caption{Radius of the shadow $R_{\text{sh}}$ and of the $\text{ISCO}$ as the variation of the parameter $q$. Schwarzschild reference values and also the EHT range admitted values for the shadow have also been entered for comparison.}
\label{shadow}
\end{figure}

First, We plot in \cite{Bonanno:2025dry} the variation of the shadows for different values of the parameter $q$ in the metric \eqref{metric}, shown in figure \ref{shadow}, together with Schwarzschild value, the EHT bounds from experimental observations \cite{EventHorizonTelescope:2021dqv,Vagnozzi:2022moj} and the radius of the ISCO.
We can observe that across the whole range of the parameter considered, their shadows shrink but remain compatible with the bounds set by the EHT.

Still in the context of asymptotic safety we collect the results for the other regular black holes in the review \cite{Spina:2025wxb}. In fact, the dependence of the shadow size on the parameters of different regular black hole metrics has been investigated in various works: for the Bonanno--Reuter solution in \cite{Lambiase:2023hng}, for the {non-singular collapse model (scheme C)} in \cite{Stashko:2024wuq}, and for Hayward-type black holes in \cite{Held:2019xde}. In Table \ref{Shadowrev} we report the values of the shadow radius computed close to the critical value of the respective parameters, in order to highlight the maximum deviation from the classical Schwarzschild case.
\begin{table}[H]
    \centering
    \begin{tabular}{l c}
        \toprule
        \textbf{Metric} & \textbf{$R_{\mathrm{Shad}}$} \\
        \midrule
        Schwarzschild & $5.196$ \\
        Bonanno--Reuter ($M = 3.505$) & $4.83312$ \\
        Hayward ($\gamma = 31/27$) & $4.91693$ \\
        Dymnikova ($l = 1.137$) & $5.19599$ \\
        Scheme B ($q = 1.38$) & $4.94024$ \\
        Scheme C ($\xi = 0.455$) & $4.86582$ \\
        \bottomrule
    \end{tabular}
    \caption{Shadow radius as seen from infinity for different regular black hole solutions evaluated close to their critical parameter values (explicitly indicated), compared with the classical Schwarzschild value. For reference, the Event Horizon Telescope (EHT) constrains the shadow size within the range $R_{\mathrm{Shad}} \sim 4.8M$--$5.2M$~\cite{Vagnozzi:2022moj,EventHorizonTelescope:2021dqv}.}
    \label{Shadowrev}
\end{table}
Among the models considered, the Bonanno--Reuter solution exhibits the most pronounced reduction of the shadow size, while the Dymnikova metric remains nearly indistinguishable from Schwarzschild (as in the fundamental mode of QNM). The other models (Hayward, {non-singular collapse solutions}) display intermediate deviations, with shadow radii (for all the models considered) consistently lying within the range constrained by the EHT observations \cite{Vagnozzi:2022moj,EventHorizonTelescope:2021dqv}. We can also point out that in all the considered cases the quantum corrections tend to reduce the size of the shadow compared to the classical Schwarzschild solution. However, given the current precision of the EHT measurements, we cannot exclude any of these AS-inspired models nor place meaningful constraints on their parameters based on the shadow size alone.

Discussing on the investigation of the shadow in quadratic gravity, general analysis and comparison with EHT observation were analyzed in \cite{Daas:2022iid,Aktar:2024akk}. 

To investigate the shadow properties of these solutions, we consider the weak-field expansion~\eqref{eq|weakfield}. Within this approximation, the radius of the photon ring obtained from Eq.~\eqref{circular} is given by~\cite{Bonanno:2025ksz,Daas:2022iid}
\begin{equation}\label{eq|lightringwf} 
\left.r-3M+S_2^-\left(3+m_2r\right)\mathrm{e}^{-m_2r}\right|_{r=r_{pr}}=0, \end{equation}
which reduces to the Schwarzschild result $r_{\rm pr}=3M$ in the limit $S_2^-=0$. The correction induced by the Yukawa charge shifts the photon-ring radius inward $r_{\rm pr}<3M$ for $S_2^->0$ and outward $r_{\rm pr}>3M$ for $S_2^-<0$. The weak-field approximation is expected to remain reliable in this context because the photon ring is located at $r\sim 3M$, where the metric already approaches its asymptotic Schwarzschild-like behavior. Moreover, Eq.~\eqref{eq|lightringwf} shows that the deviations from the Schwarzschild value are exponentially suppressed by the Yukawa term. For astrophysical black holes, these corrections are therefore parametrically small, implying that the corresponding modifications in the shadow size are expected to remain far below the current observational precision. This behavior is consistent with the results of~\cite{Daas:2022iid}, where it was shown that Schwarzschild black holes, a class of non-Schwarzschild black holes, and attractive (veiled) naked singularities can be observationally indistinguishable at the level of both the shadow size and the intensity profile. Noticeable deviations arise only in the intensity profile for specific classes of geometries, such as wormholes and some non-Schwarzschild solutions, where light can penetrate the nominal shadow region due to null geodesics with turning points below the critical impact parameter. In contrast, repulsive naked singularities do not possess a photon ring at all~\cite{Silveravalle:2023lnl}.

More generally, naked singularity spacetimes often exhibit a qualitatively different structure from black holes. In fact, black hole geometries typically possess a single unstable photon ring together with an event horizon, so that light rays crossing the photon ring inevitably fall into the horizon, generating a shadow. Horizonless compact objects, instead, frequently develop two photon rings: an outer unstable one and an inner stable one. In these cases, photons crossing the outer ring may still escape after being reflected by the inner angular-momentum barrier, leading to characteristic concentric ring structures rather than a standard black-hole shadow.

Focusing on Veiled singularities in quadratic gravity, they behave differently from both of these scenarios~\cite{Bonanno:2025ksz}. As shown in the lower-left panel of Fig.~\ref{fig|metric}, Eq.~\eqref{circular} is satisfied both at a radius outside the veil and again at the origin. In practice, the second photon ring becomes degenerate with the singularity itself. Consequently, light rays crossing the outer photon ring are not reflected back outward, but are instead inevitably driven toward the singularity. 
\begin{figure}[htb]
    \centering
\includegraphics[width=\textwidth]{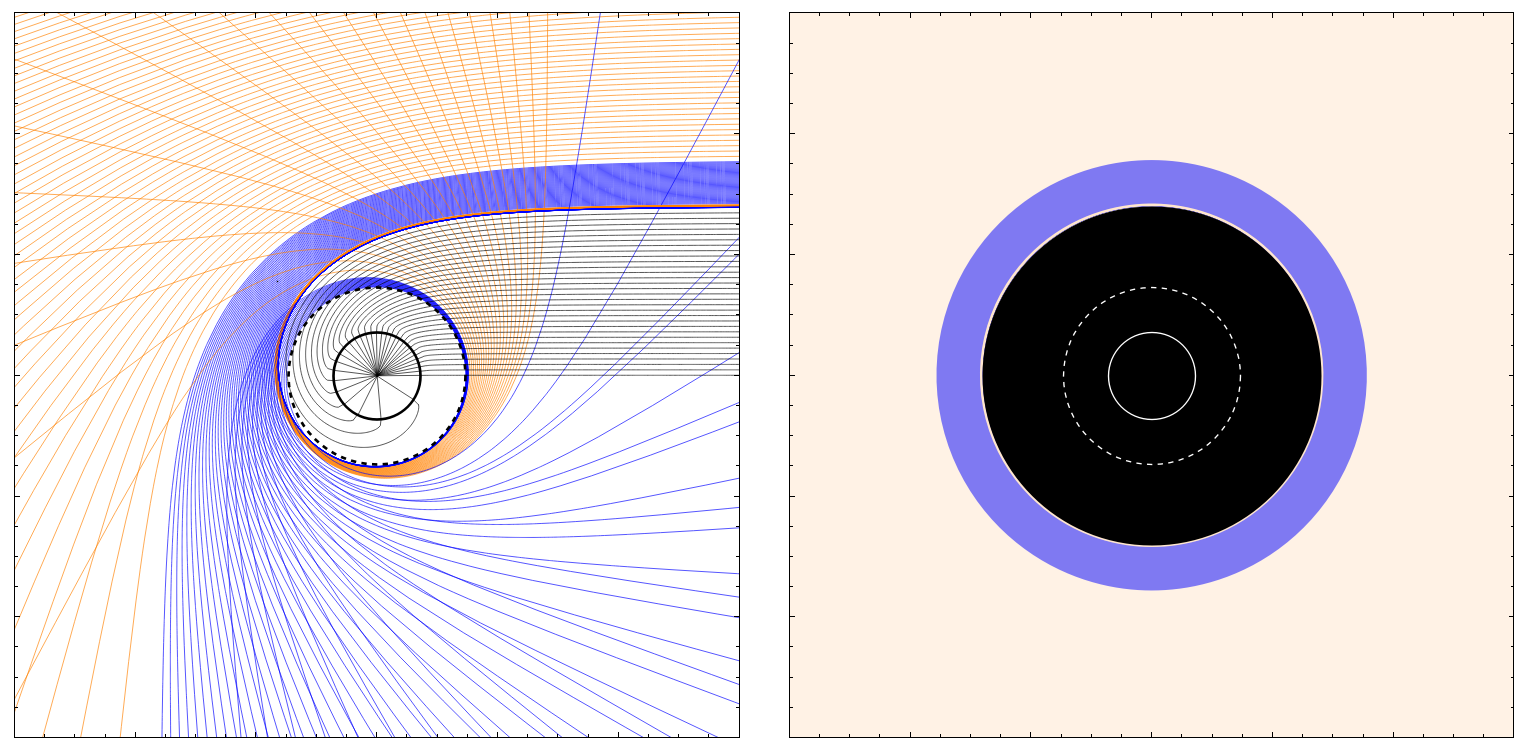}
    \caption{Null geodesics around a naked singularity in arbitrary units (for reference, we used $M=0.6$ and $S_2^-=0.07$). Left panel illustrates a ray-tracing simulation, with an observer at $x=-\infty$ and an illuminated screen assumed at $x=+\infty$. Orange lines are light rays that reach the observer, blue lines are light rays deflected back by strong lensing effects. and black lines are light rays that reach the singularity. The dashed and solid circles indicate the position of the external photon ring and the veil radius. Right panel illustrate the face-on view of the singularity from the observer position.}
    \label{fig|raytracing}
\end{figure}
This behavior has important observational consequences. Since photons entering the outer photon ring cannot escape back to infinity, veiled singularities produce a genuine shadow region, similarly to ordinary black holes, as obtained in \cite{Daas:2022iid}. A representative ray-tracing simulation is shown in Fig.~\ref{fig|raytracing}. In the left panel, the observer is placed at $x=-\infty$ while an illuminated screen is located at $x=+\infty$. Light rays passing sufficiently far from the photon ring are only deflected and can still reach the observer. As the impact parameter decreases, the deflection angle grows, and eventually photons are bent back toward the illuminated screen before reaching the observer. Trajectories passing extremely close to the photon ring undergo multiple orbits around the compact object, producing the observable bright light ring. Finally, photons crossing inside the photon ring are captured by the singularity and never re-emerge. Approaching the veil, the trajectories become effectively radial and terminate at the singularity. 
The corresponding face-on image, shown in the right panel of Fig.~\ref{fig|raytracing}, reveals that the veiled singularity produces a shadow closely resembling that of a Schwarzschild black hole, differing essentially only in its exact size, as discussed previously.

\subsection{Null geodesics and eikonal QNM correspondence}

Geodesic motion determines important features of spacetimes, among the different kinds of geodesic motion, circular geodesics are especially interesting, as we already analyze the last stable circular orbit and we focus on the unstable null geodesics which gives information on the shadow. Null geodesics can be also associated to QNM, in fact in ~\cite{Cardoso:2008bp}, a correspondence was proposed between the real and imaginary parts of quasinormal frequencies in the eikonal regime and the properties of unstable circular null geodesics. We remind that the behavior of quasinormal modes in the eikonal regime (i.e., for large multipole number \( \ell \))~\cite{Bolokhov:2025uxz,Konoplya:2023moy}, in classical general relativity, the centrifugal-like term dominates the effective potential. In this limit, the WKB method usually becomes applicable for computing quasinormal frequencies. The correspondence proposed specifically, that for any stationary, spherically symmetric, and asymptotically flat black hole spacetime, the quasinormal frequencies were conjectured to follow:
\begin{equation} \label{eikonal}
    \omega = \Omega_c\left(l+\tfrac{1}{2}\right) - i\left(n + \tfrac{1}{2}\right)|\lambda|+O(l)^{-1},
\end{equation}
where \( \Omega_c \) is the angular velocity and \( \lambda \) the Lyapunov exponent of the corresponding circular null geodesic. $\Omega_c$ is given by the expression
$$\Omega_c=\left(\frac{f'_c}{2r_c}\right)^{1/2}=\frac{f_c^{1/2}}{r_c}$$
for null geodesic, and the last equality obtained from \eqref{circular} where we called $r_c$ the solution of the equation and $f_c$ the metric function at that radius. The Lyapunov exponent, which is the inverse of the instability timescale associated with this geodesic motion, are a measure of the average rate at which nearby trajectories converge or diverge in the phase space, and for circular null geodesics is, using \eqref{circular}:
$$ 
\lambda = \frac{1}{\sqrt{2}} \sqrt{ \frac{r_c^2 f_c}{L^2} \, V''_r(r_c) }
= \frac{1}{\sqrt{2}} \sqrt{ - \frac{r_c^2}{f_c} 
\left. \frac{d^2}{dr_*^2} \left( \frac{f}{r^2} \right) \right|_{r = r_c} },$$
where
\begin{equation}
V''_r(r_c) = \frac{L^2}{r_c^4} 
\left[ 2f(r_c) - r_c^2 \left.\frac{d^2 f}{dr^2}\right|_{r=r_c} \right].
\end{equation}
However, it was later shown in~\cite{Konoplya:2017wot,Konoplya:2022gjp} that this correspondence holds reliably only for test fields in stationary, spherically symmetric black hole backgrounds. For gravitational or other non-minimally coupled perturbations — especially in theories with higher curvature corrections such as various Einstein–Gauss–Bonnet models~\cite{Konoplya:2019hml,Konoplya:2020bxa} (and we will show also in quadratic gravity \cite{Konoplya:2025afm}) — the correspondence may break down. This breakdown occurs when the effective potential is no longer a single positive-definite peak that monotonically decays toward both the event horizon and spatial infinity (or a de Sitter horizon).

For this reason we checked the correspondence for the context previously analyzed. However, in the asymptotic safety solution \eqref{metric} of our case is trivial to verify that the potential is positive definite that decays at boundaries, so we expect that the correspondence is still valid.
We test whether the null geodesic correspondence holds for \eqref{metric}, so we calculate the fundamental mode for $l=100$ with WKB in the Eikonal regime which we know is exact when $l\xrightarrow{}\infty$ with the formula
\begin{equation} \label{frequency}
    \omega M=\sqrt{V_0-i(n+1/2) \sqrt{-2V_2}}=20.341 - 0.0824361i
\end{equation}
where $V_0$ is our potential in tortoise coordinate calculated in his peak, and $V_2$ is the second derivative of the potential in the same point.
While, applying the geodesic-based formula~\eqref{eikonal} yields:
\begin{equation} 
\omega M =20.3431-0.082439i .
\end{equation} 
We can observe that the two values differ only by $0.0103\%$ for the real part and $0.00352\%$ for the imaginary part, so we can conclude that in this specific case the correspondence in Eikonal regime is valid.

The situation is different when we consider black holes solutions in quadratic gravity, which is a counterexample, in fact the effective potential for gravitational perturbations develops a deep negative gap near the event horizon, which can even diverge. In such cases, quasinormal modes clearly cannot be described by the WKB formula. Indeed, examining the expressions for the potential in eqs.~\eqref{grav_pot_Schw} for the Schwarzschild solution and eqs.~\eqref{grav_pot_NonSchw} for the non-Schwarzschild black hole, we see that in the limit \( l \to \infty \), both cases contain off-diagonal terms such as
\begin{equation} 
V_{23} = - l^2 A(r) \frac{2A - rA'}{r^2}+O(l),
\end{equation}
which introduce a negative gap structure inconsistent with the classical potential barrier picture.

To test whether the null geodesic correspondence holds in quadratic gravity, we computed quasinormal frequencies using Prony’s method (as WKB cannot be applied due to the coupled equations and the nature of the potential). For the Schwarzschild background with \( \ell = 100 \), we obtained the fundamental mode:
\begin{equation} \label{frequencyQG}
    \omega r_h = 80.62 - 0.33i.
\end{equation}
By contrast, applying the geodesic-based formula~\eqref{eikonal} yields:
\begin{equation} 
\omega r_h = \dfrac{201-i}{3\sqrt{3}}\approx 38.68 - 0.19i.
\end{equation} 
The discrepancy between these values confirms that the correspondence fails in this regime due to the different structure of the effective potential. The geodesic approach misses essential features of the spectrum, especially in the presence of higher curvature corrections.

It is important to note, however, that for test fields in quadratic gravity, the correspondence still holds, as shown in~\cite{Zinhailo:2018ska}.

\section{Hawking radiation}

The discovery of Hawking radiation in 1974 \cite{Hawking:1974rv,Hawking:1975vcx} established that black holes are not completely black, but instead emit a thermal spectrum of particles due to quantum field effects in curved spacetime. The original derivation was semiclassical in nature: gravity was treated as a fixed classical background, while quantum fields were quantized on top of it. Within this approximation, the radiation is almost perfectly thermal. When quantum gravitational corrections are included, however, deviations from the classical picture are expected, potentially altering the emission spectrum and the evaporation process, particularly in the extreme regime close to the endpoint of evaporation. These questions remain open and are an active line of research in quantum gravity.  

From a practical perspective, it is often convenient to retain a semiclassical treatment and study the evaporation of black holes whose geometry already incorporates quantum corrections. Furthermore, in the Schwarzschild case, it has been shown that the contribution of gravitons to the total massless particle emission is below the percent level, and becomes even less relevant when massive channels open at the final stages of evaporation \cite{Page:1976df,Page:1976ki}. This observation supports the use of the standard Hawking framework to estimate the flux of quantized test fields in a given corrected geometry, under the assumption that the graviton contribution is subdominant.

In this approach, the black hole is considered to be in thermal equilibrium with its environment, so that its temperature remains constant between the emission of two successive particles. The system can therefore be described in the canonical ensemble, and the Hawking temperature takes the form \cite{Hawking:1975vcx}
\begin{equation}
    T_H = \frac{f'(r)}{4\pi}\Big|_{r=r_H},
\end{equation}
where $f(r)$ denotes the lapse function of the metric and $r_H$ is the event horizon radius. The Hawking emission rate, for the number of particles of species $i$ with spin $s$ emitted per unit time and per unit energy due to Hawking radiation, is given by
\begin{equation} \label{haw_eq}
\frac{d^2 N_i}{dt\, dE_i} = \frac{1}{2\pi} \sum_{l,m} \frac{n_i\, \Gamma^{s}_{l,m}(\omega)}{e^{\omega / T_H} \pm 1},
\end{equation}
where $n_i$ is the number of internal degrees of freedom of the particle, $\omega = E_i$ is the mode frequency (in natural units), the plus (minus) sign in the denominator corresponds to fermions (bosons), and $\Gamma^{s}_{l,m}(\omega)$ are the grey-body factors (GBFs) of black holes. The latter alter the spectrum of Hawking radiation detected by a distant observer at spatial infinity. These factors quantify the fraction of the initially emitted quantum radiation that is reflected back toward the event horizon by the potential barrier. Grey-body factors can be characterized by setting up a classical scattering problem around the BH potential barrier, with boundary conditions allowing for incoming wave packets from infinity or equivalently, due to the symmetries of the scattering problem, originating from the horizon, so that the scattering boundary conditions can be expressed as
\begin{equation}
\Psi \sim e^{-i \omega r_*} + R e^{i \omega r_*}, \quad r \to r_h,
\end{equation}
\begin{equation}
\Psi \sim T e^{-i \omega r_*}, \quad r \to \infty,
\end{equation}
where $\omega$ is a real frequency, and $R$ and $T$ denote the reflection and transmission coefficients, respectively. These coefficients satisfy the conservation relation
\begin{equation}
|T|^2 + |R|^2 = 1.
\end{equation}
The grey-body factor for the multipole $l$, denoted by $\Gamma_l^s$, is defined as the transmission probability,
\begin{equation}
\Gamma_l^s \equiv |T|^2 = 1 - |R|^2.
\end{equation}
Each multipole transmission coefficient can be computed using the WKB approximation (it is one of the approaches to compute them, see for example \cite{Calza:2024fzo,Arbey:2025dnc,Calza:2025whq} for other methods and codes, which exploit direct integration, shooting methods, etc, to solve the Teukolsky equation \cite{Teukolsky:1973ha}). To compute them with WKB, we have to consider the phase factor $K$ in \eqref{Wkb formula}, which is given by \cite{Konoplya:2003ii}

\begin{equation}
K = \frac{i\left(\omega^2 - V_{\text{eff}}(r_m)\right)}{\sqrt{-2 \frac{d^2}{d r_*^2} V_{\text{eff}}(r_m)}} + \sum_{p=2}^6 \Lambda_p(\omega,V_n).
\label{Kwkb}
\end{equation}
Here, $r_m$ corresponds to the location of the peak of the effective potential, $\Lambda_p(\omega,V_n)$ are higher-order WKB correction terms which depend on the frequency $ \omega$ and on the potential and its derivatives, evaluated at the peak $r_m$. At the eikonal level, Eq.~\eqref{Wkb formula} provides the unique solution
\begin{equation}
K_{\rm eik}=
\frac{i\left(\omega^2-V_0\right)}
{\sqrt{-2V_2}}.
\end{equation}
When higher-order corrections are included, the resulting equation \eqref{Kwkb} has multiple solutions for $K$ at a fixed frequency $\omega$. The solution corresponding to the perturbative correction of the eikonal result is selected as the root closest to $K_{\rm eik}$ \cite{Konoplya:2019hlu}.

In the case of a real effective potential, $K$ is a purely imaginary quantity \cite{Iyer:1986np} and is related to the reflection and transmission coefficients as
\begin{equation}
|R|^2 = \frac{1}{1 + e^{-2 i \pi K}}, \quad |T|^2 = \Gamma_{l s}= \frac{1}{1 + e^{2 i \pi K}},
\end{equation}
where $K$ is given by \eqref{Kwkb}. Therefore, the greybody factor is obtained as the transmission probability through the effective potential barrier.

\begin{figure}[h]
    \centering
    \includegraphics[width=\columnwidth]{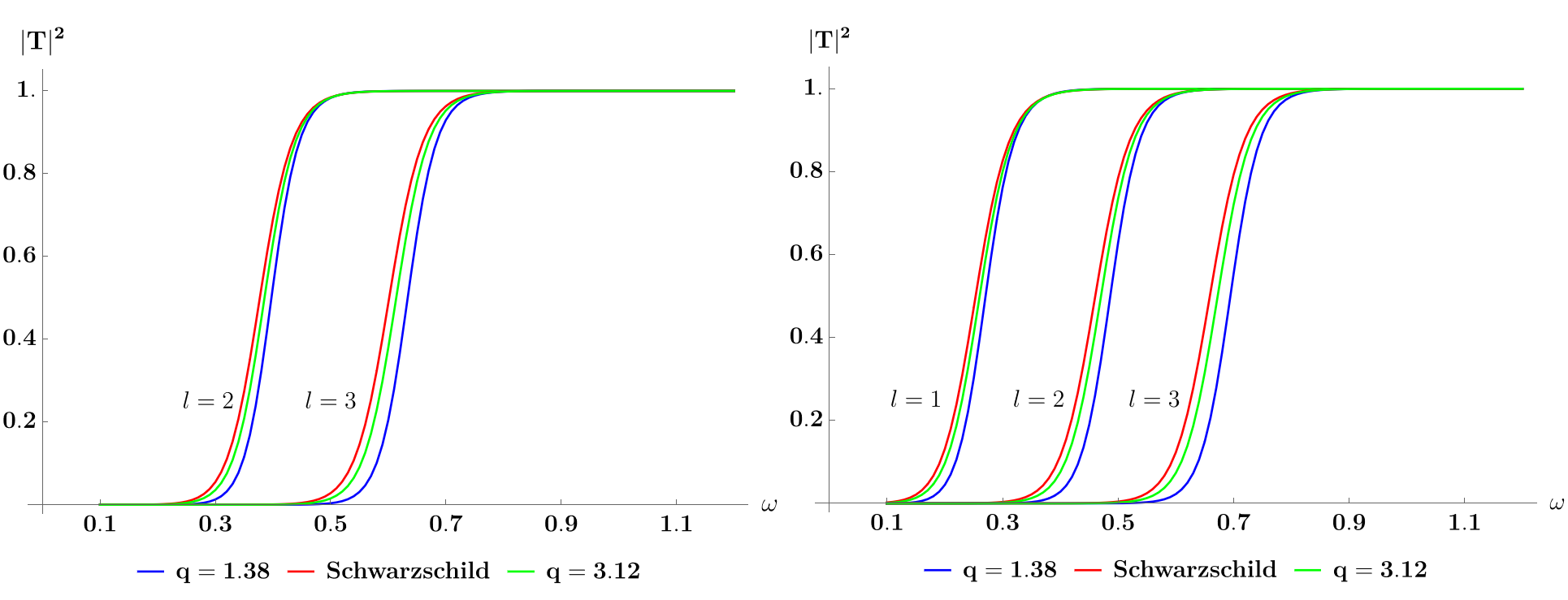}
    \caption{Grey-body factors for different values of $q$ and $l$, compared with the Schwarzschild case. The left panel shows the gravitational field, while the right panel shows the electromagnetic field.}
\label{fig9}
\end{figure}

At this point, we apply this formalism to our solution \eqref{metric} in asymptotic safety \cite{Bonanno:2025dry}. Specifically, we calculated the grey-body factors for some values of $q$ of gravitational perturbations for $l=2,3$ and electromagnetic perturbations for $l=1,2,3$ (which will be used for the radiation), whose potential is given by
\begin{equation}
    V(r)=f(r)\left(\frac{l(l+1)}{r^2}\right),
\end{equation}
and compared them with Schwarzschild, as shown in figure \ref{fig9}. 
We note that the factors decrease with the approach to the critical value of $q$, moving further away from the classical case, so that the quantum corrections induce the potential barrier to increase the absorption. It is worth noting how this behavior is similar to the case with the different choice of regularization scheme in eq.\eqref{beta} as obtained in \cite{Stashko:2024wuq,Lutfuoglu:2025ohb,Shi:2025gst}, and it is also similar to other regular black holes GBFs behavior like those shown in \cite{Konoplya:2023bpf}. It is worth noting that the grey-body factors are considerably more stable under static near-horizon deformations induced by $q$-corrections than the quasinormal overtones, despite the well-known correspondence between them \cite{Konoplya:2024lir}, in agreement with~\cite{Konoplya:2025ixm,Oshita:2024fzf,Rosato:2024arw}. This stability arises because such deformations primarily affect the low-frequency regime~\cite{Konoplya:2025ixm}, where the grey-body factors are already close to zero. Therefore, even a relatively large shift—expressed as a percentage—results in a grey-body factor that remains close to zero.

Before moving on the calculation of the emission rate, we analyze in \cite{Bonanno:2025dry} how the Hawking temperature varies with the quantum parameters of the two collapsing solutions, because is the primary contribution to the differences in the emission.
\begin{figure}[h]
    \centering
    \includegraphics[width=0.8\columnwidth]{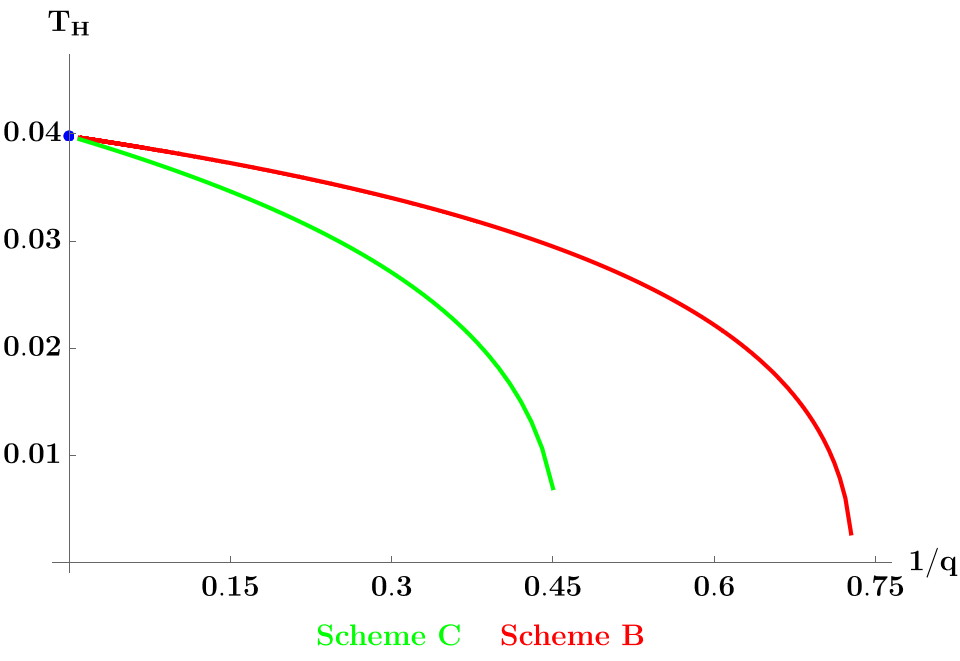}
    \caption{Plot of the Hawking temperature variation in function of the inverse of the parameter of the metric function for the two different regular black holes. In red the solution introduced here (Scheme B), in green the solution presented in \cite{Bonanno:2023rzk} with regularization scheme C. Also the Schwarzschild value is present in blue for comparison.}
    \label{Hawtemp}
\end{figure}
We see from figure \ref{Hawtemp} that deviating more from classical case by turning on the parameter, the Hawking temperature decreases and tends to vanish when approaching the critical value of the parameter for both cases.
Going further in the analysis in asymptotic safety, we evaluated in \cite{Spina:2025wxb} several solutions near their critical configuration in comparison with Schwarzschild case, where in mass units the temperature reads $T_H^{\text{Schw}} = 1/(8\pi)$. The resulting values for the different regular black hole solutions considered here are summarized in Table \ref{HawkingT}, with the near-critical values of the parameters used in the evaluation are explicitly specified in the table. These results were obtained and compared in \cite{Konoplya:2023bpf} for Bonanno-Reuter ({see also \cite{Bonanno:2000ep,Platania:2023srt} for earlier analyses including the evaporating mass and the interpretation} of remnants), Hayward and Dymnikova spacetimes and in \cite{Bonanno:2025dry} for the non-singular collapse models.
\begin{table}[H]
\centering
\begin{tabular}{l c}
\hline
Metric & $T_{\mathrm{H}}$ \\
\hline
Schwarzschild & 0.0398 \\
Bonanno--Reuter ($M = 3.505$) & 0.00071 \\
Hayward ($\gamma = 31/27$) & 0.00369 \\
Dymnikova ($l = 1.137$) & 0.00291 \\
Scheme B ($q = 1.38$) & 0.00472 \\
Scheme C ($\xi = 0.455$) & 0.00345 \\
\hline
\end{tabular}
\caption{Hawking temperature $T_{\mathrm{H}}$ in mass units for different regular black hole solutions evaluated near their critical values. Results are compared with the Schwarzschild case. As the critical configuration is approached, the solutions tend toward extremality and the temperature vanishes.}
\label{HawkingT}
\end{table}
As can be seen {in Table \ref{HawkingT}}, all the regular black hole solutions display a Hawking temperature that is significantly suppressed with respect to the classical Schwarzschild case. In particular, {near the critical parameter values, where the inner Cauchy horizon approaches the event horizon, the surface gravity approaches zero and the solutions tend to extremal configurations, for which the Hawking temperature vanishes}. This behavior implies that the corresponding evaporation process is drastically slowed down, leading to emission rates that are orders of magnitude smaller than in the classical scenario. Moreover, despite quantitative differences among the models (which obviously depends on the exact value of the parameter considered close to the extremal configuration), the qualitative picture remains robust: quantum corrections systematically lower the Hawking temperature and drive the solutions toward cold, long-lived remnants.

\subsection{Primordial black holes}

 From a phenomenological perspective, the suppression of the temperature and GBF may have important implications for the lifetime of primordial black holes and the possibility of long-lived remnants black holes. For this reason, we investigated \cite{Bonanno:2025dry} the primary photon spectrum from Hawking evaporation for black holes with mass $M_{\text{PBH}} = 10^{16}\,\text{g}$, which is comparable to that of possible evaporating primordial black holes. Moreover, in equation~\ref{haw_eq}, we considered multipoles up to $l=3$, although we observed that the dominant contribution comes almost entirely from the lowest modes.

\begin{figure}[H]
    \centering
    \includegraphics[width=0.7\textwidth]{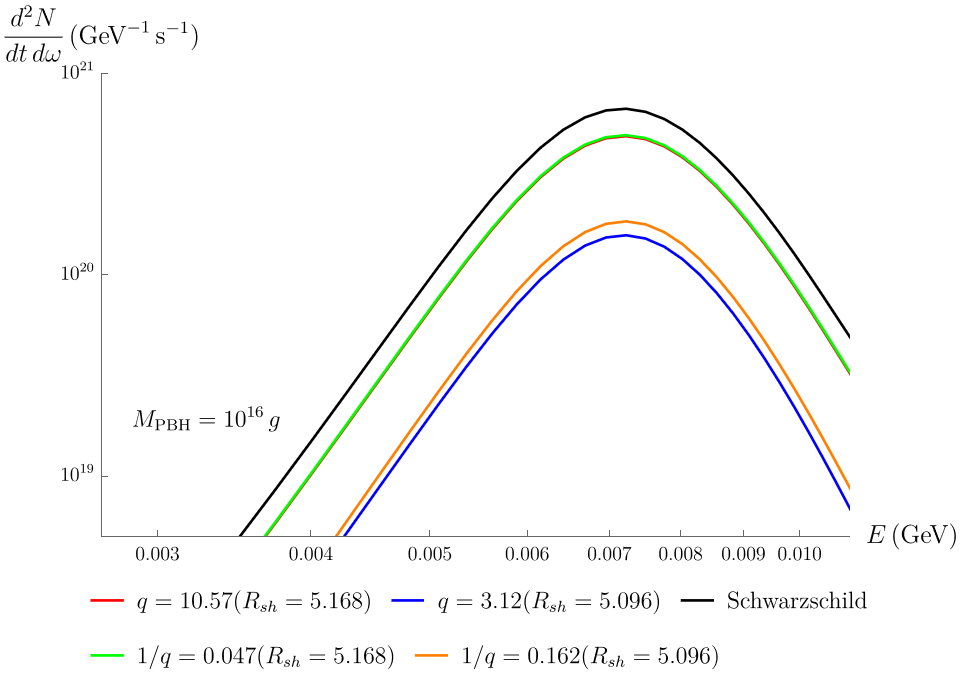}
    \includegraphics[width=0.7\textwidth]{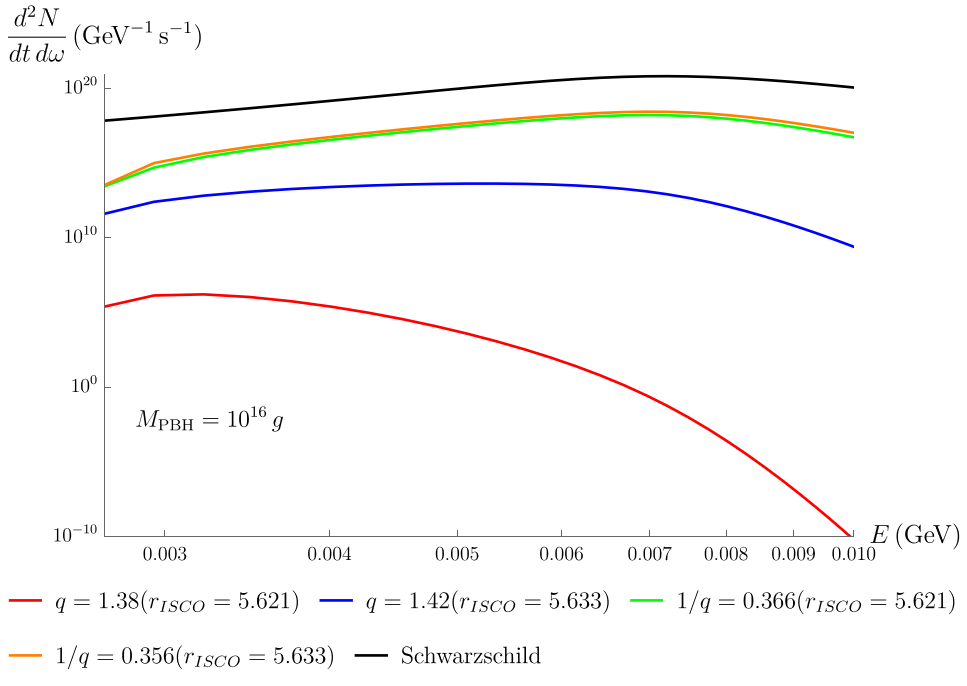}
    \caption{Evaporation spectra of the Hawking radiation emission rate for primary photon emission from a primordial black hole with $M_{\text{PBH}} = 10^{16}\text{g}$. Several values of the parameter $q$ are shown and compared with the Schwarzschild case and with the regular black hole solution of scheme C (which depends on the inverse of the parameter $1/q$).\\  
Upper panel: the solutions are compared at fixed shadow radius $R_{\text{Sh}}$ \eqref{Rshadow}.\\
Lower panel: the solutions are compared at fixed ISCO radius $r_{\text{ISCO}}$.
}
    \label{fig10}
\end{figure}

 We computed the emission rate for our regular black hole model and compared it both with the Schwarzschild case and, following the approach in~\cite{Calza:2024fzo,Calza:2025yfm,Calza:2025mwn,Calza:2024xdh} with other regular black hole solutions, the one derived using the alternative regularization scheme C given by equation~\ref{G_schemeC}, which, as previously mentioned, was introduced and studied in ~\cite{Bonanno:2023rzk,Stashko:2024wuq,Lutfuoglu:2025ohb}, but had not yet been analyzed from this specific perspective. Since the aim is to distinguish between different models in potential future observations of primordial black holes, we performed the comparison in a way that ensures their observational features are matched. Specifically, first we fix the mass as said before $M_{\text{PBH}} = 10^{16}\,\text{g}$ and we chose the quantum parameter $1/q$ of the alternative solution so that in some cases it yields the same shadow size as our model (left panel of Figure~\ref{fig10}), and in others the same ISCO radius (right panel of Figure~\ref{fig10}). In this way, we base the comparison on physically observable quantities and the resulting spectra are shown in Figure~\ref{fig10}.
The first immediate observation is that, as the quantum deviation increases (i.e., decreasing $q$), the emission rate decreases and deviates progressively more from the classical case. Furthermore, we observe that, for fixed values of the ISCO or the shadow, the solution presented in this work departs more significantly from the classical Schwarzschild case and exhibits a lower emission rate than its counterpart of regularization scheme C. This discrepancy becomes increasingly pronounced as the quantum parameters approach their critical values, where differences of several orders of magnitude emerge under equivalent conditions. Notably, we observe a dramatic drop in the emission rate—by more than 10 orders of magnitude—compared to the initial values. In these extreme cases, it is also worth noting that the peak of the emission spectrum, which initially lies around $\sim 7 - 8 \text{MeV}$ for the various configurations, shifts down to approximately $\sim 3 \text{MeV}$ for $q = 1.38$.

It is also worth noting that primordial black holes (PBHs) have been considered as viable dark matter candidates, since they are non-luminous, non-relativistic, and could have formed in the early Universe independently of baryonic physics. In the standard Schwarzschild case, PBHs with initial masses $M>10^{17}g$ are typically considered viable dark matter candidates, as they would not have emitted enough Hawking radiation to be excluded by existing observational constraints. This translates into an upper bound on the fraction $f_{\text{PBH}}$ of dark matter that PBHs can constitute, depending on their mass.
In our analysis, we have shown that regular black holes radiate less efficiently (figure \ref{fig10}) due to their lower temperature (figure \ref{Hawtemp}), which qualitatively suggests that the lower bound on their allowed mass could be relaxed. 

To quantify this effect, we also estimated the corresponding constraint on the PBH abundance with the extragalactic gamma-ray background (EGRB) \cite{Auffinger:2022khh}, following the procedure of \cite{Calza:2024fzo}. We considered a monochromatic PBH mass distribution with $M_{\rm PBH}=10^{16}\mathrm{g}$ and used the primary photon emission spectra shown in Figure~\ref{fig10}. The predicted flux was obtained by integrating the photon emission rate over the cosmological redshift distribution up to recombination and was compared with the observational upper limits used in \cite{Calza:2024fzo}. We restricted the comparison to the energy range covered by our numerical spectrum. For the Schwarzschild case, we obtain
\begin{equation}
f_{\rm PBH}^{\rm max}\simeq 4.9\times10^{-4}
\end{equation}
at $M_{\rm PBH}=10^{16}\mathrm{g}$, with the strongest constraint arising at $E\simeq 7\mathrm{MeV}$. This value is consistent with the order of magnitude of the corresponding constraints obtained in \cite{Calza:2024fzo}, providing a useful check of our procedure.

\noindent For the regular black hole solution  we considered $q=1.38$, close to the critical value, for which the predicted photon flux is instead strongly suppressed. The corresponding formal upper limit lies far above unity, $f_{\rm PBH}^{\rm max}\gg1$ and is therefore not physically restrictive. This means that, within the accuracy of our calculation, the EGRB observations do not constraint the abundance of PBHs with $M_{\rm PBH}=10^{16}\mathrm{g}$ from constituting the entirety of the dark matter in the near-critical regular case. These results provide a quantitative estimate of the effect anticipated from the comparison of the emission spectra in Figure~\ref{fig10}. In particular, while the Schwarzschild solution at $M_{\rm PBH}=10^{16}\mathrm{g}$ is constrained to contribute at most $\mathcal{O}(10^{-4})$ of the dark matter, the near-critical regular solution can, according to the constraint considered here, account for the entire dark matter abundance $(f_{\rm PBH}=1)$. 

Thus, the viable PBH dark-matter fraction is increased by more than three orders of magnitude relative to the Schwarzschild case. This behavior is consistent with the recent findings \cite{Calza:2024fzo,Calza:2024xdh,Calza:2025mwn}, where it was demonstrated that other regular PBHs can evade gamma-ray constraints more easily and thus expand the viable dark matter mass window.

\chapter{Conclusions and outlook}

The study of gravity in regimes of strong curvature represents one of the most promising avenues to probe possible extensions of General Relativity and to gain insight into the underlying structure of a quantum theory of gravity. In this work, we have investigated the perturbative dynamics and phenomenology of compact objects in quantum gravity-inspired frameworks, focusing in particular on quadratic gravity and the asymptotic safety scenario.\\

Within quadratic gravity, one of the most distinctive features with respect to General Relativity is the perturbative renormalizability of the theory, which however comes at the price of introducing an additional massive spin-2 degree of freedom associated with a wrong-sign kinetic term in the propagator. This extra mode is expected to play a crucial role in the dynamics of compact objects and their perturbative behavior.

To investigate these effects, we have analyzed tensor perturbations of static, spherically symmetric solutions, including both black holes and more exotic configurations such as veiled singularities, which naturally arise within this extended framework.

A key aspect of our analysis is the different structure of the perturbation equations depending on the multipole number. In the monopole sector ($l = 0$), the perturbations decouple, and the dynamics can be reduced to a single Schrödinger-like equation describing the massive mode. In contrast, for $l = 2$, the system becomes intrinsically coupled, leading to a more involved structure where the massless and massive modes interact non-trivially.

For black hole solutions, an important result emerging from this analysis is the revelation of a rich and non-trivial stability structure. In the $l = 0$ sector, we find that for certain ranges of the parameters, for small values of the horizon radius, the system exhibits a Gregory–Laflamme-type instability associated with the massive mode. This suggests that astrophysical black holes are expected to be effectively described by the Schwarzschild solution, while non-Schwarzschild configurations may be relevant only in regions of small masses in the parameter space.

Interestingly, the evolution of the system becomes more subtle when considering long-term processes such as black hole evaporation. In particular, near the crossing point between different branches of solutions, Schwarzschild-like configurations can become unstable, while the non-Schwarzschild solutions may enter stable regimes. However, these latter configurations exhibit unconventional behavior, such as a decrease of the mass that is not bounded from below and an increase of the horizon radius, raising questions about their physical interpretation and final fate.

These features suggest that the theory may require additional mechanisms to describe the transition between different solution branches and to determine the endpoint of the evolution, and may hint at a non-trivial phase structure of the solution space in quadratic gravity. In this context, exotic compact objects such as veiled singularities may play a relevant role. These configurations are characterized by the absence of a standard horizon structure while still satisfying a form of weak cosmic censorship, since the singularity is not directly observable. However, their perturbative analysis presents additional challenges, in particular due to the behavior of the effective potential, which may diverge and require a careful reformulation of the stability problem.

Finally, a key contribution we discovered is that the late-time behavior of perturbations exhibits asymptotic tails consistent with the universal behavior expected for massive fields, confirming that this feature persists beyond General Relativity. At the same time, our results indicate a breakdown of the correspondence between quasinormal modes and null geodesics, highlighting a further qualitative difference with respect to the classical theory.\\

Within the asymptotic safety framework, an important outcome is the possibility of constructing quantum-corrected black hole geometries that remain regular at the origin while preserving the main qualitative features of the Schwarzschild solution at large distances. This property makes them particularly appealing from a phenomenological perspective, as they provide viable alternatives to classical black holes without introducing large deviations in the weak-field regime.

In this context, a central result of this thesis is the construction and characterization of a novel regular dynamical black hole solution arising from gravitational collapse within the proper-time approach, consistently matching a collapsing interior in the asymptotic safety scenario to a static, spherically symmetric exterior geometry. A notable result is that the final metric turns out to be largely independent of non-physical choices, such as gauge fixing or parametrization ambiguities, depending only on the regularization scheme adopted in the renormalization group improvement.

We then investigated the perturbative properties of this solution, providing insight into the role of quantum corrections in the stability and dynamical response of compact objects. Our analysis shows that the configurations considered are stable under tensor perturbations and exhibit quasinormal mode spectra that deviate from the Schwarzschild case, with differences becoming more pronounced for higher overtones and depending on the value of the quantum parameter. Importantly, we find that varying the regularization scheme does not qualitatively affect the behavior of the spectrum, suggesting a certain degree of robustness of the results with respect to non-physical ambiguities in the construction.

In addition, we explored several phenomenological aspects of these quantum-corrected geometries. The shadow radius remains within current observational bounds, making these models consistent with existing data, although not yet distinguishable from General Relativity. Similarly, other observables show deviations that are currently too small to be conclusively detected. On the other hand, the analysis of Hawking radiation reveals more significant effects: the presence of quantum corrections leads to a lower temperature and a suppressed emission rate, which in turn modifies the expected constraints on primordial black holes and their possible role as dark matter candidates.

Overall, these results indicate that asymptotic safety provides a consistent and phenomenologically viable framework in which quantum effects can regularize black hole spacetimes while preserving compatibility with current observations, and at the same time offering potential signatures that may become accessible with future high-precision measurements.\\

The results obtained throughout this thesis indicate that perturbative and phenomenological analyses provide a common framework through which these theories can be confronted with observations. In this sense, compact objects act as laboratories where fundamental modifications of gravity can be translated into potentially measurable signatures.

The analyses presented in this work highlight how different approach\-es to quantum gravity, despite their conceptual differences, lead to common qualitative features in the description of compact objects. While preserving consistency with General Relativity at large scales, both quadratic gravity and asymptotic safety introduce modifications to the classical dynamics that become relevant in regimes of strong curvature, indicating that perturbative and phenomenological analyses provide a common framework through which these theories can be confronted with observations. In this sense, compact objects act as laboratories where fundamental modifications of gravity can be translated into potentially measurable signatures.

A unifying aspect emerging from our study is the central role of perturbations as a diagnostic tool. In both frameworks, the analysis of linear perturbations provides crucial information on the stability of the solutions and on their dynamical response, allowing us to identify physically viable configurations and to characterize deviations from classical predictions. In particular, quasinormal modes and late-time behavior encode signatures of the underlying theory, offering a direct connection between fundamental physics and observable quantities. From a phenomenological perspective, both approaches predict deviations from General Relativity that are, in most cases, subtle and challenging to detect with current observational capabilities. Nevertheless, specific features—such as modifications in the quasinormal mode spectrum, the behavior of Hawking radiation, and the properties of compact object shadows—may provide promising avenues for future tests. In particular, the interplay between theoretical consistency, stability, and observational signatures emerges as a key criterion to assess the viability of quantum gravity-inspired models.\\

The findings presented in this thesis naturally open several important directions and further investigation, which can build directly upon the framework developed here. A first natural extension concerns the inclusion of rotation. Astrophysical compact objects are generically rotating, and extending the analysis to axisymmetric spacetimes would provide a more realistic and comprehensive understanding of the impact of quantum corrections in both quadratic gravity and asymptotic safety. However, the increased technical complexity of rotating backgrounds makes this a highly non-trivial problem.

In quadratic gravity, further work is needed to clarify the transition between different branches of solutions and the endpoint of instabilities. Exotic compact objects, such as veiled singularities, may play a role in this context. However, their analysis is complicated by the behavior of the effective potential for perturbations, which in some cases exhibits pathological features such as divergences. Developing a consistent framework to properly define and interpret perturbations in these geometries represents an important challenge for future research.

A further natural extension concerns the inclusion of higher-derivative corrections beyond the quadratic level, which are expected to arise in a more complete effective description of quantum gravity. Understanding how these additional terms affect both the background solutions and their perturbative properties could provide deeper insight into the framework.

In the context of asymptotic safety, it would also be important to explore more general and realistic collapse scenarios, going beyond the simplified homogeneous dust models considered so far. Investigating different matter configurations and their impact on the resulting spacetime may help clarify the robustness of regular black hole formation and the possible end states of these natural extensions of General Relativity.

\appendix
\chapter{Rezzolla--Zhidenko parametrization in quadratic gravity} \label{appendix}

In this appendix we report the parametrized black-hole metric obtained in quadratic gravity using the semi-analytic expansion, as described in section \ref{NM}, and introduced by Rezzolla and Zhidenko \cite{Rezzolla:2014mua}, with coefficients computed in \cite{Kokkotas:2017zwt}.
The coefficients entering the parametrization are given by
\begin{equation}
    \begin{split}
\epsilon &= (1054 - 1203 p)\left(\frac{3}{1271} + \frac{p}{1529}\right), \\
a_0 &= 0, \\
a_1 &= (1054 - 1203 p)\left(\frac{7}{1746} - \frac{5p}{2421}\right), \\
a_2 &= -\frac{131}{102} + \frac{5p}{6} + \frac{6p^2}{17}, \\
a_3 &= \frac{\frac{4857}{29} - 385 p + \frac{9921}{31} p^2}{237 - 223 p}, \\
a_4 &= \frac{-\frac{2803}{14} + \frac{3149}{42} p + \frac{9}{14} p^2}{237 - 223 p}, \\
b_0 &= 0, \\
b_1 &= (1054 - 1203 p)\left(-\frac{2}{1585} + \frac{p}{1465}\right), \\
b_2 &= -\frac{16}{89} - \frac{109}{118} p + \frac{81}{242} p^2, \\
b_3 &= \frac{29}{56} - \frac{2}{57} p, \\
b_4 &= -\frac{121}{98} + \frac{13}{95} p.
\end{split}
\end{equation}

\section{Metric functions}

Moreover, the black-hole metric can be written as
\begin{equation}
ds^2 = -\left(1 - \frac{r_0}{r}\right) A(r)\, dt^2 
+ \frac{B(r)^2}{\left(1 - \frac{r_0}{r}\right) A(r)}\, dr^2 
+ r^2 (d\theta^2 + \sin^2\theta\, d\phi^2).
\end{equation}
where the functions $A(r)$ and $B(r)$ are given by rational functions of $r, p$ and $r_0$:
\begin{equation}
A(r) = \frac{\mathcal{P}_A(r)}{\mathcal{Q}_A(r)}, 
\qquad
B(r) = \frac{\mathcal{P}_B(r)}{\mathcal{Q}_B(r)},
\end{equation}
where $\mathcal{P}_A, \mathcal{Q}_A, \mathcal{P}_B, \mathcal{Q}_B$ are polynomials in $r$ and $r_0$ whose coefficients depend on the parameter $p$. The explicit form of the polynomials is lengthy and can be found in Appendix A of \cite{Kokkotas:2017zwt}.

The limiting value for 
\[
p \simeq \frac{1054}{1203} \approx 0.876
\]
corresponds to the merger between Schwarzschild and non-Schwarzschild branches, while for the maximum value $p \simeq1.14$ the black-hole mass approaches zero $\epsilon=-1$.

\addcontentsline{toc}{chapter}{\iflanguage{italian}{Bibliografia}{Bibliography}}
\bibliographystyle{JHEP}
\bibliography{biblio}

\chapter*{Acknowledgments}

I would like to thank my supervisor, Alfio Bonanno, for his guidance throughout these years and for having passed on to me the passion for research that I hope to carry forward in my future career.

I would also like to thank Roman Konoplya and Sasha Zhidenko for their guidance during my time in Opava, for generously sharing their knowledge and expertise, and for welcoming me with great kindness and availability.

I am deeply grateful to Samuele Silveravalle for his support, patience, and kindness, and for the fruitful collaboration we have shared throughout our research activities.

Finally, my deepest thanks go to my parents, my brother, and all my friends and colleagues, whose support, encouragement, love and friendship have accompanied me throughout this journey.

\end{document}